\documentclass[preprint,12pt,numafflabel]{elsarticle}

\usepackage{amssymb}
\usepackage{amsmath}

\usepackage{graphicx} %
\RequirePackage{amsmath}
\usepackage{mathtools} 
\usepackage{revsymb}
\usepackage{slashed}

\journal{Nuclear Physics B}

\begin{document}

\begin{frontmatter}



\title{A Fully Spin and Polarization Resolved Strong Field QED Algorithm for Particle-in-Cell Codes} 


\author[1]{Q. Qian}
\author[2,3,4]{D. Seipt}
\author[5]{M. Vranic}
\author[5]{T. Grismayer} 
\author[6]{C.P. Ridgers}
\author[1]{A. G. R. Thomas}
\affiliation[1]
{organization={G\'{e}rard Mourou Center for Ultrafast Optical Science, University of Michigan},
            addressline={2200 Bonisteel Boulevard}, 
            city={Ann Arbor},
            postcode={48109}, 
            state={Michigan },
            country={USA}}

\affiliation[2]{organization= {Helmholtz Institut Jena},  addressline={Fröbelstieg 3},  city={Jena}, postcode={07743},  country={Germany}}

\affiliation[3]{organization= {GSI Helmholtzzentrum für Schwerionenforschung GmbH}, addressline={Planckstraße 1}, city={Darmstadt}, postcode={64291}, country={Germany}}

\affiliation[4]{organization= {Institute of Optics and Quantum Electronics}, addressline={ Max-Wien-Platz 1}, city={Jena}, postcode={07743}, country={Germany}}

\affiliation[5]
{organization={GoLP/Instituto de Plasmas e Fusão Nuclear, Instituto Superior Técnico}, city={Lisbon}, addressline={Universidade de Lisboa}, postcode={1049-001},   country={Portugal} }

\affiliation[6]{organization={York Plasma Institute, Department of Physics} , city={York}, addressline={University of York}, postcode={YO10 5DD}, country={United Kingdom}}

\begin{abstract}
Modern ultra-intense laser facilities can generate electromagnetic fields strong enough to accelerate particles to near-light speeds over micron-scale distances and also approach the QED critical field, resulting in highly nonlinear and relativistic quantum phenomena. For such conditions, ab-initio modeling techniques are required that capture the electromagnetic, relativistic particle, and quantum emission processes in the plasma. One such technique is particle-in-cell (PIC) simulation. In this paper, we describe the underlying theory for and development, validation, and verification of an extension to standard QED-PIC in the OSIRIS framework to include spin- and polarization-resolved QED  processes central to next-generation laser-plasma experiments. This code can advance the current understanding of spin- and polarization-dependent QED phenomena in ultra-intense laser-plasma interactions.
\end{abstract}





\end{frontmatter}


\clearpage
\tableofcontents
\section{Introduction}

Current petawatt and even ten petawatt class lasers constructed worldwide are approaching the regime having peak intensity $I > 10^{23}\;\text{Wcm}^{-2}$, corresponding to normalized laser intensity $a_0 > 100$, where strong field quantum electrodynamics (SFQED) effects can appear. The SFQED processes occur when electromagnetic fields exceed the quantum critical field strength $E_{cr}$ \cite{Ritus_1985}. Even though the most intense laser pulse nowadays cannot reach $E_{cr}$  in the laboratory by many orders of magnitude, SFQED effects necessarily depend on Lorentz-invariant parameters, including 
$\chi_e\sim E_{RF} / E_{cr}$
which is equivalent to the magnitude of the electric field in the rest frame of an electron, where $p^\nu$ is the four-momentum of the electron and $F^{\mu\nu}$ is the electromagnetic tensor. $E_{cr}$ is the QED critical field, or the Schwinger field, which, for a static electric field, would lead to the breakdown of the vacuum. Due to the coupling of relativistic dynamics and strong electromagnetic fields, the SFQED regime may become accessible at lower intensities when relativistic plasma flows interact with the laser fields.  In this region, the electrons will be accelerated so strongly that their loss of energy to radiation must be taken into account in the equations of motion. Furthermore, the photons of that radiation will be sufficiently energetic that they can produce electron-positron pairs by colliding with photons of the laser pulse. Even if pair production is not dynamically significant, it is necessary to include quantum corrections to the radiation damping, as classical electromagnetism predicts the emission of photons with energy greater than that of the electron and so overestimates the emissivity.




SFQED effects can play a crucial role in the dynamics of relativistic plasmas. Two fundamental quantum processes—photon emission and electron-positron pair production—significantly impact plasma behavior. When a charged particle emits an energetic photon, the radiation back-reaction alters its trajectory and energy. Conversely, the absorption of high-energy photons via pair production increases the plasma density by generating electron-positron pairs. In environments where electromagnetic fields are sufficiently strong and the plasma is highly energetic, these quantum effects can continuously modify basic plasma parameters, ultimately influencing the collective behavior of the system. As a result, the plasma dynamics may diverge sharply from classical expectations. Such systems are often referred to in the literature as ``QED plasmas'' \cite{Zhang_POP, melrose_2008_quantum, melrose_2013_quantum, Uzdensky_2014_RPP, Uzdensky_2019_arXiv}.

Research in SFQED has reached an advanced stage: theoretical frameworks and simulation tools for studying particle dynamics in strong fields are now well developed \cite{Elkina_11, Ridgers_14, Gonoskov_15, Grismayer_POP_2016}, and experimental investigations in this regime are emerging \cite{Ta_NP_2012, Sarri_PRL_2014, Poder_PRX_2018, Cole_PRX_2018, Pirozhkov_2024}. This progress motivates a critical reassessment of the standard SFQED models, which typically rely on specific approximations. In particular, traditional SFQED calculations often assume unpolarized particles by averaging (summing) over initial (final) spin states \cite{GonoskovA_RevModphy_2022}. In this work, we relax this approximation to examine how the spin and polarization of particles evolve within intense fields.

Spin is an intrinsic form of angular momentum present in elementary particles, composite particles (such as hadrons), and atomic nuclei \cite{Griffiths_2018_introduction}. Unlike classical angular momentum—which arises from actual spinning or rotation—quantum spin is a fundamental property of particles. It does not correspond to any physical spinning, but is instead a purely quantum mechanical phenomenon with no classical equivalent. A pivotal moment came with the 1922 Stern-Gerlach experiment \cite{Stern_1922_spin}, which demonstrated that electrons and other particles exhibit a quantized form of angular momentum that cannot be explained by orbital motion alone. This experiment exposed the inherently quantum nature of spin. Dirac’s relativistic equation \cite{Dirac_1930_QM} for the electron later provided a theoretical foundation for spin, naturally incorporating it into the quantum description and, importantly, predicting both electron spin and the existence of antimatter.

The QED theory, developed by Dirac and others \cite{Dirac_1930_QM}, also formalizes the quantum description of photons and their polarization—a key intrinsic property. An individual photon exhibits polarization, which can be described in terms of right- or left-handed circular polarization, or as a superposition of both. Alternatively, photon polarization can be expressed as horizontal or vertical linear polarization, or any superposition of both.

The fundamental processes of photon emission and pair creation are intrinsically dependent on the spin and polarization states of the participating particles. Consequently, these processes can shape the distribution of spin and polarization in the plasma, leading to the generation of polarized particle beams. Such polarized beams are not only scientifically interesting in their own right, but may also serve as powerful experimental probes—offering new diagnostics for the structure and properties of strong-field environments. The polarization state of the particles will also influence the SFQED processes, thereby modifying the rates of photon emission and pair creation. 

Spin and polarization-resolved SFQED has recently received increased interest. Novel methods for generating highly polarized lepton and gamma ray beams on a femtosecond timescale and under a compact full-optics setup are proposed \cite{Del_Sorbo_17, Li_PRL_2019, Seipt_PRA_2019}. Spin-polarized lepton beams are crucial tools in contemporary physics. They are used extensively to probe the properties of matter, reveal atomic and molecular structures \cite{kessler_1985_polarized, gradmann_1991_surface}, and, in the relativistic domain, study nuclear structure \cite{Abe_PRL_1995, Alexakhin_PLB_2007}, generate polarized photons \cite{Olsen_PR_1959, Martin_PRL_2012} and positrons \cite{Olsen_PR_1959, Abbott_PRL_2016}, investigate parity violation \cite{Anthony_PRL_2004}, and explore phenomena beyond the Standard Model \cite{Moortgat_PR_2008}. In conventional synchrotron facilities, this polarization process is slow due to the relatively weak magnetic fields (around 1 Tesla). In the context of high-intensity laser-plasma interaction, as the laser field strength can be many orders of magnitude higher than that in synchrotrons, enabling polarization timescales on the order of a few femtoseconds, comparable to the duration of high-intensity, ultrashort laser pulses.

Walking out of the laser lab and looking up to the sky, the light from the star is like a messenger coming to us with a story waiting to be interpreted. The scientific potential of polarized high-energy X-ray signals has long been recognized by the astrophysics community \cite{Chattopadhyay_JAA_2021}. It has been used to understand the magnetic configuration near the black holes \cite{Akiyama_APJL_2021}, crab nebulae \cite{Bucciantini_NatureAstro_2023}, and to analyze the particle acceleration in blazars’ jets \cite{Liodakis_Nature_2022}. Measuring the polarization of high-energy photons is much more challenging, but significant progress has been made recently \cite{Bernard_GammaPolarimetry_2022, Tomsick_COSI_2023, Kole_POLAR2_2019}. The emission mechanism of this high-energy gamma-ray could be relevant to the SF QED process; polarization-resolved QED simulation tools could help explore the origin of the polarization signal in this new energy regime \cite{Gong_PRL_astro_2023}. 

The inclusion of spin and polarization effects is also important in the context of QED plasma. The studies have shown that the spin/polarization-distinguished QED code can more accurately simulate multi-staged processes, such as avalanche and shower-type electron-positron pair production cascades \cite{Seipt_PRA_2020, Seipt_NJP_2021, King_PRA_2013}. The spin and polarization signal can also serve as an indicator of some extreme plasma processes like ultrarelativistic plasma current filamentation instabilities \cite{Gong_PRL_2023}, as well as radiation reaction-dominated magnetic reconnection, which may be realized in extreme power laser facilities and intrinsically exist in extreme astrophysical reconnection scenarios  \cite{Gong_PRL_2025}. 

In this paper, we discuss in detail the implementation of our fully spin- and polarization-resolved SFQED processes within a Monte Carlo algorithm, which is integrated into the OSIRIS particle-in-cell (PIC) code platform \cite{Fonseca_Osiris_note, Fonseca_Osiris_PPCF}. The algorithm is rigorously verified under various conditions and thoroughly validated by reproducing notable benchmark results.

\section{{Inclusion of Spin and Polarization in SFQED}}

The SFQED emission rates depend on the spin of the electrons and positrons, as well as on the polarization of the photons, as discussed by Ilderton, King, and Tang (2020) \cite{Ilderton_2020_PRD} and Seipt and King (2020) \cite{Seipt_PRA_2020}. This can result in the generation of polarized electron, positron, and photon beams [see Chen et al. (2019) \cite{Chen_PRL_2019}, Seipt et al. (2019 )\cite{Seipt_PRA_2019}, King and Tang (2020) \cite{King_PRA_2020}, and Tang, King, and Hu (2020)\cite{Tang_PLB_2020}, which can be important for high-energy physics applications. We begin with the definition of basis for spin and polarization proposed by G. Torgrimsson \cite{Torgrimsson_NJP_2021}, as well as Seipt and King \cite{Seipt_PRA_2020}. Under the high-intensity laser plasma interaction scenario, a convenient basis for transverse photon polarization vectors is given as:

\begin{equation}
\epsilon^\mu_{1} =e_1^{\mu}  - \frac{\kappa\cdot e_1}{\kappa \cdot k } k,\ \ \ \ \epsilon^\mu_{2} =e_2^{\mu} - \frac{\kappa\cdot e_2}{\kappa \cdot k } k
\label{eq:pol_basis_4v}
\end{equation}

Which $e_1$ and $e_2$ are two space-like unit vectors in the transverse plane, $\kappa$ is the laser four-wave vector. $e_1\cdot k = 0$, $ e_2\cdot k = 0$. The vectors $\epsilon_j$ for j = 1, 2 are not only mutually orthogonal, but also fulfill $k\cdot \epsilon_j = \kappa\cdot \epsilon_j = 0$. If the photon is linearly polarized, one calls $j = 1 (j = 2)$ the $E\ (B)$ polarization. For a circularly polarized laser, one conveniently also defines the complex
left/right-handed basis vectors $\epsilon_{\pm} = (\epsilon_{1}\pm i\epsilon_{2})/\sqrt{2}$

The spin 4-vector $\alpha^{\mu}$ of an electron in a pure state is related to the spinors
\begin{equation}
u\bar u = \frac{1}{2}(\slashed{p}+m)(1+\gamma^5\slashed{\alpha})
\end{equation}
Where $\alpha _{\mu}$ refers to the asymptotic spin state of the particle (with $\alpha^2 = -1$,  $p\cdot \alpha = 0$). A convenient spin basis was given by \cite{Dinu_PRD_2019, Seipt_PRA_2018, Dinu_PRD_2020, Torgrimsson_NJP_2021}:

\begin{equation}
\alpha^\mu_{(1)} = e_{1}^{\mu} -  \frac{p\cdot \varepsilon}{p \cdot k } k^{\mu},\ \ \ \ \alpha^\mu_{(2)} = e_2^{\mu} -  \frac{p\cdot \beta}{p \cdot k } k^{\mu},\ \ \ \ \alpha^\mu_{(3)} = p^{\mu} -  \frac{1}{p \cdot k } k^{\mu},
\label{eq:spin_basis_4v}
\end{equation}

$\alpha_{(i)}\cdot \alpha_{(j)}=\delta_{ij}$. The third vector $\alpha_{(3)}$ is called the lightfront helicity vector. The vectors $e_1$, $e_2$ are mutually orthogonal and perpendicular to $k_{\mu}$. The polarisation vector $\alpha^{\mu}$ of a partially polarised electron in a mixed state, $-1<\alpha^2\leq 0$ can be expanded in the basis as

\begin{equation}
\alpha^{\mu} = \sum_{i}n_i\alpha_{(i)}^{\mu}
\end{equation}

which defines the Stokes vector for electrons $\pmb{n}=(n_1,n_2,n_3)$. For an electron at rest, $\pmb{n}$ points in the spin direction \cite{Dinu_PRD_2019, Dinu_PRD_2020, Torgrimsson_NJP_2021},
\begin{equation}
\pmb{n} = \frac{1}{2}u^{\dagger}\Sigma u\ \ (\text{when }\pmb{p}=0)
\end{equation}
where $\Sigma=\textbf{I}\{\gamma^2\gamma^3,\  \gamma^3\gamma^1,\  \gamma^1\gamma^2 \}$, which $\textbf{I}$ is the identity marix. A fully polarised electron has $\pmb{n}^2 = 1$, an unpolarised has $\pmb{n} = \pmb{0}$, and $0 < \pmb{n}^2 < 1$ represents a partially polarised state.  

The probability of spin and polarization dependent nonlinear Compton scattering and nonlinear Breit–Wheeler pair production can be expressed as \cite{Torgrimsson_NJP_2021}:

\begin{equation}
\begin{split}
{\mathbb{P}} = & \langle {\mathbb{P}} \rangle + \textbf{n}_{\gamma}\cdot \textbf{P}_{\gamma} + \textbf{n}_{1}\cdot \textbf{P}_{1} + \textbf{n}_{0}\cdot \textbf{P}_{0} + \textbf{n}_{\gamma}\cdot \textbf{P}_{\gamma 1} \cdot \textbf{n}_{1} + \textbf{n}_{\gamma}\cdot \textbf{P}_{\gamma 0} \cdot \textbf{n}_{0}\\ & + \textbf{n}_{1}\cdot \textbf{P}_{1 0} \cdot \textbf{n}_{0}  + \textbf{P}_{\gamma 1 0, ijk} \textbf{n}_{\gamma i}  \textbf{n}_{1 j}  \textbf{n}_{0 k},
\end{split}
\end{equation}

where $\pmb{n}_\gamma$ is the Stokes vector for the photon, and $n_{1,0}$ are the Stokes vectors for the fermions. $\langle {\mathbb{P}}\rangle$ gives the spin and polarization averaged probability, $\textbf{P}_{i}$ gives the dependence on the spin or polarization of one particle when averaging over the spin or polarization of the other two particles, $\textbf{P}_{ij}$ describes the correlation between the spin or polarization of two particles, and $\textbf{P}_{ijk}$ describes the correlation between the spin and polarization of all three particles.

In \cite{Dinu_PRD_2020, Torgrimsson_NJP_2021}, the authors present a gluing approach, where higher-order processes are approximated by linking together the spin- and polarization-dependent probabilities of nonlinear Compton scattering and Breit-Wheeler pair production. They demonstrate that this approach works under the regime where the formation length is small, so the field can be treated as approximately constant during particle production. This is the basis of particle-in-cell (PIC) codes. They presented general results for the Mueller matrices of nonlinear Compton scattering and Breit–Wheeler pair production, in this method, the first-order probabilities can be expressed as:
\begin{equation}
{\mathbb{P}} = \textbf{M}_{ijk}\textbf{N}_{1i}\textbf{N}_{0j}\textbf{N}_{jk}
\end{equation}

where M is a $ 4\times4\times4$ matrix and $i, j, k = 1,\ \cdots,\ 4$. $N = \{1, \pmb{n}\}$ is the 4D Stokes vector. The Stokes vectors are used in conjunction with strong-field QED Mueller matrices for treating spin sums in higher-order processes. 
 The ‘N-step’ can be obtained by matrix multiplication. For example, if a photon is emitted at step $m$ and decays at step $n$, then the sum over its polarization is included via $\textbf{M}_{i_m j_m k}^{(m)}\textbf{M}_{i_n j_n k}^{(n)}$. The detailed discussion of this method can be found in \cite{Dinu_PRD_2020, Torgrimsson_NJP_2021}.

\section{On the Lepton Spin and Photon Polarization Basis}
\label{spin_basis}

In the quantum calculation of our QED algorithm, the particle spin or polarization state will be projected onto a spin basis. In other words, the code measures the particle spin or polarization state after every quantum event. The choice of spin and polarization basis is crucial to retaining important information about particle polarization after the projection. There are various ways to construct the spin and polarization basis. For example, in some literature \cite{BKS, Chen_PRD_2022, Li_PRL_2020_helicity_transfer}, the particle acceleration direction $ \pmb {\hat a}$, velocity direction  $ \pmb {\hat v}$, and their cross-project $\pmb {\hat v} \times \pmb {\hat a}$ is used to construct the basis. In the SF QED model under the local constant field approximation (LCFA),the direction of the electric and magnetic fields in the particle's rest frame and the momentum vector should be close to mutually perpendicular. Here, for consistency with the SF QED framework, this inherently orthogonal system is used to construct a spin and polarization basis  (as the spin is also evaluated in the rest frame of the particle, this provides another reason to use the rest frame EM field direction to construct the basis). Because the PIC code uses three-vector objects, we express the bases in three-vector form instead of four-vector form. Note that in Eq.~\ref{eq:spin_basis_4v}, if we set $e_1^{\mu} = (0,\hat{\pmb{\varepsilon}}_0)$, $e_2^{\mu} = (0,\hat{\pmb{\beta}}_0)$, where $\hat{\pmb{\varepsilon}}_0$ and $\hat{\pmb{\beta}}_0$ are electric field the magnetic field direction in the lab frame, when we transform the basis 
to the rest frame of an electron $\alpha_{(i)}\xrightarrow{\Lambda(-{p})}\alpha_{(i)}^{RF}$, the time-like component will be zero, and we get three mutually orthogonal three-vectors ($\hat{\pmb \varepsilon}$, ${\hat{\pmb \beta}}$, $\hat{\pmb k}$). They are exactly the direction of the electric, magnetic, and wave vectors of the background field in the rest frame of the particle \cite{Seipt_PRA_2018}. 

\begin{equation}
\alpha^{RF}_{(1)} = (0,\hat{\pmb{\varepsilon}}),\ \ \ \ \alpha^{RF}_{(2)} = (0,\hat{\pmb{\beta}}),\ \ \ \ \alpha^{RF}_{(3)} = (0,\hat{\pmb{k}})\;,
\end{equation}
where ${\pmb{\hat\varepsilon}} = {{\pmb{E}_{RF}}}/|{\pmb{E}_{RF}|}$, $\pmb{\hat\beta} = {\pmb{B}_{RF}}/{|\pmb{B}_{RF}|}$, $\hat{\pmb k}$ is the lepton momentum direction, which under the collinear emission approximation, is also the direction of the radiated photon momentum. As a result, we choose the three vectors ($\hat{\pmb \varepsilon}$, ${\hat{\pmb \beta}}$, $\hat{\pmb k}$) to construct the spin basis for our algorithm. 

\subsection{Lepton Spin Basis}

{For leptons, the logical choice for the spin quantization axis is} along the rest frame B-field direction $\hat{\pmb\beta}$ and only considers the spin component along $\hat{\pmb\beta}$ \cite{Seipt_PRA_2020}. This is because {this is the non-precessing component for the rest frame spin vector $\pmb{s}$. Furthermore,} radiative processes only increase the spin component in the $\hat{\pmb \beta}$ direction and only cause the $\hat{\pmb \varepsilon}$ and $\hat{\pmb \kappa}$ components to decay away. This basis is suitable for calculating an unpolarized lepton beam interacting with a constant magnetic field or a linearly polarized plane wave. We need to resolve all three spin components during a quantum process for more complicated field configurations or a polarized lepton beam. Seipt et al. \cite{Seipt_PRA_2018} calculated radiation-induced spin changes for all three spin components along  ($\hat{\pmb \varepsilon}$, $\hat{\pmb \beta}$, $\hat{\pmb k}$). Yuhui Tang \cite{Tang_PRA_2021} introduces the variation of radiative spin polarization due to instantaneous no-photon emission and proposes a recipe to implement this radiation-induced spin change into the Monte-Carlo quantum radiation algorithm. Cain's manual \cite{Chen_1995_CAIN} also offers a recipe for constructing a spin basis. They use transverse acceleration direction $\pmb {\hat e_1} = \pmb a -  (\pmb a \cdot \pmb v)\pmb v$  particle velocity direction $\pmb {\hat e_v} = \pmb {\hat v}$ and their cross product $\pmb {\hat e_2} =  \pmb {\hat v} \times \pmb {\hat e_1}$. It is easy to prove that under ultra-relativistic approximation, the basis used in Cain's manual is essentially the same as the basis ($\hat{\pmb \varepsilon}$, ${\hat{\pmb \beta}}$, $\hat{\pmb k}$).

From the spin- and polarization-resolved spectrum, we can calculate the probability of spin state in the $\hat{\pmb \varepsilon}$, ${\hat{\pmb \beta}}$, or $\hat{\pmb k}$ direction after every quantum process. The QED-PIC code requires the use of a vector quantity to represent the spin state for the classical spin precession calculation. The expected spin vector $\langle \pmb{s}_f\rangle$ becomes the best candidate to keep most of the spin information. In some literature, the unit vector along this direction is called the spin quantization axis \cite{Chen_PRD_2022, Li_PRL_2020_helicity_transfer}. It is constructed from the expected radiation-induced spin change for all three spin components  ($\hat{\pmb \varepsilon}$, ${\hat{\pmb \beta}}$, $\hat{\pmb k}$): $\langle \pmb{s}_f\rangle  = \langle {S}_{E}\rangle \hat{\pmb \varepsilon} + \langle {S}_B\rangle \hat{\pmb \beta} + \langle {S}_K\rangle \hat{\pmb k}$ (Fig.~\ref{fig:spin_measurement} ). The detailed derivation of the expected spin direction for both NLC and NBW processes is shown in Appendix~\ref{appA}. The expression for the expected spin direction of an electron in the NLC processes is:

\begin{figure}[h!]
\includegraphics[width=0.5\textwidth]{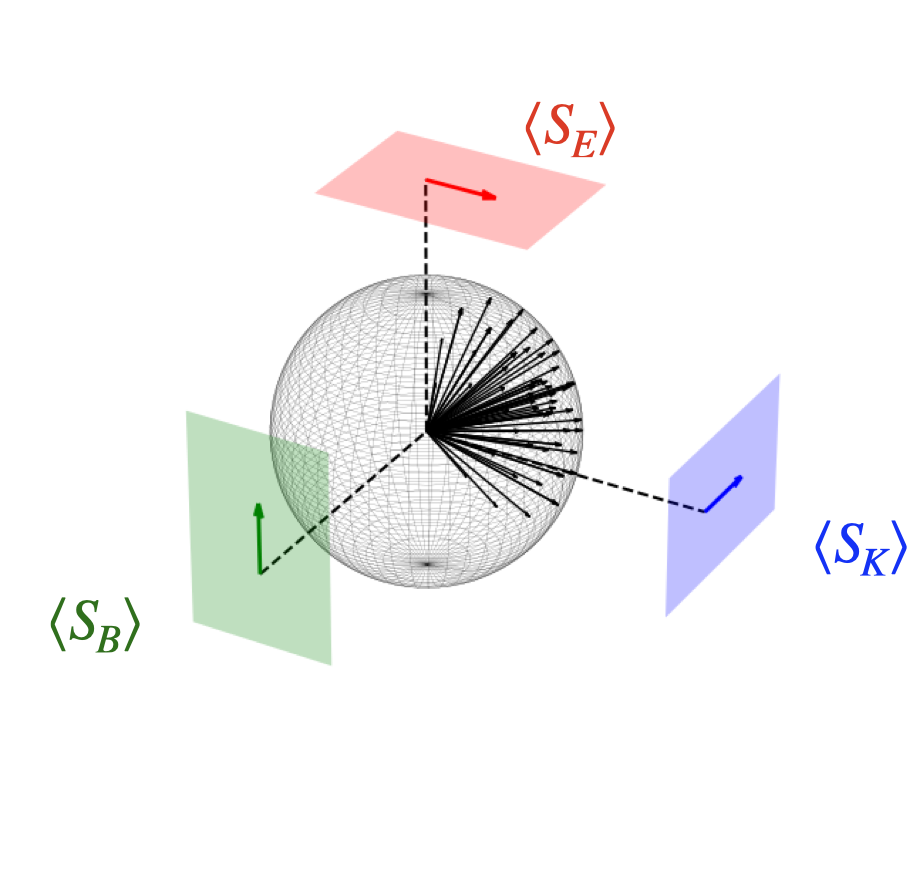}
\centering
\caption{Schemetic of spin measurement under the basis ($\pmb{\hat \varepsilon}$, ${\pmb{\hat \beta}}$, $\pmb{\hat k}$)}
\label{fig:spin_measurement}
\end{figure}

\begin{equation}
 \langle \pmb{s}_f\rangle_{NLC} = \frac{\pmb{g}}{w_0}
\label{eq:SQA_NLC}
\end{equation}

which

\begin{equation}
    w =-\text{IntK}_{1/3}(u')+\kappa\text{K}_{2/3}(u')+\lambda(\textbf{s}_i\cdot \hat{\pmb{\beta}})\text{K}_{1/3}(u'),
\end{equation}

\begin{equation}
\begin{split}
    \textbf{g} = &u \text{K}_{1/3}(u') \hat{\pmb{\beta}} + \left(-\text{IntK}_{1/3}(u') + 2\text{K}_{2/3}(u')\right)\textbf{s}_i  \\ &  + \lambda u\left(-\text{IntK}_{1/3}(u') + \text{K}_{2/3}(u')\right)(\textbf{s}_i\cdot\hat{\textbf{k}}) \hat{\textbf{k}}
\end{split}
\end{equation}

Here, $\lambda$ is the momentum ratio between the radiated photon and the initial energy of the electron, $u = \lambda/(1-\lambda)$ is the momentum ratio between the radiated photon and the final energy of the electron. $\kappa = 1/(1-\lambda)+ (1-\lambda)$, $\tilde \kappa =  1/(1-\lambda) - (1-\lambda)$, $u' = \frac{2u}{3\chi_{e}}$. 

The expected spin direction for positron in the NBW process is:

\begin{equation}
 \langle \pmb{s}_p\rangle_{NBW}=\frac{\pmb{j}_p}{w_0}
\label{eq:SQA_NBW}
\end{equation}

which:
\begin{equation}
w_0 =  \text{IntK}_{1/3}(\tilde u')+\rho\text{K}_{2/3}(\tilde u') - \xi_3\text{K}_{2/3}(\tilde u')
\end{equation}

\begin{equation}
\begin{split}
\pmb{j}_p = & \left(-\frac{1}{\tilde \lambda}+\xi_3\frac{1}{1-\tilde\lambda}\right)\text{K}_{1/3}(\tilde u')\hat{\pmb{\beta}}-\xi_1\frac{1}{1-\tilde \lambda}\text{K}_{1/3}(\tilde u')\hat{\pmb{\varepsilon}} \\ &
 +\xi_2\left( \frac{1}{\tilde\lambda}\text{IntK}_{1/3}(\tilde u')-\tilde\rho\text{K}_{2/3}(\tilde u')\right)\hat{\pmb{k}}
 \end{split}
\end{equation}

Here, $\tilde\lambda$ are the momentum ratios between the positron and the initial photon. $\rho =  (1-\tilde\lambda)/\tilde\lambda+\tilde\lambda/(1-\tilde\lambda) $, $\tilde \rho = (1-\tilde\lambda)/\tilde\lambda -\tilde\lambda/(1-\tilde\lambda)$, $\tilde u = 1/\tilde\lambda+1/(1-\tilde\lambda)$ and $\tilde u' = \frac{2}{3\chi_{\gamma}}\tilde u$. 


\subsection{Photon Polarization Basis}
\label{chpt:pol_vec}

Describing the polarization of the emitted photons requires choosing a basis for this polarization. The photon polarization state can be represented as
$\hat{\pmb \psi} = a_1 \hat{\pmb{\psi}}_1 + a_2 \hat{\pmb{\psi}}_2$. 

\begin{figure}[h!]
\includegraphics[width=0.5\textwidth]{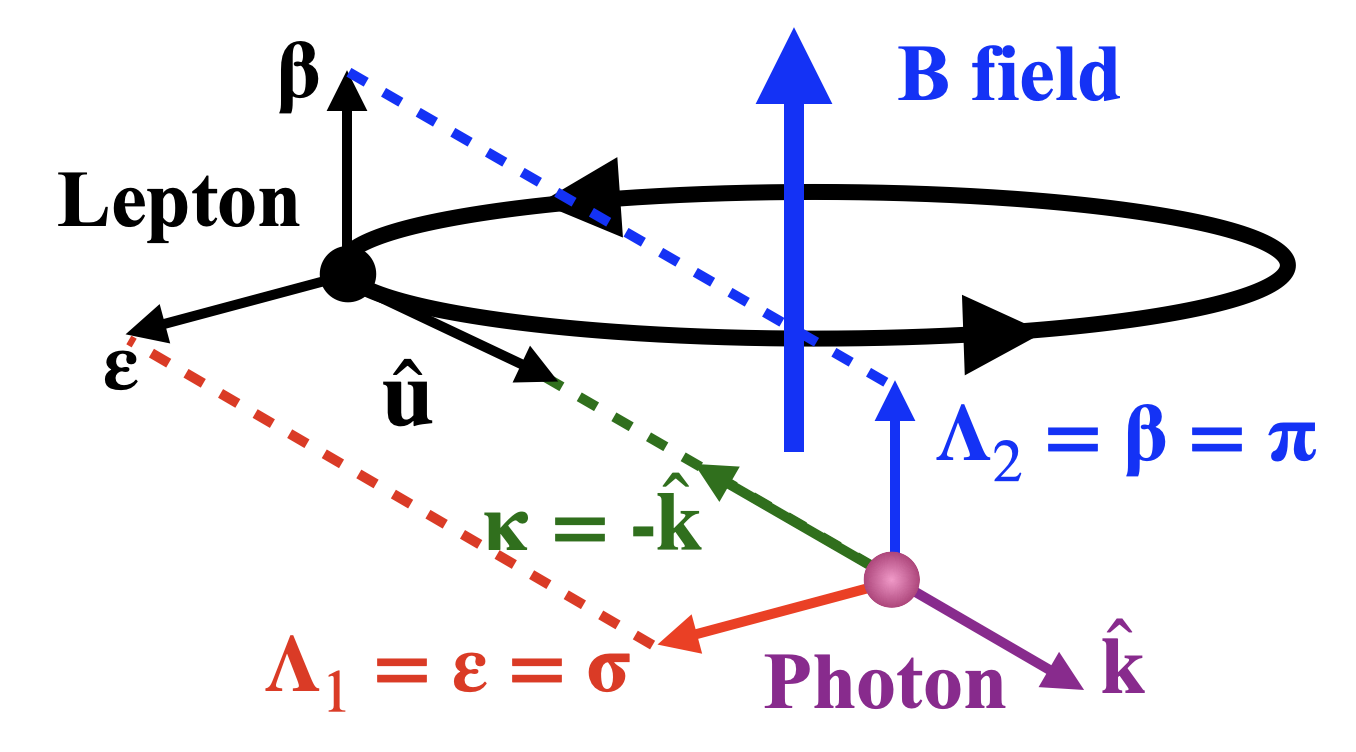}
\centering
\caption{Photon polarization basis. A high-energy lepton with momentum in direction $\pmb{\hat u} $ emits a photon whose momentum direction $\pmb{\hat k} = \pmb{\hat u}$. The polarization basis vector for the photon is $\pmb {\hat\varepsilon}$, $\pmb{\hat\beta}$. When $\pmb {\hat\kappa} = -\pmb {\hat k}$, $\pmb\sigma$ and $\pmb\pi$ are in the same direction as $\pmb{\hat\varepsilon}$ and $\pmb{\hat\beta}$.}
\label{fig:synchrotron}
\end{figure}

When the photon is born, $\hat{\pmb{\psi}}_1 = \hat{\pmb \varepsilon} $ and $\hat{\pmb{\psi}}_2 =\hat{\pmb \beta}$ are the same as the directions of the rest frame electric and magnetic field experienced by the incoming lepton, respectively. The components of the Stokes vector $\pmb{\xi} = (\xi_1,\xi_2,\xi_3)$ corresponding to this basis are  $\xi_1 = a_1a_2^* + a_1^*a_2$, $\xi_2 = i(a_1a_2^*-a_1^*a_2)$, $\xi_3 = |a_1|^2-|a_2|^2$. Fig.~\ref{fig:synchrotron} demonstrates the polarization basis under the simple synchrotron radiation geometry. An electron gyrating inside a constant magnetic field will undergo synchrotron radiation. The two polarization directions $\pmb\sigma$ and $\pmb\pi$ generally used in synchrotron radiation are the same as the directions $\pmb{\hat\varepsilon}$ and $\pmb{\hat\beta}$.  
 Similar to how the lepton spin basis is constructed, the photon-polarization Stokes vector is set to the  expected photon Stokes vector: $\pmb \xi_{ave} = (\langle\xi_1\rangle, \langle\xi_2\rangle, \langle\xi_3\rangle)$. The detailed expression for the expected photon Stokes vector is given in the appendix \ref{appA}, in equation \ref{eq:NLC_stokes_F1}, \ref{eq:NLC_stokes_F2}, and \ref{eq:NLC_stokes_F3}. 
 
For observation, however, the basis will be different because of the different field configurations with which the photon interacts. The observation basis, in general, can be expressed as: 
\begin{equation} 
\hat{\pmb  o}_1 = \hat{\pmb \psi}_1\cos(\phi) + \hat{\pmb \psi}_2\sin(\phi),\  
\hat{\pmb  o}_2 = - \hat{\pmb \psi}_1\sin(\phi) + \hat{\pmb \psi}_2\cos(\phi)
\end{equation}
The Stokes parameter under the new observation basis $\pmb\xi^{(o)} $ is then:
\begin{equation} 
\xi_1^{(o)}= \xi_1\cos(2\phi) - \xi_3\sin(2\phi),\ 
\xi_2^{(o)} = \xi_2,\  
\xi_3^{(o)} = \xi_1\sin(2\phi) +\xi_3\cos(2\phi)
\label{eq:observe_stokes_parameter}
\end{equation}

In the actual code implementation, we can obtain the Stokes parameter when the photon is born from our polarisation-resolved NLCS rate calculation. However, we must translate the Stokes parameter from the frame when it is born to the observation frame when the photon interacts with other EM fields. This means that besides information about the Stoke parameters, we also need to record information about the basis when a photon is born. We want to reduce the number of parameters required to record the full information of the photon polarisation state to reduce the computational memory cost. Here, we find that it is possible to use one vector quantity to represent the polarisation state and record all information about the photon polarisation state.: $\pmb{P} = (P_{\psi_1}$, $P_{\psi_2}$, $P_{{k}})$, where $P_{\psi_1}$, $P_{\psi_2}$, $P_{{k}}$ is the component in the $({\hat{\pmb\psi}_1}, {\hat{\pmb\psi}_2}, {\hat{\pmb k}})$ coordinate. We begin by transforming the coordinate to the circular polarization basis:
\begin{equation}
	\pmb{\hat e_+} = \frac{1}{\sqrt{2}}(\pmb{\hat e_1} + i\pmb{\hat e_2}),\ 
	\pmb{\hat e_-} = \frac{1}{\sqrt{2}}(\pmb{\hat e_1} - i\pmb{\hat e_2})
\end{equation}

The photon polarization state can be rewritten under the new coordinate 
\begin{equation}
\psi = A_{+} \pmb{\hat e_+} +  A_{-} \pmb{\hat e_-}
\end{equation}

Which, $A_{+} = (a_1-ia_2)/\sqrt{2}$, $A_{-} = (a_1+ia_2)/\sqrt{2}$.The Stokes parameter can be represented, under the circular polarization basis, as:
\begin{equation}
	\xi_1 = (A_+^*A_- - A_+A_-^*)/i,\ 
	 \xi_2 = |A_+|^2 - |A_-|^2,\  
	\xi_3 = (A_+^*A_- + A_+A_-^*)
\label{eq:stokes_circ_base}
\end{equation}

We can simplify the expression by rewriting the complex parameter $A_+$ and $A_-$ as: $A_+ = |A_+|e^{i\theta_+}$, $A_- = |A_-|e^{i\theta_-}$, so that:
\begin{equation}
\xi_1 = 2|A_+||A_-|\frac{e^{i(\theta_- -\theta_+)}-e^{i(\theta_- -\theta_+)}}{2i}=2|A_+||A_-|\sin 2\Delta\theta
\end{equation}
Which $\Delta\theta = (\theta_- - \theta_+)/2$.  
Also, we can have
\begin{equation}
\xi_3 = 2|A_+||A_-|\frac{e^{i(\theta_- -\theta_+)}+e^{i(\theta_- -\theta_+)}}{2}=2|A_+||A_-|\cos 2\Delta\theta
\end{equation}

Since the intensity is normalized to unity: $|A_{+}|^2 + |A_{-}|^2 = 1$, and from Eq.~\ref{eq:stokes_circ_base}, $|A_{+}|^2 - |A_{-}|^2 = \xi_2$, we can obtain the magnitude of $A_{+}$ and $A_{-}$:
\begin{equation}
|A_{+}| = \sqrt{\frac{1+\xi_2}{2}},\ |A_{-}| = \sqrt{\frac{1-\xi_2}{2}}.
\end{equation}

The linear Stokes parameter can be written in terms of the circular Stokes parameter $\xi_2$ and a phase term $\Delta\theta$:
\begin{equation}
	\xi_1 = \sqrt{1-\xi_2^2}\sin 2\Delta\theta,\ 
	\xi_3 = \sqrt{1-\xi_2^2}\cos 2\Delta\theta
\end{equation}

We can then reconstruct a unit vector to represent the polarization state: $\pmb{P} = (P_{\psi_1}$, $P_{\psi_2}$, $P_{{k}})$. The component along the momentum direction is set to be the circularly polarized Stokes parameter $P_{{k}} = \xi_2$. We provide a geometric meaning for angle $\Delta\theta$, which is the angle between $\pmb P$ component in $\hat{\pmb \psi}_1 - \hat{\pmb \psi}_2$ plane: $\pmb{P}_{\psi_1-\psi_2}$ = ($P_{ \psi_1}$, $P_{ \psi_2}$, 0) and $\hat{\pmb \psi}_1$. The component $P_{ \psi_1}$ and $P_{ \psi_2}$ can be written as:  

\begin{equation}
	P_{\psi_1} = \sqrt{1-P_k^2}\cos(\Delta\theta) = \sqrt{1-\xi_2^2}\cos(\Delta\theta),\ 
	P_{\psi_2} = \sqrt{1-\xi_2^2}\sin(\Delta\theta)
\end{equation}

The relationship between $P_{\psi_1}$, $P_{\psi_2}$, $P_{k}$ and the Stokes parameter can be written as:
\begin{equation}
	\xi_1 = \frac{2P_{\psi_1} P_{\psi_2}}{\sqrt{1-P_{{k}}^2}},\ \ 
	\xi_2 = P_{{k}},\ \ \xi_3 = \frac{P_{\psi_1}^2 -  P_{\psi_2}^2}{\sqrt{1-P_{{k}}^2}}
\label{eq:pvector_stokes_relation}
\end{equation}

Under the observation frame, we can obtain the observed Stokes parameters by projecting the vector $\pmb P$ to the observation basis $(\hat{\pmb o}_1, \hat{\pmb o}_2, \hat{\pmb k})$.
\begin{equation}
 {P}_{{ o_1}} = \pmb{P} \cdot \hat{\pmb o}_1 =  {P}_{\psi_1} \cos(\phi) +   {P}_{\psi_2} \sin(\phi)
\end{equation}
\begin{equation}
P_{{o_2}} = \pmb{P} \cdot \hat{\pmb o}_2= -P_{\psi_1} \sin(\phi) + P_{\psi_2} \cos(\phi)
\end{equation}

Because the direction of the photon momentum does not change during propagation, $\xi_2^{(o)} = P_{{k}} = \xi_2$ remains unchanged. The Stokes parameter under the observation basis, calculated using Eq.~\ref{eq:pvector_stokes_relation} can be written as:

\begin{equation}
	\xi_{1}^{(o)} = -\sin(2\phi)\xi_3 + \cos(2\phi)\xi_1
, \ \ \xi_{3}^{(o)} = \cos(2\phi)\xi_3 + \sin(2\phi)\xi_1
\end{equation}

Which is the same as Eq.~\ref{eq:observe_stokes_parameter}. As a result, we can just use the reconstructed vector $\pmb{P}$ with the proper transformation rule to represent the full photon polarization states. Here, the photon polarization is measured on the basis of $(\hat{\pmb o}_2, \hat{\pmb o}_2, \hat{\pmb k})$, which $\hat{\pmb o}_1$ is the polarization direction of the linearly polarized laser, $\hat{\pmb{k}}$ is the photon momentum direction, $\hat{\pmb{o}}_2$ is the obtained by taking the cross product of $\hat{\pmb{o}}_1$ and $\hat{\pmb{k}}$. 

\section{Overview of QED-PIC Codes}

We start with a general overview of  QED-PIC codes before introducing our spin and polarization-resolved QED-PIC module. This well-developed method provides a solid foundation for our work. When estimating the scale length of quantum processes in high-intensity laser-plasma interactions, it is found that the formation length of QED processes can be significantly smaller than the typical inhomogeneities of the EM field. Thus, the calculation of quantum processes can be greatly simplified by treating them as occurring instantaneously, and the EM field strength can be viewed as a constant during these processes. The simulation of SFQED phenomena is generally based on the local constant field approximation (LCFA), which is discussed in detail in the previous section. The LCFA approximation enables the calculation of quantum processes as discrete events that occur with a certain probability, which allows the Monte Carlo method to be used to calculate the SFQED processes. This method accounts for the probabilistic nature of quantum processes. Integrating this Monte Carlo Quantum calculation together with the PIC code is commonly referred to as QED-PIC.  The QED-PIC code splits the electromagnetic field into two parts: a coherent classical field sampled on the grid of the simulation domain and a population of high-frequency photons that originate from the incoherently summed synchrotron emission of the particles. In much the same way that charged particles are treated in ordinary PIC codes, the distribution of high-frequency photons is sampled with an ensemble of so-called macro photons. These macro photon particles are created through the NLC process and propagate ballistically at the speed of light.  The only interaction experienced by these photons in the QED-PIC algorithm is the NBW process, which, when photons travel through an intense EM field, has a certain probability of decaying into electron-positron pairs. In the following subsection, we will discuss the general framework of how this QED-PIC works.  

\subsection{Decision Making in QED-PIC}

The QED-PIC method extends the traditional PIC loop in that, at every step, each energetic particle inside the strong field is evaluated to determine whether quantum events happen. Based on the information provided by the PIC code, such as particle momentum and the instantaneous electromagnetic field experienced by the particle, we can calculate the probability of the quantum events for each particle at every step. In the most common case, this probability is based on spin- and polarization-averaged rates $R_f(\chi) = \int_{0}^{1}R_f'(\chi, f')df'$, which depend only on the instantaneous quantum parameter $\chi$ calculated through the particle moment and the local field strength experienced by the particle. Here, $R_f'=\partial R/\partial f$ is the differential rate of the relevant process. One has to choose a sufficiently short time step, such that the probability of multiple events occurring within a single time step is negligible, i.e., $N_f = \Delta t R_f\ll 1$. For the two quantum processes considered in this work, the NLCS process typically has a much higher rate than the NBW. This maximum time step $\Delta t_{QED}$ that satisfies the NLC rate sets this condition. It may be compared to the maximum time step permitted by the Courant-Friedrichs-Lewy condition for finite-difference-time-domain methods $\Delta t_{CFL}$ and the need to resolve the Debye length $\Delta t_{D}$ as follows \cite{Ridgers_14}:
\begin{equation}
\frac{\Delta t_{QED}}{\Delta t_{CFL}}\approx \frac{10 N}{a_0},\ \frac{\Delta t_{QED}}{\Delta t_{D}}\approx \frac{100}{a_0}\sqrt{\frac{n_emc^2}{n_{cr}k_BT}}
\end{equation}

where N is the number of grid cells used to resolve the laser wavelength, $n_e$ is the electron number density, $n_{cr}$ is the critical density, and $T$ is the plasma temperature.

Whether or not the event happens can be decided by simply generating a uniformly distributed pseudo-random value $r_1\in[0,1]$, and the event is accepted if $r_1 < N_f$. The probability $N_f$ must be less than 1 for all possible values of $\chi$, and the smaller the time step, the more accurate the calculation. An alternative way is based on the computation of the cumulative path. The cumulative probability of a quantum event after the particle travels through an EM field of optical depth $\tau_{em}$ is $P(t)=1-\exp[{-\tau_{em}}]$. Each electron, positron, and gamma-ray photon is assigned an optical depth $\tau_{em} =\exp[-\int_{0}^{t}R_f(t')dt']$. In the actual numerical simulation, this integral is solved by first-order Eulerian integration, i.e., $\tau_{em}(t+\Delta t) = \tau(t) + R_f(t) \Delta t$. A pseudo-random value is first generated from the unit interval $P_1 \in [0,1)$. Then, it is inverted according to the equation of cumulative probability $\tau_1 = -ln(1-P_1)$ and compared with the optical depth of the particle $\tau_{em}$. The event is accepted when $\tau_1 \leq \tau_{em}$.

\subsection{Sampling the Spectrum}

Once the algorithm accepts the event, the energy and momentum of the generated photons or electron-positron pairs are sampled through the quantum spectrum. A method based on inverse transform sampling was developed. The momentum fraction $f$ is determined by solving $r_2 = C(f,\chi) = \int_{0}^t R'_f(\chi, f')df'$;  where $r_2$ is another pseudorandom value from $r_2 \in [0, 1]$ and C is the cumulative distribution function. This equation can be solved by obtaining the inverse cumulative distribution function $C^{-1} (f,\chi)$ and applying $r_2$ as the input of the function to obtain the momentum fraction $f$. This inverse function can be precalculated and tabulated for computation efficiency. For nonlinear Compton scattering, the differential rate $R_f'\rightarrow \infty$ as $f \rightarrow 0$ as $f^{-2/3}$, which is integrable. A low $f$ cutoff is commonly applied to exclude this region: a cutoff equivalent to $2mc^2$ is sufficiently small not to affect subsequent pair production and does not affect the magnitude of radiation emission too much if $\gamma$ is sufficiently large. The need for a cutoff can be avoided by augmenting the tabulated values of $C(f,\chi)$ (or its inverse) with the asymptotic analytical expressions.

One may instead use rejection sampling, which bypasses the calculation of the cumulative distribution function and can determine whether the event happens and the momentum of the generated particle simultaneously. One uniformly distributed random value $r_1 \in [0, 1]$ defines the candidate value of the energy ratio $f$. This procedure requires that $R_f'(\chi, f)\Delta t < 1$ for all values of $f \in [0, 1]$ and all possible values of $\chi$. As a result, a flat function is usually preferred for this method, and we cannot sample a distribution that diverges with $f$ or $\chi$. Due to the shape of the spectrum, we can easily apply this method to the Breit-Wheeler process. However, the infrared divergence for nonlinear Compton scattering means this will be violated for some vicinity of $f = 0$. This problem can be eliminated (following Gonoskov et al. (2015) \cite{Gonoskov_15}). The candidate $f$ can be generated as $f = r_1^3$ (i.e., with small values) more often, such that the corresponding acceptance probability $3r_1^2R_f'(\chi,r_1^3)$ becomes bounded from above.

\subsection{Wrap up the Calculation}

After the NLCS/NBW calculation, radiation reaction is incorporated by subtracting the photon’s momentum from that of the emitting electron (or positron). In pair production events, the photon transfers all its momentum to the resulting electron-positron pair and is subsequently removed from the simulation. A common challenge in large-scale QED simulations is the massive number of particles generated during extremely strong-field QED phenomena, such as QED cascades. This proliferation of particles greatly increases both memory usage and computation time, scaling exponentially with laser duration and intensity. Consequently, this severely restricts the range of parameters that can be studied for extreme QED processes. 

An effective strategy to address this issue is the particle merging algorithm. This technique reduces memory requirements by combining particles that are close to each other in both position and momentum (phase space) into a single representative particle. This approach has already been utilized in OSIRIS for unpolarized strong-field QED models, enabling more efficient exploration of high-intensity regimes \cite{Vranic_CPC_2015}.

\section{Spin and Polarization-Resolved QED-PIC }

The spin- and polarization-resolved QED-PIC module builds upon the same framework as the conventional, spin- and polarization-averaged QED-PIC codes. However, explicitly tracking spin and polarization increases the dimensionality of the problem, introducing many additional variables. Consequently, the underlying algorithms become considerably more complex, necessitating certain approximations to maintain computational efficiency—a topic we will explore in detail in this section.

Traditionally, the spin-averaged, unpolarized QED algorithm focuses on calculating the spectrum of generated particles—either photons in nonlinear Compton scattering (NLCS) or electron-positron pairs in nonlinear Breit-Wheeler (NBW) processes—described by  $S(\chi, \lambda)$, where  $\lambda$  is the momentum transfer ratio. The quantum parameter  $\chi_e/\chi_\gamma$ for the initial particle is known from its momentum and the local electromagnetic field. The Monte Carlo method is then used to sample this spectrum over various momentum transfer ratios for given values of $\chi_e/\chi_\gamma$.

Incorporating spin and polarization adds significant complexity to this procedure. For example, the NLCS spectrum in the spin- and polarization-resolved algorithm is written as $S(\chi_e, \pmb{s}_i, \lambda, \pmb{s}_f, \pmb{\xi})$, where the spectrum now depends on the initial quantum parameter $\chi_e$ and spin state  $\pmb{s}_i$ of the lepton, the momentum transfer ratio $\lambda$ , the final lepton spin $\pmb{s}_f$, and the polarization state $\pmb{\xi}$ of the emitted photon. In practice, only $\chi_e$ and $\pmb{s}_i$ are directly known; the other variables must be sampled. The amplitude of $S(\chi_e, \pmb{s}_i; \lambda, \pmb{s}_f, \pmb{\xi})$ prescribes the probability of outcomes for $\lambda$, $\pmb{s}_f$, and $\pmb{\xi}$.

Ideally, direct sampling from this full spectrum would be possible; however, the resulting function is nine-dimensional: $\chi_e$ and $\lambda$ are scalar variables, while $\pmb{s}_f$ and $\pmb{\xi}$ are vector quantities (accounting for six additional variables), and only the component of the initial spin along the magnetic field in the particle’s rest frame (a scalar) is relevant—giving nine variables total. Pre-tabulating such a high-dimensional function would require an impractically large amount of storage and memory. Moreover, accurately sampling this multidimensional distribution via Monte Carlo methods would necessitate an unfeasibly large number of macro-particles, rendering the algorithm computationally intractable.

\begin{figure}[h]
\includegraphics[width=\textwidth]{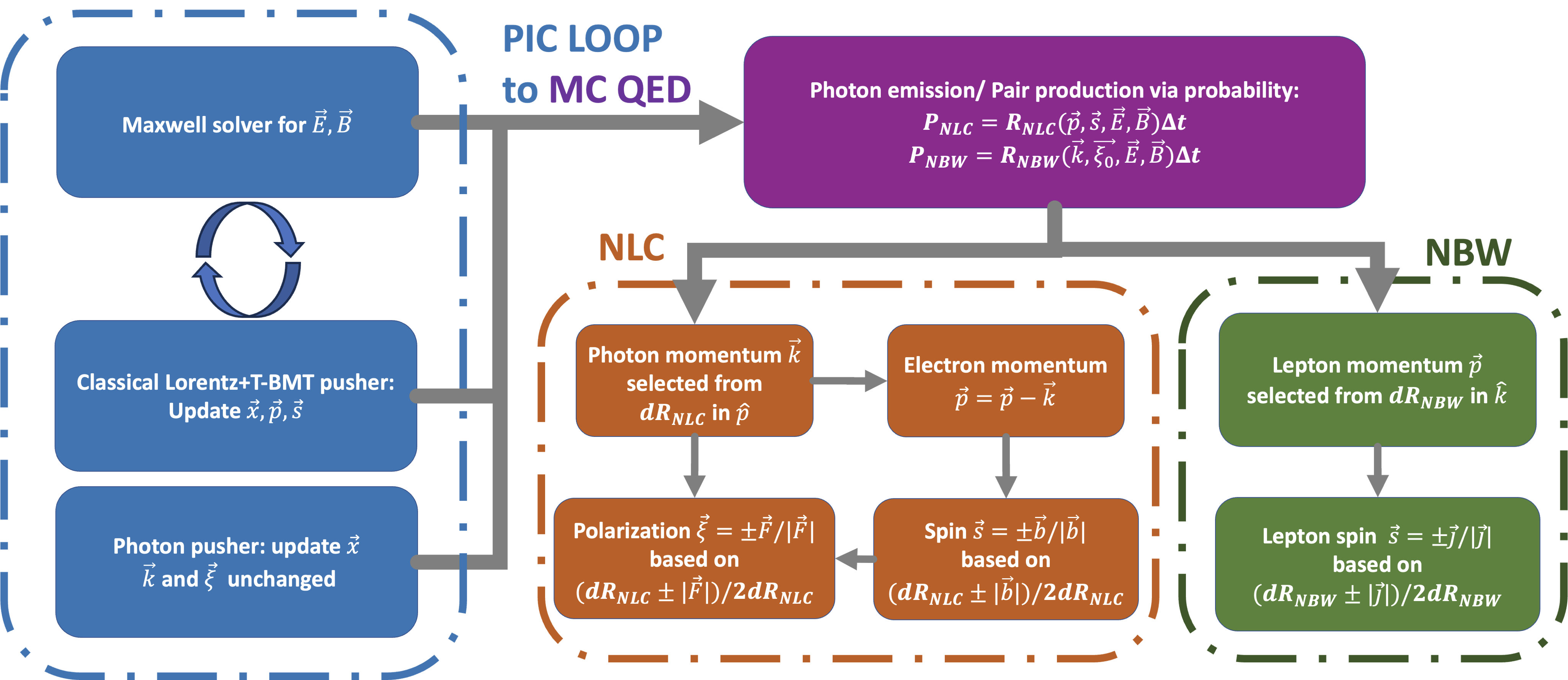}
\centering
\caption{Flow chart for spin-polarized QED algorithm}
\label{fig:flow_chart_spin_QED}
\end{figure}

To address these challenges, our approach employs conditional probabilities to determine unknown quantities sequentially, rather than sampling all variables simultaneously. Specifically, the algorithm first determines the momentum transfer ratio $\lambda$, given the initial $\chi_e$ and spin $\pmb{s}_i$, using a spectrum averaged over all possible final lepton spins and photon polarizations, $S(\chi_e, \pmb{s}_i, \lambda)$. Next, having fixed  $\lambda$, we determine the final lepton spin $\pmb{s}_f$ (still averaged over photon polarizations), using $S(\chi_e, \pmb{s}_i, \lambda, \pmb{s}_f)$. Finally, the photon polarization state  $\pmb{\xi}$ is selected based on the full spectrum $S(\chi_e, \pmb{s}_i, \lambda, \pmb{s}_f,\pmb{\xi})$, given the previously determined quantities. This hierarchical, conditional sampling process is outlined in Figure~\ref{fig:flow_chart_spin_QED}, with detailed implementation discussed in the following section.

\subsection{Spin and Polarization Dependent NLCS Implementation }
\label{chpt:S_basis_NLCS}
A detailed flow chart of the algorithm for the quantum radiation process with spin and polarization effects is presented in Fig.~\ref{fig:flow_chart_NLCS}. The code integrates a Monte Carlo-based quantum process into the main PIC loop. At each time step, the position, momentum, and spin of each lepton are updated through the classical Lorentz pusher and the T-BMT spin pusher. The quantum parameter $\chi_e$ is then calculated for each lepton. Additionally, the direction of the magnetic field in the particle’s rest frame, denoted by $\pmb{\hat\beta}$, is determined, since the spin-dependent quantum emission rate depends solely on the component of the initial spin along this direction. We only precompute spin-dependent quantum emission rates for cases where the initial spin is fully aligned or anti-parallel ($W_{\uparrow}/W_{\downarrow}$) to the rest frame magnetic field. The emission rate for an arbitrary spin projection $s_{{\beta}}$ can then be obtained as a linear combination of these two pre-tabulated rates:

\begin{figure}[h!]
\includegraphics[width=\textwidth]{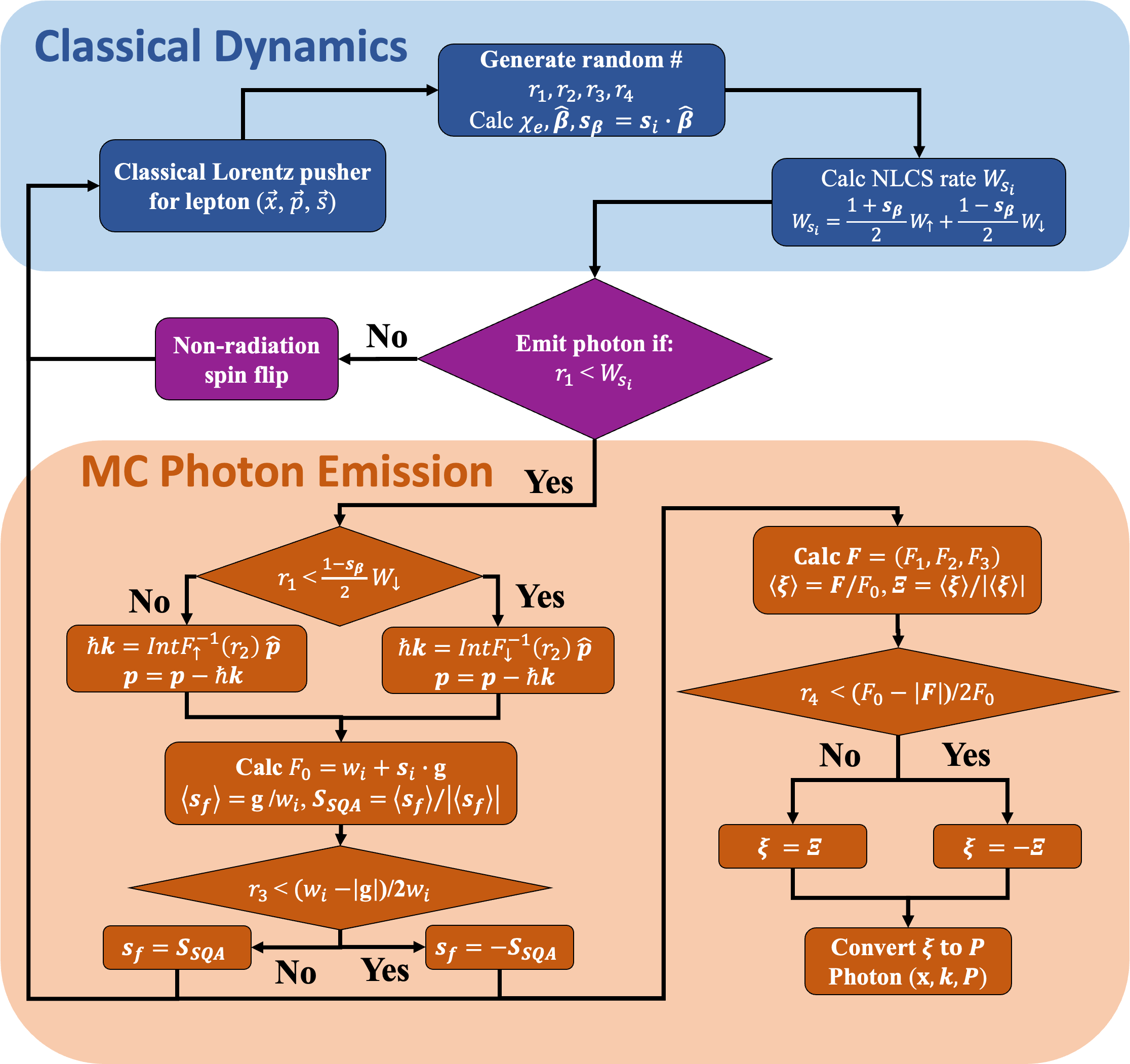}
\centering
\caption{Flow chart for spin-polarized NLCS algorithm}
\label{fig:flow_chart_NLCS}
\end{figure}

\begin{equation}
	W_{{s}_{i}} = W_{\uparrow}\frac{1+\pmb{s}_i\cdot\pmb{\hat \beta}}{2} + W_{\downarrow}\frac{1-\pmb{s}_i\cdot\pmb{\hat \beta}}{2}
\end{equation}

A random number $r_1 \in [0,1]$ is generated and compared to the quantum radiation probability to determine whether a quantum radiation event occurs. If the event takes place, the next step is to sequentially determine the momentum transfer ratio $\lambda$, the post-radiation spin state, and the polarization state of the emitted photon, as previously described.

The first quantity to be sampled is the momentum transfer ratio, $\lambda = \hbar\omega/E$. The spectrum employed for this step incorporates the dependence on the initial spin projection along the rest-frame magnetic field, $s_{\beta}$, but averages over the final spin and photon polarization states. For efficient sampling, we tabulate the inverse cumulative distribution function (ICDF) of the photon spectrum. A second random number, $r_2 \in [0,1]$, is generated and used as the input to the ICDF, representing a cumulative probability. The ICDF then returns the value of $\lambda$ corresponding to the specified $r_2$.

Ideally, we would tabulate the ICDF for all possible values of ${s}_{\beta}$, yielding a three-dimensional table. However, this approach would substantially increase memory requirements. To optimize storage and computational efficiency, we instead precompute the ICDF only for the cases where the initial spin is fully aligned or anti-parallel to the rest-frame magnetic field ($IntF^{-1}_{\uparrow}/IntF^{-1}_{\downarrow}$). Unlike the emission rates, the ICDF for an arbitrary ${s}_{\beta}$ cannot be reconstructed as a linear combination of these two cases. To resolve this, we make use of the previously generated random variable $r_1$, employing it as a criterion to select between the two precomputed ICDF spectra—aligned or anti-parallel—for the sampling of $\lambda$:

\begin{equation}
    \begin{cases}
	\text{if}\ \ r_1 <  W_{\downarrow}\frac{1-\pmb{s}_i\cdot\pmb{\hat \beta}}{2},\ \  \lambda = IntF^{-1}_{\uparrow}(r_2)\\
	\text{if}\ \ W_{\downarrow}\frac{1-\pmb{s}_i\cdot\pmb{\hat \beta}}{2} < r_1 < W_{s_i},\ \ \lambda = IntF^{-1}_{\uparrow}(r_2)
    \end{cases}       
\end{equation}

While this selection scheme is not fully rigorous from a statistical standpoint, it produces results that do not significantly deviate from the actual ICDF spectrum for a given $s_{\pmb{\beta}}$.

Once the momentum transfer ratio $\lambda$ has been determined, the momentum of the emitted photon can be assigned. Under the ultra-relativistic approximation, the direction of the emitted photon is assumed to align with the lepton's initial momentum, which gives $\hbar \pmb{k} = \pmb{p}_i \lambda$. Consequently, the post-emission lepton momentum is simply $\pmb{p}_f = \pmb{p}_i (1 - \lambda)$.

The next step is to determine the final spin state of the lepton following emission. Given the quantum parameter $\chi_e$, the direction of the rest-frame magnetic field, the initial spin, and the momentum transfer ratio $\lambda$, we obtain the probability distribution for the lepton's final spin state:

\begin{equation}
F_0(\chi, \pmb{s}_i, \lambda; \pmb{s}_f) = w_i + \pmb{s}_f \cdot \pmb{g}
\end{equation}

The explicit forms of $w_i$ and $g$ are provided in Appendix~\ref{appendix_NLCS}. It is important to note that the function $F_0$ depends not only on the initial spin component along the rest-frame magnetic field, but also on the components $\pmb{s}_{\varepsilon}$ and $\pmb{s}_{\kappa}$. The value of $F_0$ determines the likelihood of each possible final spin state: a higher $F_0$ corresponds to a higher probability of realizing that particular spin state. From this distribution, $F_0$, we can then compute the expectation value (mean) of the final spin state:

\begin{equation}
\langle \pmb{s}_f \rangle = \frac{\int{\pmb{s}_f’F_0}d \pmb{s}_f’}{\int{F}d\pmb{s}_f’} = \pmb{g}/w_i
\end{equation}

Ideally, one would sample the complete distribution of possible final spin states as a function of $\chi_e$, $\pmb{s}_i$, and $\lambda$. However, accurately capturing this distribution requires a substantial number of simulation particles, which can become computationally prohibitive given that sampling the main spectrum is already resource-intensive.

To address this, we simplify the determination of the final spin state by assigning it the expectation value under the distribution $F_0$; that is, we set $\pmb{s}_f = \langle \pmb{s}_f \rangle$. If it is desirable to maintain the spin length at unity throughout the simulation, we instead use the normalized expectation value as the spin quantization axis, namely $\pmb{S}_{\mathrm{SQA}} = \langle \pmb{s}_f \rangle / |\langle \pmb{s}_f \rangle|$. A random number $r_3 \in [0,1]$ is then generated, and used to decide whether the spin after emission is aligned or anti-aligned with this quantization axis:

\begin{equation}
    \begin{cases}
	\text{if}\ \ r_3 < (w_i - |g|)/2w_i,\ \ \pmb{s}_f = - \pmb{S}_{SQA}  \\
	\text{if}\ \ r_3 > (w_i - |g|)/2w_i,\ \ \pmb{s}_f =  \pmb{S}_{SQA}
    \end{cases}       
\end{equation}

Finally, we decide what the polarization state is, represented by the Stokes parameter $\pmb{\xi}$ of the photon.  Given the quantum parameter of lepton $\chi_e$, initial spin $\pmb{s_i}$, final spin $\pmb{s_f}$, and the momentum transfer ratio $\lambda$, we can obtain the probability distribution of the photon polarization state:

\begin{equation}
			F_{full}(\chi, \pmb{s}_i, \pmb{s}_f, \lambda; \pmb{\xi}) = F_0 + \pmb{\xi} \cdot \pmb{F}
\end{equation}

The detailed expression of $\pmb{F}$ can be seen in the appendix. Similarly to how we decide the final spin state, we first look at the mean of all possible photon polarization states under this distribution $F_{full}$:

\begin{equation}		
\langle \pmb{\xi} \rangle = \frac{\int{F_{full} \pmb{\xi}’}d\pmb{\xi}’}{\int{F_{full} }d\pmb{\xi}’} = \pmb{F}/F_0
\end{equation}

For similar reasons as with the final spin state assignment, we set the photon polarization state to the expectation value under the distribution $F$, that is, $\pmb{\xi} = \pmb{F}/F_0$. If it is necessary for each macro-photon to be $|\pmb{\xi}| = 1$, we instead define the polarization quantization axis as $\pmb{\Xi}_{\mathrm{PQA}} = \langle \pmb{\xi} \rangle / |\langle \pmb{\xi} \rangle|$, which is a superposition of linear, diagonal, and circular polarization components, but as a pure state. A random number $r_4 \in [0,1]$ is then generated to determine whether the photon polarization state aligns with $\pmb{\Xi}_{\mathrm{PQA}}$ or $-\pmb{\Xi}_{\mathrm{PQA}}$:

\begin{equation}
    \begin{cases}
	\text{if}\ \ r_4 < \frac{F_0 - |\pmb{F}|}{2F_0},\ \  \pmb{s}_f = - \pmb{\Xi}_{PQA}  \\
	\text{if}\ \ r_4 > \frac{F_0 - |\pmb{F}|}{2F_0},\ \ \pmb{s}_f = \pmb{\Xi}_{PQA}
    \end{cases}       
\end{equation}

The polarization vector $\pmb{\xi}$ determined above specifies the photon's polarization state in the emission basis $(\pmb{\hat{\varepsilon}}, \pmb{\hat{\beta}}, \pmb{\hat{k}})$ in which the photon is generated. However, to analyze the photon’s polarization at the point of observation—particularly when it interacts with external electromagnetic fields—it is necessary to transform the Stokes parameters from the emission frame to the observation (or laboratory) frame. This transformation requires knowledge of both the Stokes parameters at emission and the polarization basis at that instant.

To minimize computational memory usage, we aim to reduce the number of parameters needed to fully specify the photon polarization state. We find that the polarization state can be uniquely represented by a single vector quantity, the photon polarization vector $\pmb{P} = (P_{\psi_1}, P_{\psi_2}, P_{{k}})$, whose components correspond to the $({\hat{\pmb\psi}_1}, {\hat{\pmb\psi}_2}, {\hat{\pmb k}})$ coordinate axes. The definitions of these coordinates and their connection to the Stokes parameters have already been discussed in Section~\ref{chpt:pol_vec}.

Through the procedure described above, we have developed a fully spin- and polarization-resolved NLCS algorithm compatible with the PIC loop. Finally, in accordance with the conditional probability structure of our Monte Carlo approach~\cite{Chen_1995_CAIN, Chen_PRD_2022}, the electron’s spin is also carefully updated in cases where emission does not occur. In a later section, we will provide a detailed discussion of the treatment of spin for non-radiative events.

\subsection{Spin and Polarization Dependent NBW Implementation }

\begin{figure}[h!]
\includegraphics[width=\textwidth]{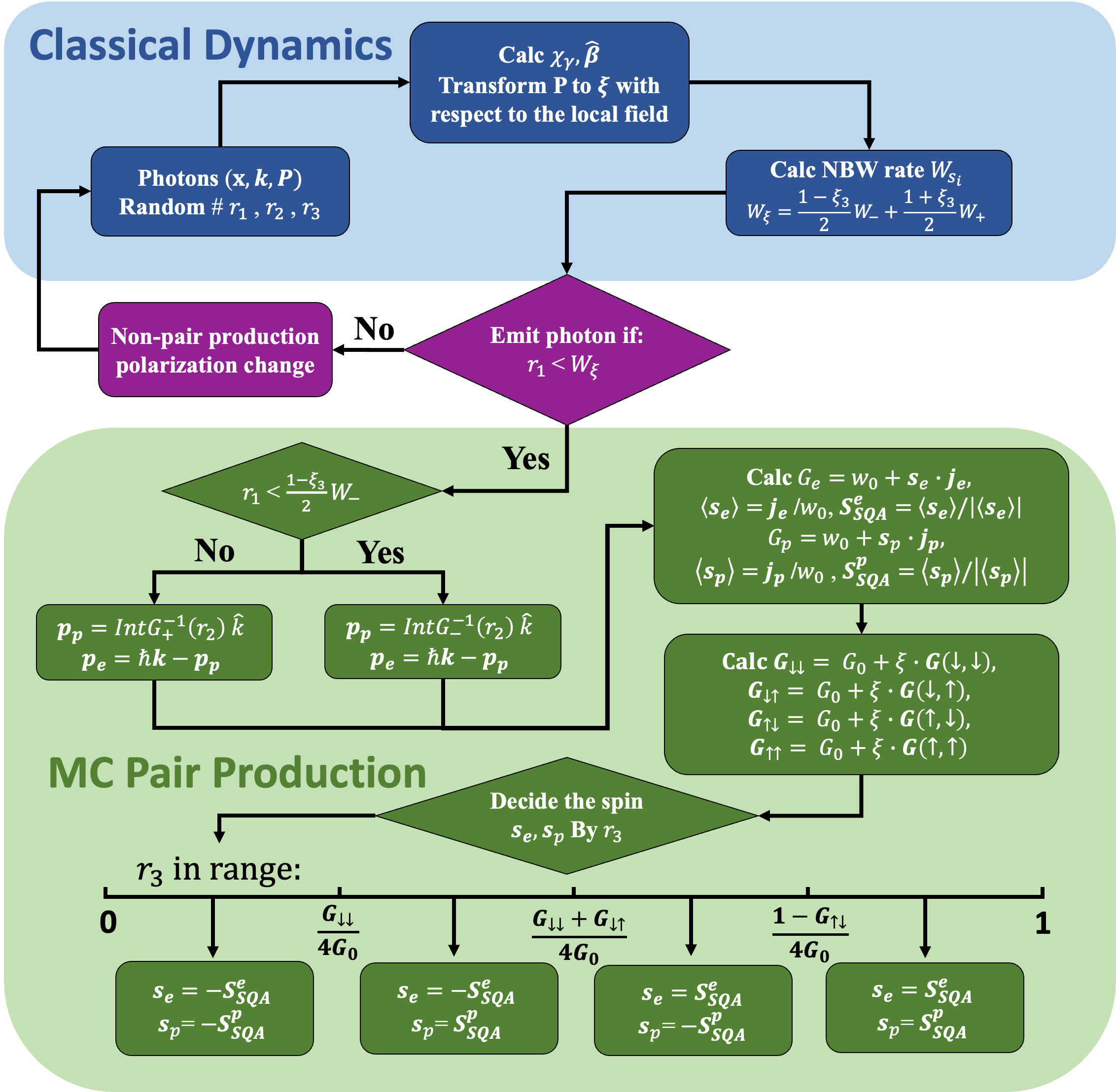}
\centering
\caption{Flow chart for spin-polarized NBW algorithm}
\label{fig:flow_chart_NBW}
\end{figure}

A detailed flow chart illustrating the spin- and polarization-dependent quantum radiation process for the NBW algorithm is presented in Fig.~\ref{fig:flow_chart_NBW}. The dynamics of photons in the classical PIC loop are considerably simpler than those for charged particles: each photon is treated as effectively a point particle ({a wavepacket with small extent}) that propagates along its momentum direction at the speed of light in vacuum, while both its momentum and polarization vectors remain unchanged.

In the quantum calculations, the electromagnetic field experienced by each photon is monitored at every simulation step, allowing the quantum parameter $\chi_\gamma$ to be computed. For each photon, the polarization basis vectors $\pmb{\hat{\varepsilon}}$, $\pmb{\hat{\beta}}$, and $\pmb{\hat{\kappa}}$ are constructed with respect to the local field. The photon polarization vector is then projected onto this basis to extract the Stokes parameters corresponding to the photon’s polarization state in the given field.

The polarization-resolved NBW pair-production rate depends on both the quantum parameter $\chi_\gamma$ and the linear Stokes parameter $\xi_3$ of the photon, with $\xi_3$ taking values between $-1$ and $+1$. For computational efficiency, we pre-tabulate the NBW rates only for the cases where the linear Stokes parameter is either $+1$ or $-1$ ($W_{+}$ and $W
_{-}$). The rate for arbitrary $\xi_3$ is then obtained by a linear combination of these two pre-tabulated rates:

\begin{equation}
W_{\xi} = \frac{1-\xi_3}{2}W_{-} + \frac{1+\xi_3}{2}W_{+}
\end{equation}

A random number $r_1 \in [0,1]$ is generated and compared with the NBW pair production probability to determine whether a pair creation event occurs. If the event takes place, it is then necessary to determine the momentum transfer ratio $\lambda$ between the positron and the photon, as well as the spin states of the generated electron and positron.

Analogously to the procedure for NLCS, the first quantity to be sampled is the momentum partition ratio, defined as $\lambda = |\pmb{p}_p|/\hbar k$, where $\pmb{p}_p$ is the positron momentum and $\hbar k$ is the incident photon momentum. The energy spectrum utilized in this step incorporates the dependence on the photon's polarization—specifically, only the linear Stokes parameter $\xi_3$—and is averaged over all possible spin states of the generated pair.

For efficient sampling, we tabulate the inverse cumulative distribution function (ICDF) for the positron spectrum. A second random number $r_2 \in [0,1]$ is generated and serves as the input for the ICDF, representing the cumulative probability. The ICDF yields the corresponding value of the momentum transfer ratio $\lambda$ for the chosen $r_2$.

To minimize the computational memory requirements, we precompute the ICDF only for photons with linear polarization $\xi_3 = +1$ and $\xi_3 = -1$ ($IntG_{+}^{-1}/IntG_{-}^{-1}$). Following a procedure analogous to that used in the NLCS algorithm, we use the previously generated random value $r_1$ to select which ICDF -corresponding to $\xi_3 = +1$ or $\xi_3 = -1$ - will be used to sample the momentum transfer ratio $\lambda$.

\begin{equation}
    \begin{cases}
	\text{if}\ \ r_1 <  \frac{1-\xi_3}{2}W_{-},\ \   \lambda = IntG_{-}^{-1}(r_2) \\
	\text{if}\ \ \frac{1-\xi_3}{2}W_{-}< r_1 <   W_{\xi},\ \   \lambda = IntG_{+}^{-1}(r_2)
    \end{cases}       
\end{equation}

Once the momentum transfer ratio $\lambda$ has been determined, the momenta of the generated electron and positron can be assigned. Under the ultra-relativistic approximation, the directions of the produced pairs are assumed to align with the incident photon’s momentum. Thus, the positron momentum is given by $\pmb{p}_{p} = \hbar \pmb{k} \lambda$, while the electron momentum is $\pmb{p}_{e} = \hbar \pmb{k} (1-\lambda)$. In the NBW process, the photon transfers all of its momentum to the electron-positron pair and decays away.

The subsequent step is to determine the spin states of the generated electron and positron. Since their spin states are inherently correlated, the algorithm samples both quantities simultaneously. This is achieved by first evaluating the probability distribution of the electron’s spin (averaged over all possible positron spin states) and the probability distribution of the positron’s spin (averaged over all possible electron spin states), both conditioned on the previously determined momentum transfer ratio.

\begin{equation}
G_e(\chi, \pmb{\xi}, \lambda; \pmb{s}_e) = w_0 + \pmb{s}_e \cdot \pmb{j}_e
\end{equation}
\begin{equation}
G_p(\chi, \pmb{\xi}, \lambda; \pmb{s}_p) = w_0 + \pmb{s}_p \cdot \pmb{j}_p
\end{equation}

The detailed expression for $w_0$, $\pmb{j}_e$, $\pmb{j}_p$ can be found in the appendix \ref{appendix_NBW}. One thing worth to mention is that function $\pmb{j}_e$ and $\pmb{j}_p$ depend not only just the linear polarization component $\pmb{\xi}_3$, but other component $\pmb{\xi}_1$ and $\pmb{\xi}_2$ as well. The expectation value of the electron and positron spin is:

\begin{equation}
\langle \pmb{s}_e \rangle = \frac{\int{G_e  \pmb{s}_e’}d \pmb{s}_e’}{\int{G_e}d\pmb{s}_e’} = \pmb{j}_e/w_0
\end{equation}
\begin{equation}
\langle \pmb{s}_p \rangle = \frac{\int{G_p  \pmb{s}_p’}d \pmb{s}_p’}{\int{G_p}d\pmb{s}_p’} = \pmb{j}_p/w_0
\end{equation}

The spin quantization axes for the electron and positron in the NBW process are defined as $\pmb{S}_{\mathrm{SQA}}^e = \langle \pmb{s}_e \rangle/|\langle \pmb{s}_e \rangle|$ and $\pmb{S}_{\mathrm{SQA}}^p = \langle \pmb{s}_p \rangle/|\langle \pmb{s}_p \rangle|$, respectively. For computational simplicity, we assign the electron and positron spins to be either fully aligned or anti-aligned with their respective expectation values along these axes. This approach allows us to efficiently sample the correlated spin states according to their associated probabilities:

\begin{equation}
		G_{\downarrow \downarrow} = G_0(\downarrow,\downarrow) + \pmb{\xi}\cdot\pmb{G}(\downarrow,\downarrow)
\end{equation}
\begin{equation}
		G_{\downarrow \uparrow} = G_0(\downarrow,\uparrow) + \pmb{\xi}\cdot\pmb{G}(\downarrow,\uparrow)
\end{equation}
\begin{equation}
		G_{\uparrow \downarrow}= G_0(\uparrow,\downarrow) + \pmb{\xi}\cdot\pmb{G}(\uparrow,\downarrow)
\end{equation}
\begin{equation}
		G_{\uparrow\uparrow} = G_0(\uparrow,\uparrow) +  \pmb{\xi}\cdot\pmb{G}(\uparrow,\uparrow)
\end{equation}

Here, the first arrow denotes the spin state of the generated positron, while the second arrow corresponds to the spin state of the generated electron. The labels $\uparrow$ and $\downarrow$ indicate whether the spins are aligned or anti-aligned with respect to their respective spin quantization axes. The Monte Carlo method is then employed to determine the spin states of the electron-positron pair. Specifically, a random number $r_3 \in [0,1]$ is generated:

\begin{equation}
    \begin{cases}
	\text{if}\ \ r_3 < \frac{G_{\downarrow\downarrow}}{4G_0},\ \ \pmb{s}_p = -\pmb{S}_{SQA}^p,\ \pmb{s}_e = -\pmb{S}_{SQA}^e \\
	\text{if}\ \ \frac{G_{\downarrow\downarrow}}{4G_0}<r_3 < \frac{G_{\downarrow\downarrow}+G_{\uparrow\downarrow}}{4G_0},\ \ \pmb{s}_p = -\pmb{S}_{SQA}^p,\ \pmb{s}_e = \pmb{S}_{SQA}^e \\
    \text{if}\ \ \frac{G_{\downarrow\downarrow}+G_{\uparrow\downarrow}}{4G_0}<r_3 < \frac{1-G_{\uparrow\uparrow}}{4G_0},\ \ \pmb{s}_p = \pmb{S}_{SQA}^p,\ \pmb{s}_e = -\pmb{S}_{SQA}^e\\
    \text{if}\ \ r_3 > \frac{1-G_{\uparrow\uparrow}}{4G_0},\ \ \pmb{s}_p = \pmb{S}_{SQA}^p,\ \pmb{s}_e = \pmb{S}_{SQA}^e
    \end{cases}       
\end{equation}

At the end of the calculation, the photon involved in the process is removed from the simulation, and the newly generated electron-positron pair is incorporated into the PIC loop. Through the procedure detailed above, we establish a fully spin- and polarization-resolved NBW algorithm that is seamlessly integrated with the PIC framework.

It is also important to note that, analogous to the treatment in the NLCS process, the photon polarization distribution must be updated in the case of a non-pair creation event. This adjustment is necessary because the NBW process preferentially depletes photons with linear polarization $\xi_3 = 1$. Accordingly, the photon’s Stokes parameter is projected along or opposite to the polarization quantization axis whenever pair production does not occur. The detailed treatment of photon polarization in non-pair production events will be discussed in a later section.

\section{Quantum Non-Transition Events}

In the Monte Carlo algorithm, for a given initial spin/polarization state, the spin/polarization resolved rate can be expressed as:
\begin{equation}
{\mathbb{P}} = \langle {\mathbb{P}} \rangle + \textbf{P}_i\cdot \pmb{\zeta}_i \;,
\end{equation}
where $\langle {\mathbb{P}} \rangle$ is the average spin/polarization rate, $\textbf{P}_i \cdot \pmb{\zeta}_i $ is the spin/polarization-related transition rate for a given initial state. One crucial point that is somewhat counterintuitive is that the selection effect of the spin/polarization-dependent rate will actually change the polarization state for the macroparticles that do not have the transition. It is easy to assume that the spin state for non-transition particles is just their initial spin state. However, this is not correct. We need to treat carefully  the macroparticles that do not undergo quantum transitions:

The probability that the transition does not occur in a time interval $dt$  is $1-(\langle {\mathbb{P}} \rangle + \textbf{P}_i\cdot \pmb{\zeta}_i)$. Consider an ensemble (one macro-particle) of N (real) particles having the polarization vector $\pmb{\zeta}_{i, n},\ \text{which } n = 1,2, …, N$. Each of this is a unit vector $|\pmb{\zeta}_{i, n}| = 1$. The initial spin/polarization for this macro-particle is then $\pmb{\zeta}_i = \langle \pmb{\zeta}_{i, n} \rangle$.

We can rewrite the transition rate related to the initial polarization state $\pmb{P}_i = P_i \pmb{Q_i}$. $\pmb{Q_i}$  is the quantization axis, which for a lepton is the rest frame B field direction $\pmb{\hat \beta}$, and for a photon it is the linear polarization component $\pmb{\xi}_3$. The final polarization state for the Marco particle for the non-quantum transition event along $\pmb{Q_i}$ is then: 

\begin{equation}
\begin{split}
\sum_n\sum_{\pm}\pm\frac{1\pm \pmb{Q}_{i}\cdot\pmb{\zeta}_{i,n}}{2}&[1-(\langle{\mathbb{P}} \rangle \pm \textbf{P}_i\cdot \pmb{Q_i})\Delta t] \\ & = \sum_{n}[\pmb{Q_i}\cdot \pmb{\zeta}_{i,n}(1-\langle{\mathbb{P}} \rangle\Delta t) - \textbf{P}_i\cdot \pmb{Q_i}] \\ & = N \pmb{Q_i}\cdot [\pmb{\zeta}_i(1-\langle{\mathbb{P}} \rangle \Delta t)-\pmb{P}_i\Delta t]
\end{split}
\end{equation}

The total number of real particles inside that macroparticle without transition is $N[1-(\langle {\mathbb{P}} \rangle + \textbf{P}_i\cdot \pmb{\zeta}_i)\Delta t]$. The final polarization vector for the macroparticle is then:

\begin{equation}
\pmb{\zeta}_{f,\ \text{no trans}} = \frac{\pmb{\zeta}_i(1-\langle{\mathbb{P}} \rangle \Delta t)-\textbf{P}_i\Delta t}{1-(\langle {\mathbb{P}} \rangle + \textbf{P}_i\cdot \pmb{\zeta}_i)\Delta t}
\end{equation}

\begin{figure}[h!]
\includegraphics[width=\textwidth]{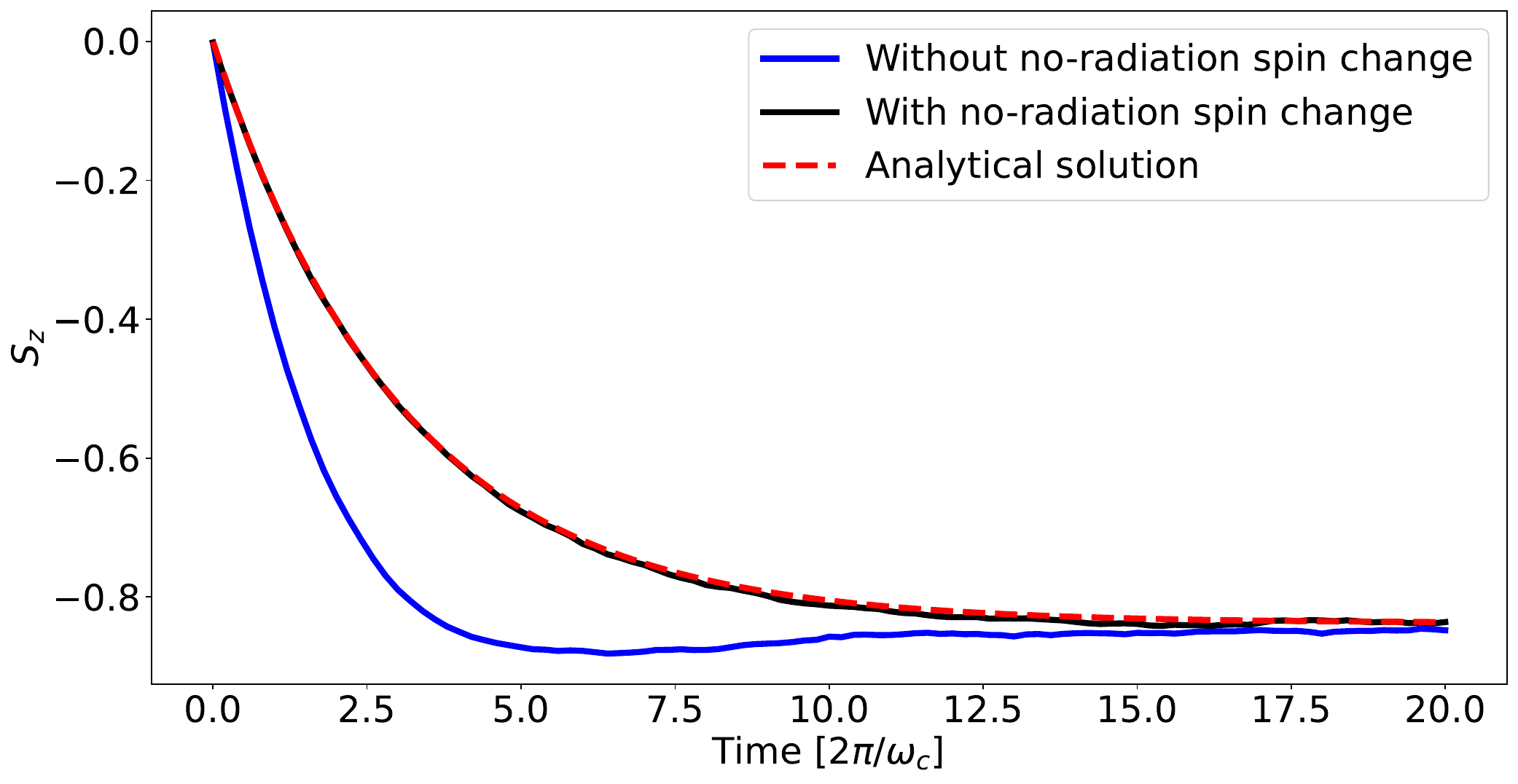}
\centering
\caption{The influence of no-quantum radiation correction to the lepton spin evolution}
\label{fig:nospin_transition_nlc}
\end{figure}

\begin{figure}[h!]
\includegraphics[width=\textwidth]{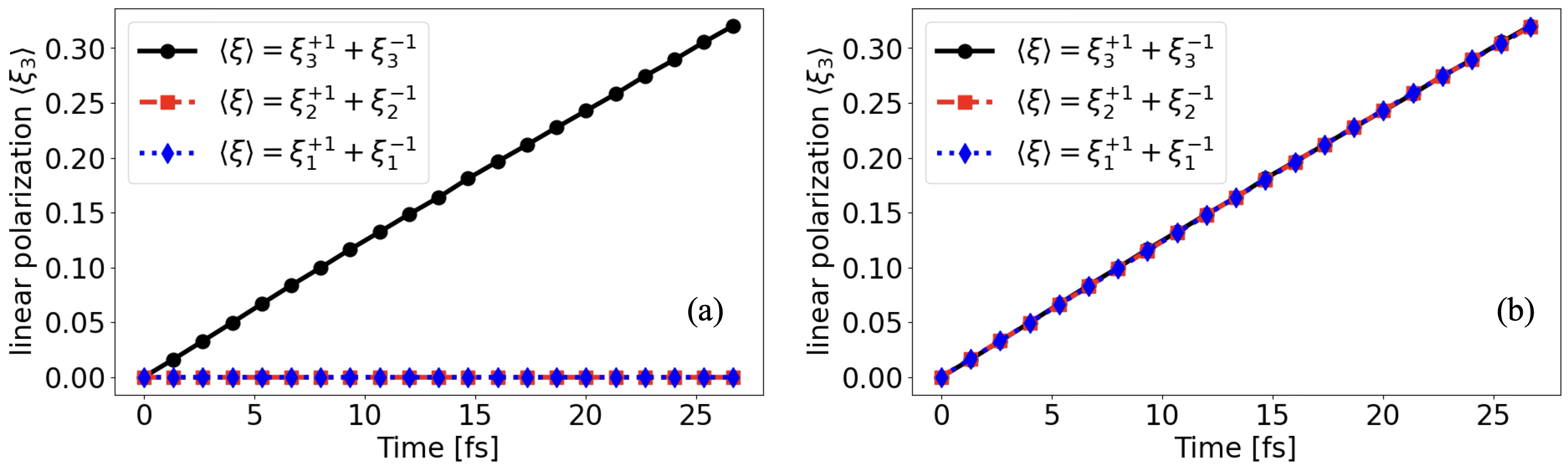}
\centering
\caption{(a) The evolution of linear polarization state, calculated without non-pair creation polarization state change. (b) The evolution of the linear polarization state, calculated with non-pair creation polarization state change. The unpolarized beams in the simulation were constructed here from two identical parts with opposite polarization states. The black curve is for opposite linear polarization states, the red curve is for opposite circular polarization states, and the blue curve is for opposite diagonal polarization states initially. }
\label{fig:nopol_transition_nbw}
\end{figure}

The influence of the non-quantum transition event can be illustrated by two examples here. In the first example, an unpolarized, mono-energetic, 1 GeV lepton beam gyrates inside a constant magnetic field with a field strength equal to 2.26 MT. The field and the momentum are perpendicular, the electron quantum parameter $\chi_e=1$. We ignore radiation back-reaction for simplicity. The evolution of spin along the magnetic field direction is shown in Fig.~\ref{fig:nospin_transition_nlc}. The blue curve shows the spin evolution without including the non-emission spin change; the black curve is the result when the non-emission spin change is taken into account. The red dashed curve is from the analytical solution of the Sokolov-Ternov effect, shown in the next section in Eq.~\ref{eq:sok_ter}. We can see that the blue curve without the non-emission spin change deviates from the analytical solution, whereas the black curve, including the non-emission spin change, agrees well with the analytical solution. 

In the second example, we construct an unpolarized photon beam with two equivalent parts that are at the same location and have the same momentum distribution and beam profiles. These two parts are either fully linearly polarized, diagonally polarized, or circularly polarized, but have opposite polarization states initially. As a result, the average polarization state for these two parts in total will be zero. These constructed beams are mono-energetic with an initial energy equal to 1 GeV and travel inside a constant magnetic field with a field strength equal to 2.26 MT. The field and the momentum are perpendicular, the photon quantum parameter $\chi_\gamma=1$. For simplicity, the generated electron-positron pair is excluded from the quantum process calculation. The evolution of the degree of linear polarization is shown in Fig.~\ref{fig:nopol_transition_nbw} (a) and (b), where the black curve is for when this unpolarized beam is constructed by two parts with opposite linear polarization states, the red curve is when the two parts have opposite circular polarization states, and then the blue curve is when the two parts have opposite diagonal polarization states, initially. Fig.~\ref{fig:nopol_transition_nbw} (a) is when we don't include the non-pair creation polarization state change, and we can see that the degree of linear polarization for three "unpolarized beams" evolves very differently with time. Fig.~\ref{fig:nopol_transition_nbw} (b) is when we include the non-pair creation polarization state change, and we can see that the degree of linear polarization for the three "unpolarized beams" evolves exactly the same with time. This result underscores the importance of employing a model in which the time evolution of "unpolarized beams" is consistent, regardless of their initial construction---at least in an average sense. Ensuring such consistency is, in our view, essential for developing a reliable and robust model.

\section{Code Verification}

Ensuring the correct numerical implementation of all quantum processes is essential. To validate this, we first compare the NLC and NBW spectra generated by the code for various quantum parameters $\chi_e/\chi_\gamma$ and initial lepton spin or photon polarization states against the corresponding closed-form expressions. In the figures, the orange and blue curves represent simulation results, while the purple and red dashed curves show the analytical, spin- and polarization-resolved NLC and NBW spectra.

\subsection{NLCS Photon Spectrum for Different Spin-Polarized Lepton}

We examine the gamma-ray spectrum produced by a monoenergetic electron beam in a constant magnetic field via the NLCS process. To isolate the quantum effects, we neglect all classical dynamics—including the Lorentz force and spin precession—as well as the effects of radiation reaction and radiation-induced lepton spin changes. Our analysis begins with unpolarized electron beams.

\begin{figure}[h!]
\includegraphics[width=\textwidth]{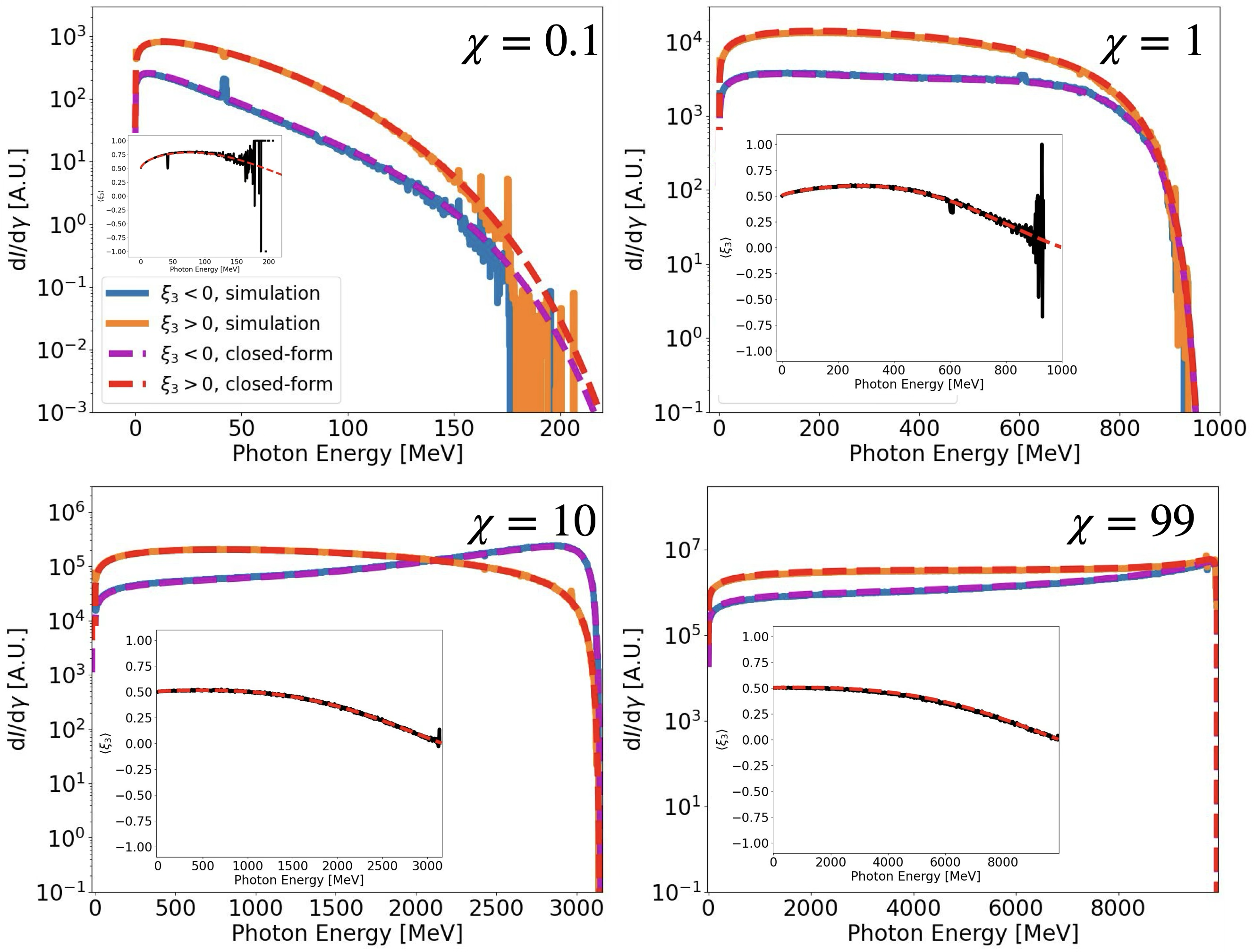}
\centering
\caption{Benchmark of linearly polarized ($\xi_3$) photon spectrum from the NLC process for an unpolarized lepton beam inside a constant magnetic field. The classical particle dynamics, radiation back reaction, and radiation-induced spin change are all switched off for simplicity. We scan the spectrum through $\chi = 0.1$, $\chi = 1$, $\chi = 10$ and $\chi = 99$. }
\label{fig:benchmark_unpol_nlc}
\end{figure}

We tested the spectrum for electron quantum parameters $\chi = 0.1, 1, 10$, and $99$, and compared our simulation results to the corresponding closed-form expressions, as shown in Fig.~\ref{fig:benchmark_unpol_nlc}. The case for $\chi = 0.01$ is discussed in the next section \ref{chpt:classical_spect}, since this low value allows comparison with the polarization-resolved classical radiation spectrum.

Under these conditions, the emitted photons exhibit only linear polarization. In the figure, the blue curve corresponds to photons with linear polarization component $\xi_3 < 0$, while the orange curve represents $\xi_3 > 0$, both derived from our simulations. Both curves show excellent agreement with the analytical results.

\subsection{NBW Pair Production Spectrum for Different Polarized Photon}

\begin{figure}[h!]
\includegraphics[width=\textwidth]{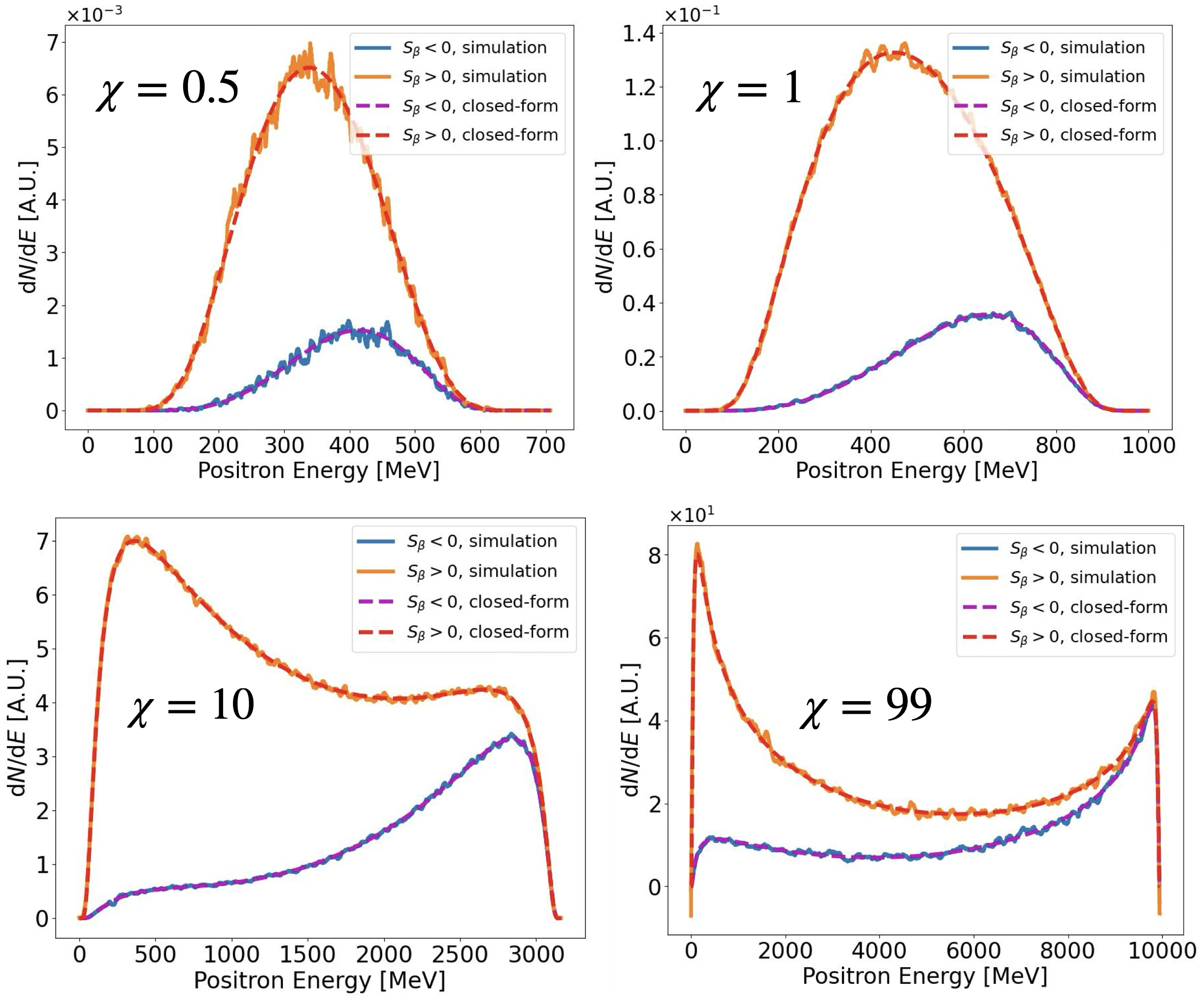}
\centering
\caption{Benchmark of positron spectrum for spin component $S_\beta$ from the NBW process for an unpolarized photon beam inside a constant magnetic field. The classical particle dynamics, radiation back reaction, and radiation-induced spin change are all switched off for simplicity. We scan the spectrum through $\chi = 0.5$, $\chi = 1$, $\chi = 10$ and $\chi = 99$. }
\label{fig:benchmark_unpol_nbw}
\end{figure}

For the NBW process, we examine the positron spectrum generated by a monoenergetic photon beam propagating in a constant magnetic field. To facilitate direct comparison, we disregard classical dynamics—such as the Lorentz force and spin precession—as well as the effects of radiation reaction and radiation-induced spin changes in the generated leptons. Our analysis begins with an unpolarized photon beam.

We evaluated the positron spectra for photon quantum parameters $\chi_\gamma = 0.5, 1, 10$, and $99$, and compared the simulation results to the corresponding closed-form expressions, as shown in Fig.~\ref{fig:benchmark_unpol_nbw}. For an unpolarized photon beam, the generated positron’s spin can only be in the direction of the magnetic field in its rest frame, according to Eq.~(\ref{eq:SQA_NBW}), so we only plot the spectrum for spin component $S_\beta < 0$ or $S_\beta > 0$. The blue curve depicts the positron spectrum with spin component $S_\beta < 0$ along the magnetic field direction, while the orange curve corresponds to $S_\beta > 0$, as obtained from the simulations. Both spectra show excellent agreement with the analytical results.

\section{Code Validation}
In this section, we present benchmark tests of our spin- and polarization-resolved QED code across a variety of scenarios. The simulation results show excellent agreement with analytical predictions as well as with previously published results. These benchmarks provide further validation of the accuracy and robustness of our algorithm.

\subsection{Sokolov Ternov Effect}

To benchmark our code’s performance, we first reproduce the well-known Sokolov-Ternov effect, which describes the self-polarization process of relativistic electrons or positrons traversing a magnetic field at high energies. This effect arises because the transition rates between spin states aligned and anti-aligned with the magnetic field differ, resulting in net polarization of the lepton beam either parallel or antiparallel to the field.

\begin{figure}[h!]
\includegraphics[width=\textwidth]{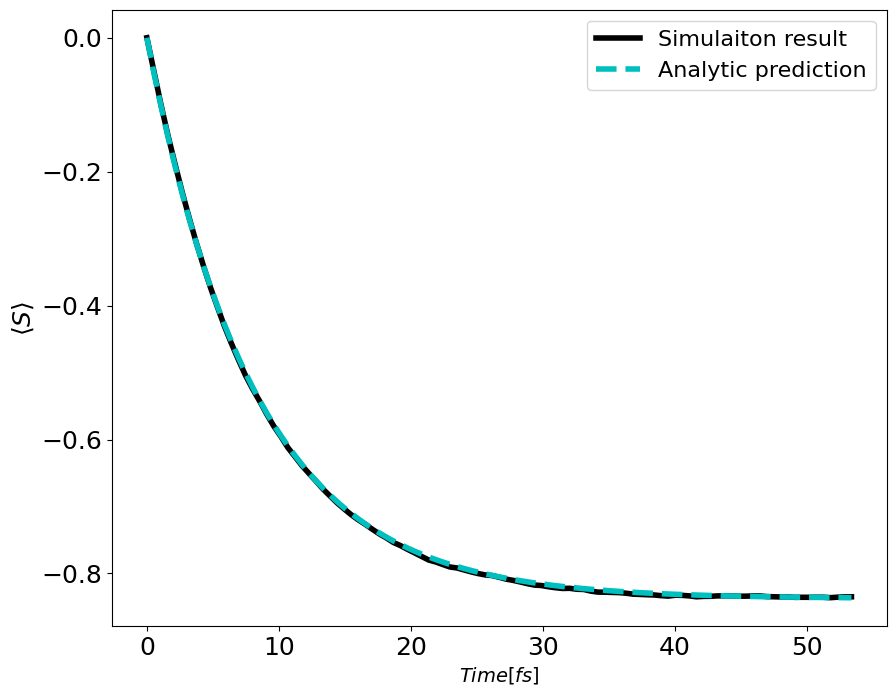}
\centering
\caption{The spin evolution along magnetic field direction for a 1 GeV beam inside a 2.26 MT constant magnetic field. Radiation back-reaction is switched off for simplicity. The simulation result is compared with the well-known Sokolov-Ternov effect prediction.}
\label{fig:sok-ter}
\end{figure}

We begin with a two-dimensional PIC simulation: an initially unpolarized, monochromatic 1 GeV electron beam is initialized moving in the x-y plane. A constant magnetic field of 2.26 MT is applied in the z-direction throughout the simulation domain. For this benchmark, radiation reaction and pair production are disabled, so the particle energy remains constant, allowing for direct comparison between the simulation results and the analytical prediction given by Eq.~\ref{eq:sok_ter}.

\begin{equation}
\frac{d S_{\beta}}{dt} = -P(\psi_1(\chi)S_{\beta} + \psi_3(\chi))
\label{eq:sok_ter}
\end{equation}

Here, $P = \frac{\alpha_f m_e c^2}{\sqrt{3}\gamma \hbar \omega_L}$, $\psi_1(\chi) = \int_0^{\infty} u'' du, K_{2/3}(u')$, $\psi_3(\chi) = \int_0^{\infty} u'' duK_{1/3}(u')$, with $u'' = u^2/(1+u)^3$, $u' = 2u/(3\chi)$, and $u = \hbar \omega/(\gamma - \hbar \omega)$. The Sokolov-Ternov effect specifically describes spin polarization along the magnetic field direction in the particle's rest frame, so our analysis focuses on the time evolution of the $S_\beta$ component. As shown in Fig.~\ref{fig:sok-ter}, the simulation results exhibit excellent agreement with the analytical prediction.


\subsection{Comparing with Classical Polarization Spectrum}
\label{chpt:classical_spect}

\begin{figure}[h!]
\includegraphics[width=\textwidth]{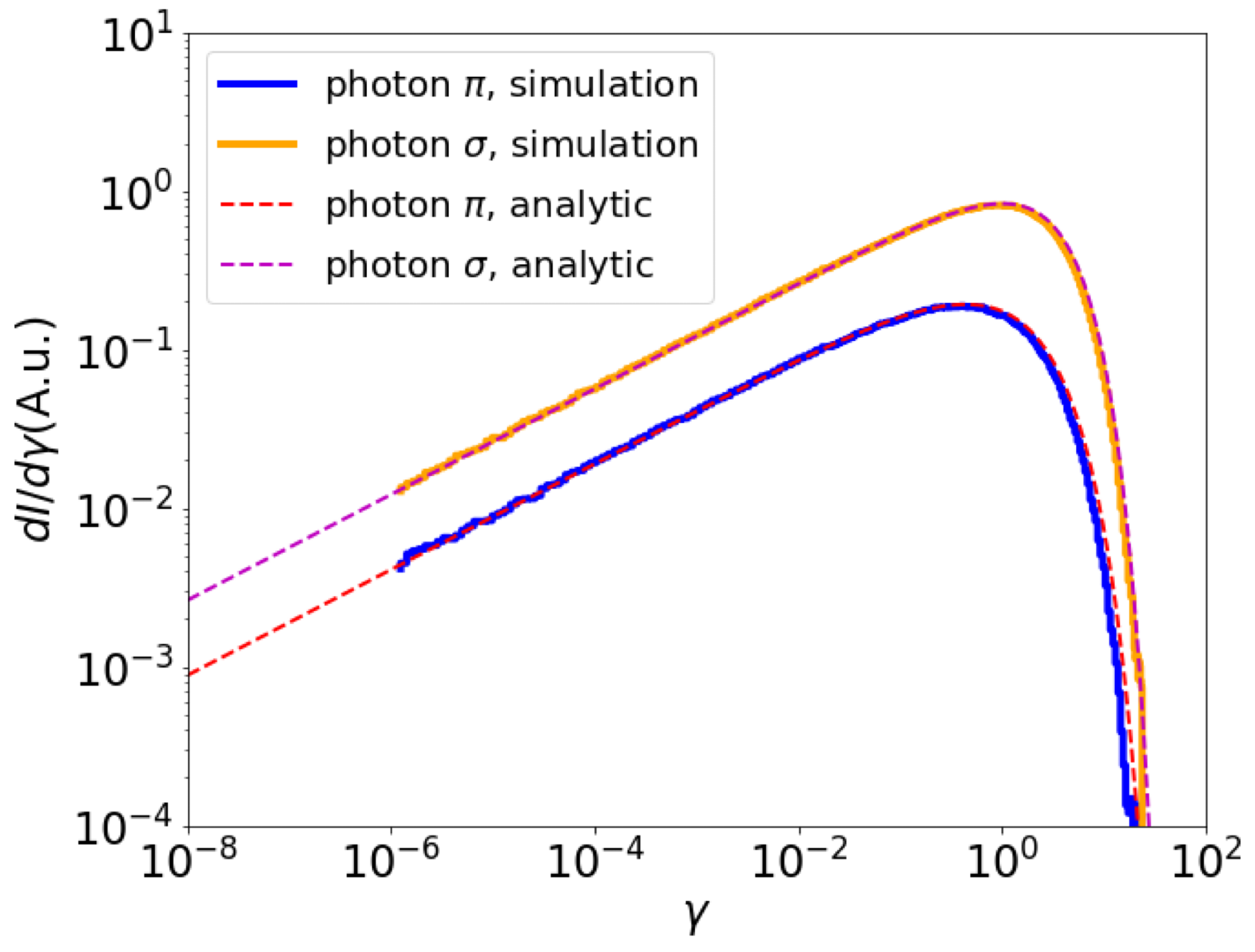}
\centering
\caption{$\pi$ and $\sigma$ polarized photon spectrum from the simulation (blue and yellow solid line) compared with the classical radiation equations (red and purple dashed line) when $\chi = 0.01$.}
\label{fig:class-pol}
\end{figure}

The polarization-resolved spectrum for classical synchrotron radiation is well established. It is therefore natural to ask whether the quantum calculation converges to the classical result in the limit of a low quantum parameter, $\chi$. Figure~\ref{fig:class-pol} presents the polarized photon spectrum obtained from our simulation and compares it with the analytical classical synchrotron radiation spectrum for $\chi_e = 0.01$.

The simulation is performed in two dimensions, using a monoenergetic, unpolarized electron beam propagating in a constant magnetic field. We neglect all classical dynamics—including the Lorentz force and spin precession—as well as radiation reaction and radiation-induced lepton spin changes.

The classical radiation spectrum is given by:

\begin{equation}
W_{\sigma, \pi} = \frac{9\sqrt{3}}{16\pi}u'W^{cl}\left[(l_{\sigma}^2 + l_{\pi}^2)\int_{u'}^{\infty} K_{5/3}(x)dx + (l_{\sigma}^2 - l_{\pi}^2) K_{2/3}(u')\right]
\end{equation}

Which if photon is polarized in $\sigma$ direction,  $l_{\sigma} =1$ and $l_{\pi} = 0$, otherwise if it is polarized in $\pi$ direction, $l_{\sigma} =0$ and $l_{\pi} = 1$. $W^{cl} = \frac{2}{3}\frac{e^2c}{R}\gamma^4$ is the total radiation energy of an ultra-relativistic electron. $R$ is the gyro-radius. The simulation result shows good agreement with the analytic calculation.

\begin{figure}[h!]
\includegraphics[width=\textwidth]{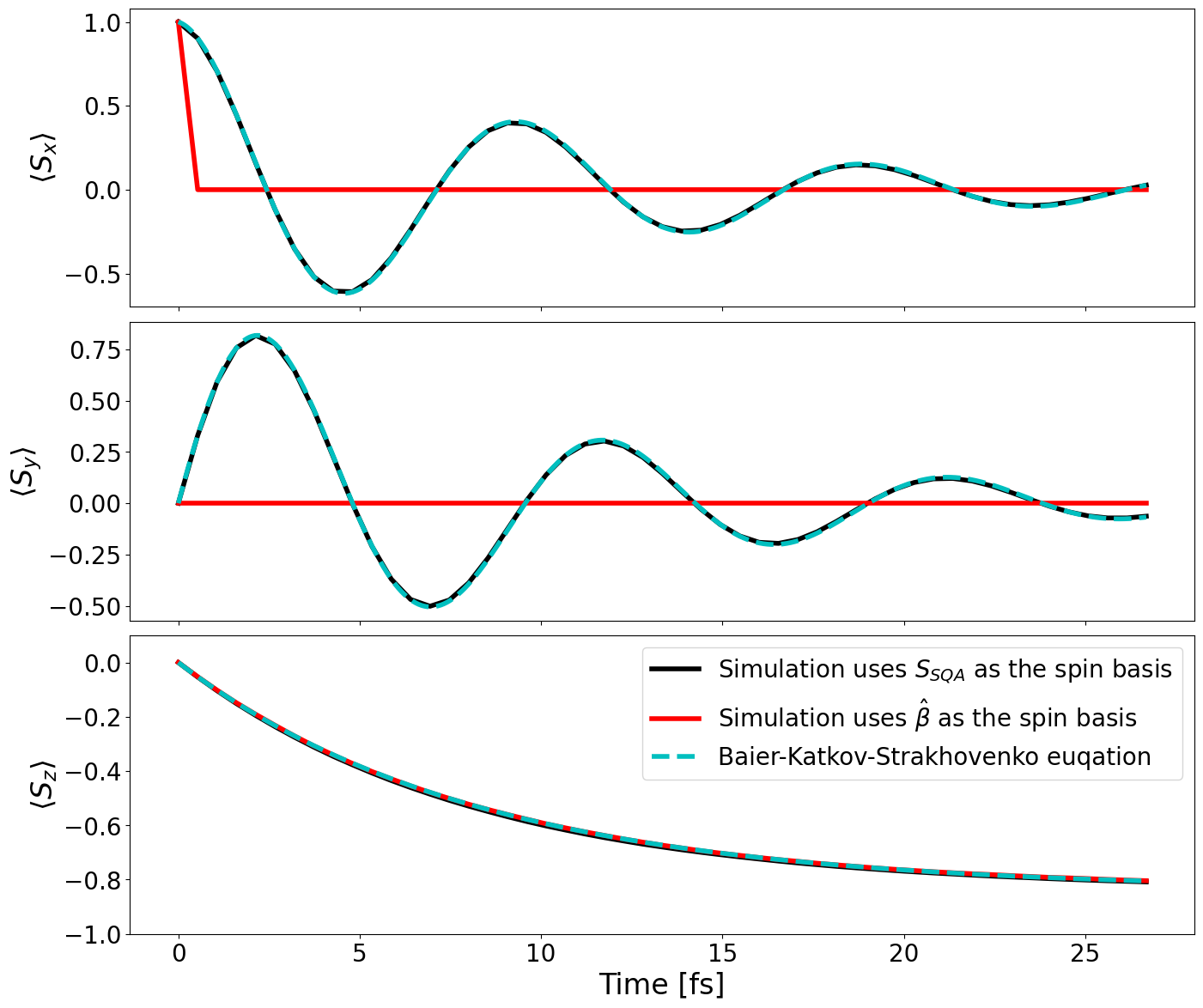}
\centering
\caption{The evolution spin component along the x, y, z direction for a fully polarized electron beam inside a constant B field. The beam is initially polarized along the x direction, while the magnetic field is along the z direction. The radiation back-reactions are switched off for simplicity. The red curve shows the simulation result using rest frame B field $\pmb{\hat \beta}$ as the spin basis. The black curve shows the simulation result using the spin quantization axis $\pmb{S}_{SQA}$ defined in Eq.~\ref{eq:SQA_NLC} as the spin basis. The dashed cyan curve is the analytic prediction using the B-K-S equation.} 
\label{fig:inital_pol_bks}
\end{figure}

\subsection{Spin Evolution for Arbitrary Initial Conditions and Field Configurations}

In previous tests, we examined the spin evolution of an unpolarized beam in a static field. For more complex scenarios, however, the choice of spin basis becomes increasingly important. Here, we present two benchmark tests: one involving an initially polarized lepton beam in a constant magnetic field, and another featuring an unpolarized electron beam in a time-varying magnetic field. Simulations were conducted using both the spin quantization axis, $\pmb{S}_{SQA}$, and the rest-frame magnetic field direction, $\pmb{\hat\beta}$.

In the first test, we use a two-dimensional PIC simulation with an initially polarized, monochromatic 1GeV electron beam propagating in the $x$-$y$ plane. A constant magnetic field of 2.26MT is applied along the $z$-direction throughout the simulation domain, and the electron beam is initially polarized along the $x$-direction. Radiation back-reaction and pair production are disabled to facilitate direct comparison with analytical predictions.

Simulation results are benchmarked against the extended Baier-Katkov-Strakhovenko (B-K-S) equation\cite{BKS}, shown in Eq.\ref{eq:B-K-S}, which incorporates both radiation-induced spin dynamics and classical spin precession.

\begin{equation}
\frac{d\pmb S}{dt} = \left( \frac{d\pmb S}{d t}\right)_{T} - P\left[\psi_1(\chi)\pmb S + \psi_2(\chi) (\pmb S \cdot \pmb{\hat v}) \pmb{\hat v} + \psi_3(\chi) \pmb{\hat\beta}\right]
\label{eq:B-K-S}
\end{equation}

Here, $\psi_1(\chi) = \int_0^{\infty} u''du K_{2/3}(u')$, $\psi_2(\chi) =\int_{0}^{\infty} u''du\int_{u'}^{\infty}dx K_{1/3}(x)- \psi_1(\chi)$, $\psi_3(\chi) = \int_{0}^{\infty} u''duK_{1/3}(u)$, $P = \frac{\alpha_f m_ec^2}{\sqrt{3}\gamma \hbar \omega_L}$, and $u'' = u^2/(1+u)^3$. $\left( \frac{d\pmb S}{d t}\right)_{T}$ is the T-BMT equation:

\begin{equation}
\left( \frac{d\pmb S}{d t}\right)_{T} = -\frac{e}{m_ec}\left[ \left(a_e + \frac{1}{\gamma}\right)(\pmb B - \pmb v \times \pmb E) - \pmb v \frac{a_e\gamma}{\gamma+1}\pmb v \cdot \pmb B \right] \times \pmb S = \pmb \Omega \times \pmb S 
\end{equation}

The evolution of the three spin components along the $(x, y, z)$ axes is shown in Fig.~\ref{fig:inital_pol_bks}. I performed simulations using two different spin bases. The red curves represent results obtained with $\pmb{\hat\beta}$ as the spin basis, where the simulated spin evolution matches the analytic results only along the magnetic field direction ($z$-axis). In contrast, the black curves show results using the spin quantization axis $\pmb{S}_{SQA}$ as the basis. In this case, the evolution of all three spin components agrees well with the analytic calculations.

\begin{figure}[h!]
\includegraphics[width=\textwidth]{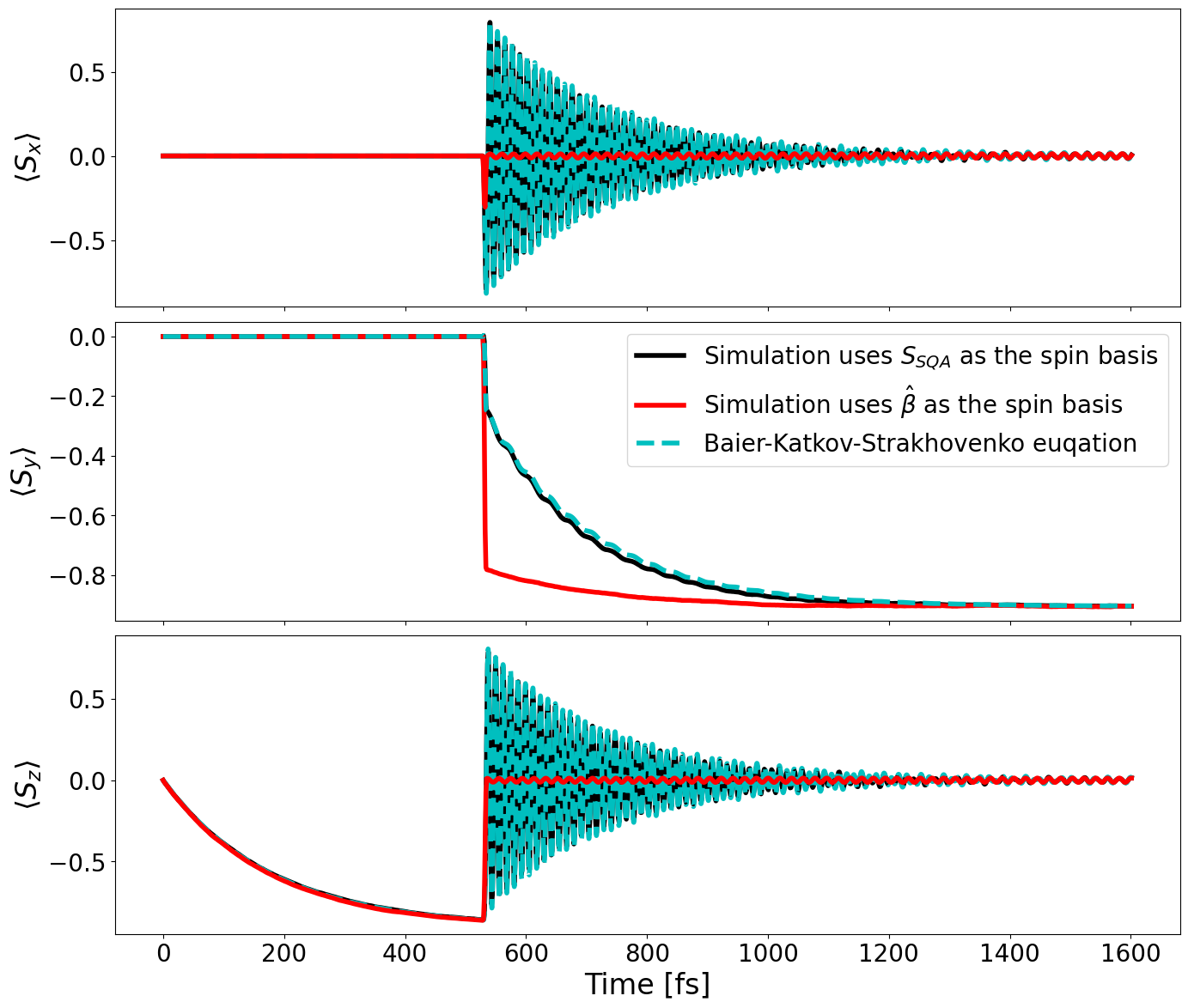}
\centering
\caption{The evolution spin component along the x, y, z direction for an unpolarized electron beam inside a changing B field. The magnetic field is initially along the z direction and stays in z for a while. It then quickly rotates to the y direction at a rotation speed close to the electron gyro frequency. The amplitude of the magnetic field remains the same. The radiation back-reactions are switched off for simplicity. The red curve shows the simulation result using rest frame B field direction $\pmb{\hat \beta}$ as the spin basis. The black curve shows the simulation result using the spin quantization axis $\pmb{S}_{SQA}$ defined in Eq.~\ref{eq:SQA_NLC} as the spin basis. The dashed cyan curve is the analytic prediction using the B-K-S equation.} 
\label{fig:rotate_field_bks}
\end{figure}

The second test examines the spin dynamics of an unpolarized electron beam in a time-varying magnetic field. Here, an initially unpolarized 100 MeV lepton beam is placed inside a 2.26 MT magnetic field. The field strength remains constant, but its direction is initially aligned with the $z$-axis before rotating towards the $y$-axis at a rate comparable to the gyrofrequency. Simulation results are compared with predictions from the B-K-S equation. The spin components' evolution is shown in Fig.~\ref{fig:rotate_field_bks}. Again, simulations were conducted with both spin bases. The red curves, using $\pmb{\hat \beta}$ as the basis, align with analytical predictions only prior to the field rotation; deviations appear when the field direction rotates from $z$ to $y$. However, for the black curves, calculated using $\pmb{S}_{SQA}$ as the basis, the evolution of all spin components maintains excellent agreement with analytical results throughout the simulation.

\subsection{Benchmark against Ptarmigan}

We also benchmark our code against other well-known polarization-resolved QED algorithms. Specifically, we utilize the Monte Carlo particle-tracking code Ptarmigan, developed by Tom Blackburn et al.~\cite{Blackburn_POP_2023}, which recently implemented polarization dependence in both photon emission and pair creation. Since Ptarmigan does not include lepton spin effects, we conduct a simple test case involving an unpolarized, high-energy electron beam colliding with a circularly polarized laser pulse.

In this setup, the laser has a peak intensity of $a_0 = 64$, a wavelength of $\lambda_0 = 0.8\mu$m, a pulse duration of $\tau_0 = 40\text{fs}$, and a focal radius of $w_0 = 32\mu$m. The electron beam is mono-energetic with an initial energy of 5 GeV, and its density follows a Gaussian distribution transversely ($\sigma_y = 4.8\mu$m) and longitudinally ($\sigma_x = 0.8\mu$m).

\begin{figure}[h!]
\includegraphics[width=\textwidth]{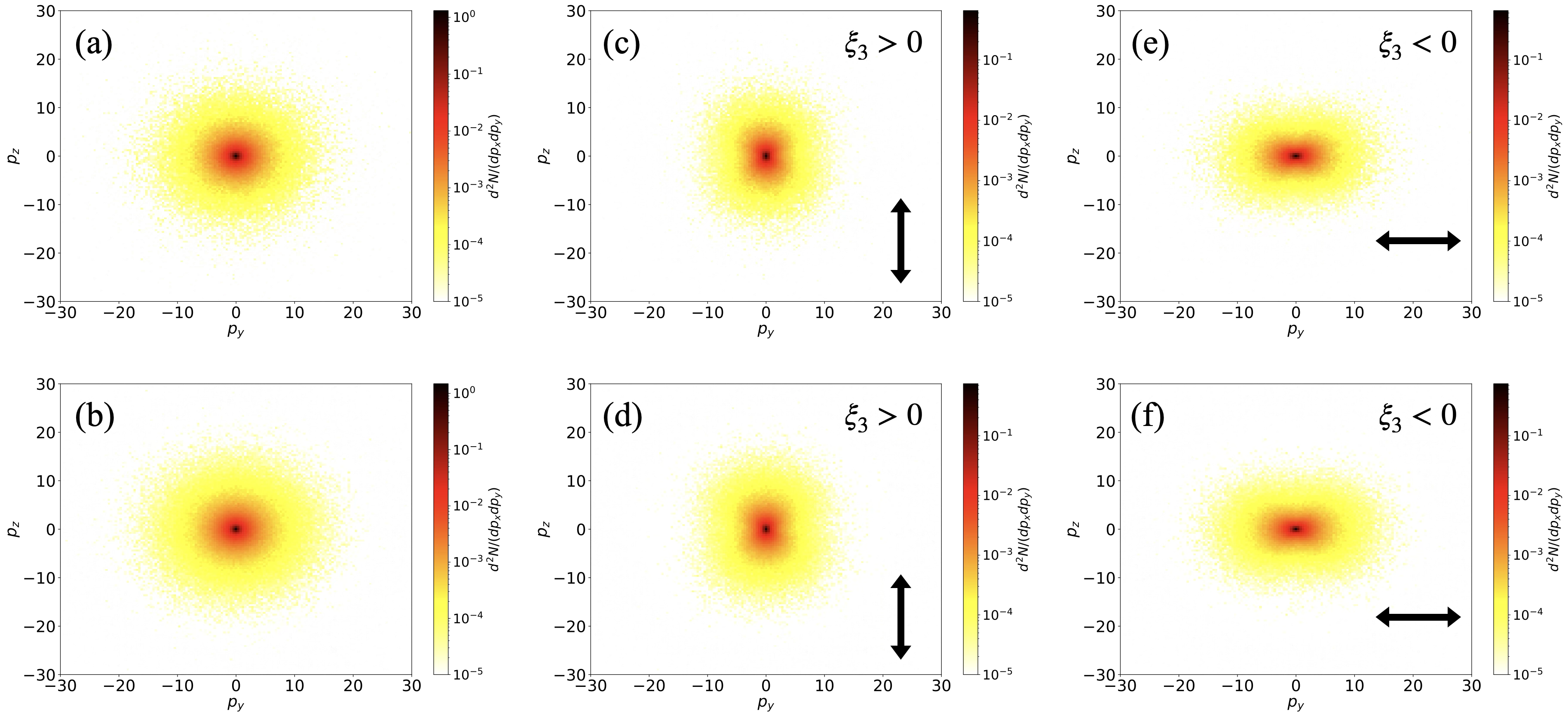}
\centering
\caption{Photon polarization spectrum from Osiris compared with Ptarmigan. (a), (c), (e), Simulation result from ptarmigan. (a), (b) The total photon
density distribution in py-pz space, (c),(d) The density distribution for photons with
polarization state $\xi_3>0$. (e)(f) The density distribution for photons with polarization
state $\xi_3<0$. (a),(c),(e) are the results from Ptarmigan. Our code results are shown in (b), (d), (f).}
\label{fig:ptarmigan-Osiris}
\end{figure}

We track the transverse distribution of linearly polarized photons produced during the interaction. The results from Ptarmigan are shown in Fig.\ref{fig:ptarmigan-Osiris}(a), (c), and (e): (a) depicts the total photon density distribution in $p_y$-$p_z$ space, (c) corresponds to photons with polarization state $\xi_3>0$, and (e) to photons with $\xi_3<0$. Our code’s results are shown in Fig.\ref{fig:ptarmigan-Osiris}(b), (d), and (f). The polarized photon density distributions produced by both codes are nearly identical, demonstrating strong agreement between the two approaches.

\subsection{Benchmark against Previously Published Results}

The previous benchmarks clearly demonstrate the accuracy of our code, although they were performed under highly idealized conditions that are challenging to replicate in real-world experiments. To further validate our approach, we now aim to reproduce results recently reported in several publications, which propose experimentally accessible setups involving a lepton beam colliding with a laser pulse. The parameters suggested in these works are realistic and could be realized in the near future.

These studies systematically explore various characteristics of spin- and polarization-resolved NLCS and NBW processes by varying the properties of the lepton beam and laser pulse. We divide the scenarios into three main categories:

\begin{itemize}
\item \emph{Spin Polarization along the Rest Frame Magnetic Field:}
This examines how the NLCS and NBW processes tend to polarize lepton spins along the rest frame magnetic field. For example, the collision of an unpolarized electron beam with a linearly polarized laser pulse generates linearly polarized photons.

\item \emph{Interplay Between Quantum Radiation and Classical Spin Precession:}
Here, the interaction between quantum-induced spin alignment and classical spin precession is explored—demonstrating, for instance, how spin, initially aligned by quantum processes, can be rotated to the longitudinal direction through precession.

\item \emph{Spin Polarization Along the Longitudinal Direction:}
This case focuses on helicity transfer in spin-polarized NLCS and NBW processes, where longitudinally polarized lepton beams and circularly polarized photons exchange angular momentum, as predicted by polarization-resolved QED.
\end{itemize}
Most existing works employ Monte Carlo-based particle tracking algorithms. In what follows, we employ our own spin- and polarization-resolved QED code to reproduce and validate the key results reported in these recent studies.

\subsubsection{Spin Polarization from Colliding with a Bichromatic Laser}

As predicted by the Sokolov-Ternov effect, particle spins tend to align either parallel or antiparallel to the magnetic field in the particle’s rest frame. However, simply colliding an unpolarized lepton beam with a strong, symmetric laser field does not produce a polarized beam, as the net effect of the field averages to zero over an optical cycle. To achieve polarization, one approach is to break the symmetry of the laser field, for example, by using a bichromatic laser pulse.

In this work, we reproduce the bichromatic laser pulse setup described in recent literature \cite{Chen_PRL_2019}. The pulse consists of two copropagating lasers with wavelengths $\lambda_1 = 1\mu\text{m}$ and $\lambda_2 = 0.5\mu\text{m}$, both with a pulse duration $\tau = 33\text{fs}$ and a focal radius $w_0 = 5\mu\text{m}$. The ratio of peak field strengths is $R = a_{01}/a_{02} = 2$, with a total field strength of $a_0 = a_{01} + a_{02} = 83$. The electron beam is initialized with a transverse Gaussian width of $\sigma_y = 0.4\mu\text{m}$, a longitudinal length of $\sigma_x = 10\mu\text{m}$, and an initial energy of 2~GeV with a $2\%$ energy spread.

\begin{figure}[h!]
\includegraphics[width=\textwidth]{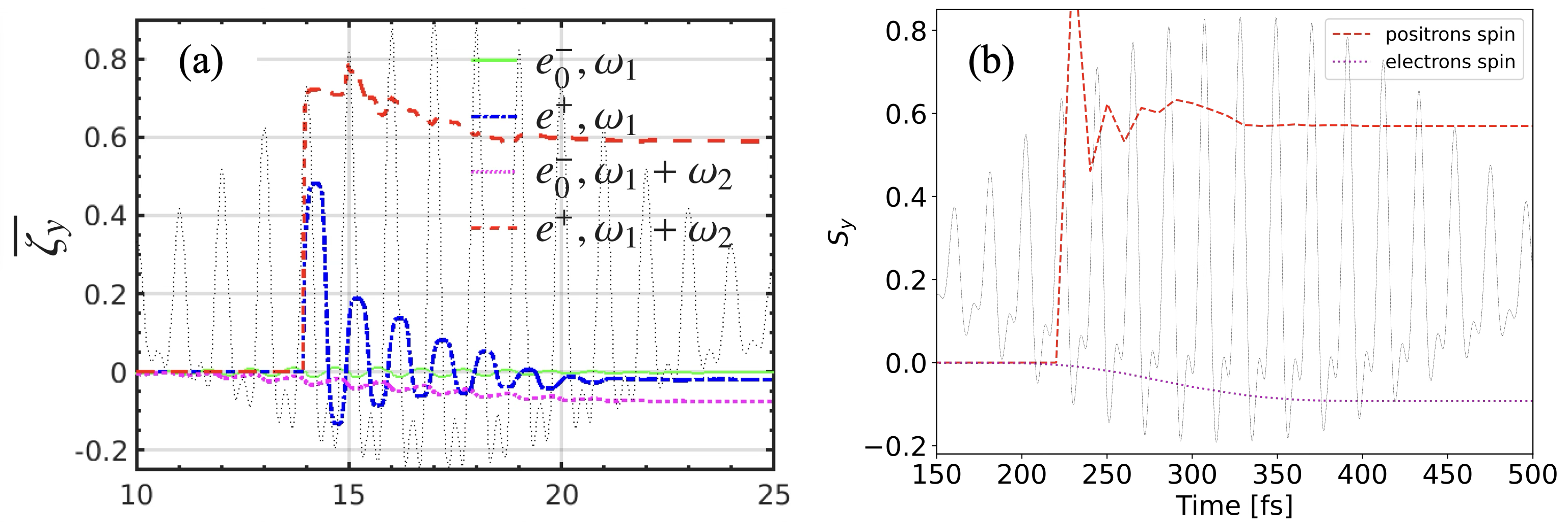}
\centering
\caption{Electron and positron beam spin polarization evolution in a bichromatic laser. (a) From the paper \cite{Chen_PRL_2019} (b) from the OSIRIS spin and polarization-resolved QED simulation.
Subplot (a) reproduced from Yue-Yue Chen et al. Phys. Rev. Lett. \textbf{123}, 174801 (2019).{\textcopyright} American Physical Society.
} 
\label{fig:bichromatic_model_compare}
\end{figure}

We compare the results of our spin- and polarization-resolved QED simulations to those reported in the reference paper, as shown in Fig.~\ref{fig:bichromatic_model_compare}. The paper reports an average positron polarization of about $58\%$; our simulation yields a value of $57.9\%$. The average lepton spin polarization is reported to be $7.7\%$, while our simulation gives $8.8\%$, indicating strong agreement. The small discrepancy may arise from comparing our 2D PIC simulation with the paper's 3D particle tracking simulations; however, we believe that two-dimensional simulations are sufficient to capture the essential physics of this process.

\begin{figure}[h!]
\includegraphics[scale=0.3]{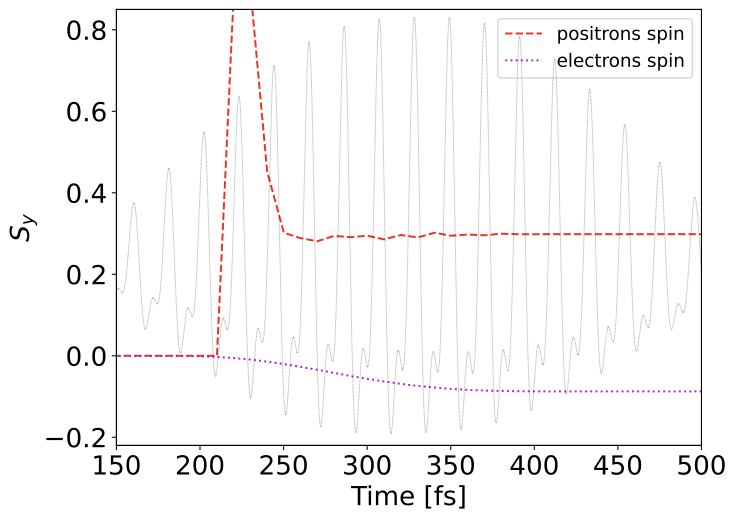}
\centering
\caption{Spin polarization in a bichromatic laser, with the full spin and polarization model in our code} 
\label{fig:bichromatic_full_model}
\end{figure}

It is important to note that the reference paper does not incorporate the full spin- and polarization-resolved QED effects in its calculations; the emitted photons are assumed to be unpolarized. Our algorithm, by contrast, includes a flexible switch that allows specific effects to be toggled on or off. In the earlier comparison, we adopted the same simplified physics model as the paper to ensure consistency, resulting in comparable outcomes.

However, this simplified model does not accurately represent the polarization dynamics. In reality, photons generated during the laser-beam collision are predominantly linearly polarized, and this polarization significantly affects the resulting positron polarization. When we activate the full spin- and polarization-resolved QED effects in our code, as shown in Fig.~\ref{fig:bichromatic_full_model}, the average positron polarization is reduced to just $31\%$. This highlights the importance of properly accounting for polarization effects in simulations of these processes.

\subsection{Spin Polarization from Colliding with an Elliptically Polarized Laser}

Yan-fei Li et al. \cite{Li_PRL_2019} propose a clever and straightforward method for generating a polarized electron beam. Their approach involves colliding a slightly elliptically polarized laser beam with an unpolarized lepton beam. The lepton acquires polarization along the direction of the magnetic field in its rest frame, while the ellipticity of the laser introduces a correlation between the electrons’ transverse momentum and their spin. This momentum–spin correlation allows a polarized electron beam to be obtained by selectively collecting electrons emitted at specific angles.

\begin{figure}[h!]
\includegraphics[width=\textwidth]{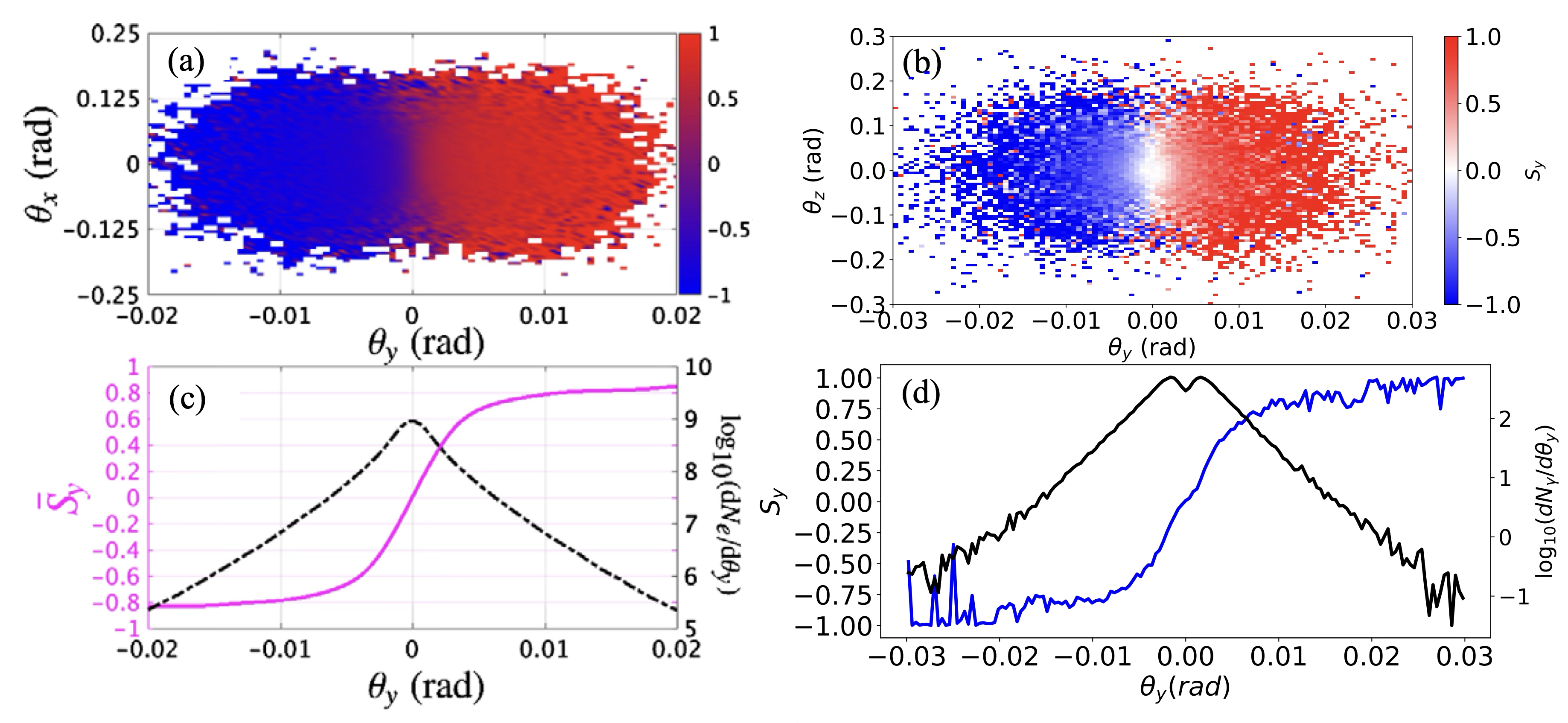}
\centering
\caption{Spin polarization from colliding with an elliptically polarized laser. (a), (b) The spin distribution in the $\theta_y-\theta_z$ base space after the interaction. (c), (d) The spin and density distribution with respect to the deflection angle $\theta_y$. (a),(c) is from the paper \cite{Li_PRL_2019}. (b), (d) from the OSIRIS spin and polarization-resolved QED simulation. Subplots (a) (c) reproduced from Yan-Fei Li et al. Phys. Rev. Lett. \textbf{122}, 154801 (2019).{\textcopyright} American Physical Society.} 
\label{fig:elliptical-polarized_laser_model_compare}
\end{figure}

In their setup, the laser beam has a peak intensity of $a_0 = 100$, a wavelength of $\lambda_0 = 1\ \mu\text{m}$, a pulse duration of $\tau_0 = 16.7\ \text{fs}$, and a focal radius of $w_0 = 5\ \mu\text{m}$. The electron beam profile follows a Gaussian distribution in both transverse and longitudinal directions, with a width of $\sigma_y = 1\ \mu\text{m}$ and a length of $\sigma_x = 5\ \mu\text{m}$. The initial energy of the electron beam is 4 GeV, with an energy spread of $6\%$.

Figure~\ref{fig:elliptical-polarized_laser_model_compare} shows the results of our spin- and polarization-resolved QED simulations compared to those reported in \cite{Li_PRL_2019}. The simulated spin distribution in the $\theta_y$–$\theta_z$ phase space closely matches their findings. The results confirm that the degree of polarization increases with the electron deflection angle along the minor axis of the elliptically polarized laser. Our simulations yield a maximum polarization of about $80\%$ for $|\theta_y| < 0.02$, which agrees well with the predictions in the referenced paper.

\subsubsection{Linear Polarization Gamma-Ray in Laser-Solid Target Interaction}

We next examine the generation of dense, linearly polarized gamma-ray beams in the context of laser–solid interactions. In this study, a strong linearly polarized (LP) laser pulse is directed onto a near-critical-density (NCD) hydrogen plasma, backed by an ultra-thin planar aluminum (Al) target. Electrons in the plasma are accelerated to ultra-relativistic energies and subsequently collide head-on with the reflected laser pulse from the Al target, resulting in copious emission of linearly polarized gamma photons.

\begin{figure}[h!]
\includegraphics[width=\textwidth]{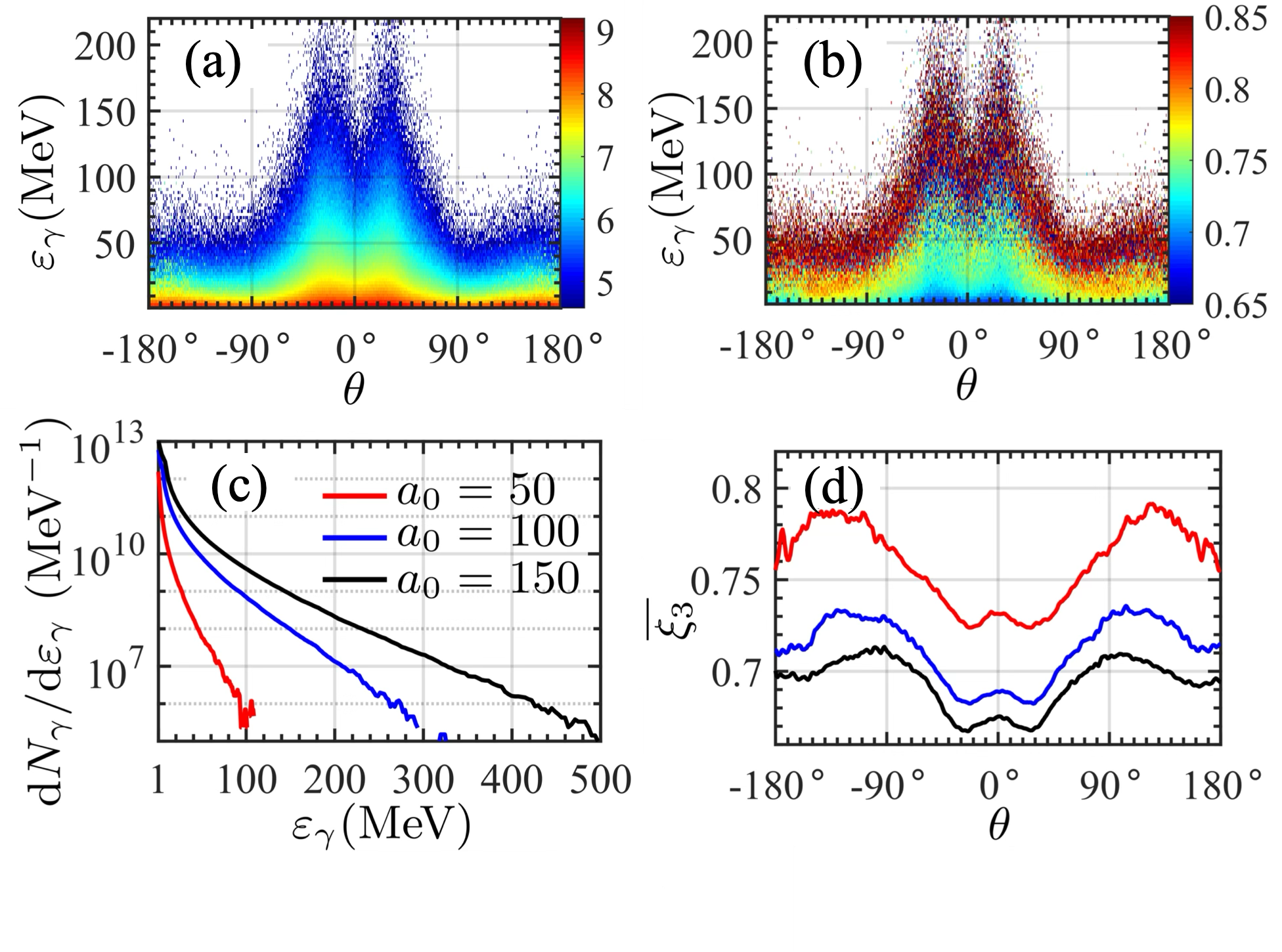}
\centering
\caption{Linear polarization gamma-ray in laser-solid target interaction, from paper \cite{Xue_MRE_2022}. (a) Photon density distribution in the photon energy $\varepsilon_\gamma$ -deflection angle $\theta$ phase space. (b) Photon linear polarization distribution in the $\varepsilon_\gamma -\theta$ phase space. (c) Photon density distribution with respect to the photon energy $\varepsilon_\gamma$ for laser $a_0 = 50,\ 100,\ 150$. (d) Photon linear polarization distribution with respect to the deflection angle $\theta$ for lasers $a_0 = 50,\ 100,\ 150$. This figure is from ``Generation of highly-polarized high-energy brilliant $\gamma$-rays via laser-plasma interaction'' Matter Radiat. Extremes \textbf{5}, 054402 (2020)  by Kun Xue et al., which is licensed under CC BY 4.0.} 
\label{fig:linear-polarization_from_paper}
\end{figure}

The peak intensity of the laser, $a_0$, is varied between 50, 100, and 150, while the wavelength is fixed at $\lambda_0 = 1,\mu\text{m}$, pulse duration at $\tau_0 = 30,\text{fs}$, and focal radius at $w_0 = 5,\mu\text{m}$. The planar Al target has an electron density of $n_e^{\mathrm{Al}} = 702 n_c$ and thickness $d_{\mathrm{Al}} = 1,\mu\text{m}$, where the plasma critical density is $n_c = m\omega_0^2/4\pi e^2 \approx 1.1\times 10^{21},\text{cm}^{-3}$. On the laser-incident side of the Al target is the NCD hydrogen plasma with electron density $n_e = 5 n_c$ and thickness $d_p = 10,\mu\text{m}$.

\begin{figure}[h!]
\includegraphics[width=\textwidth]{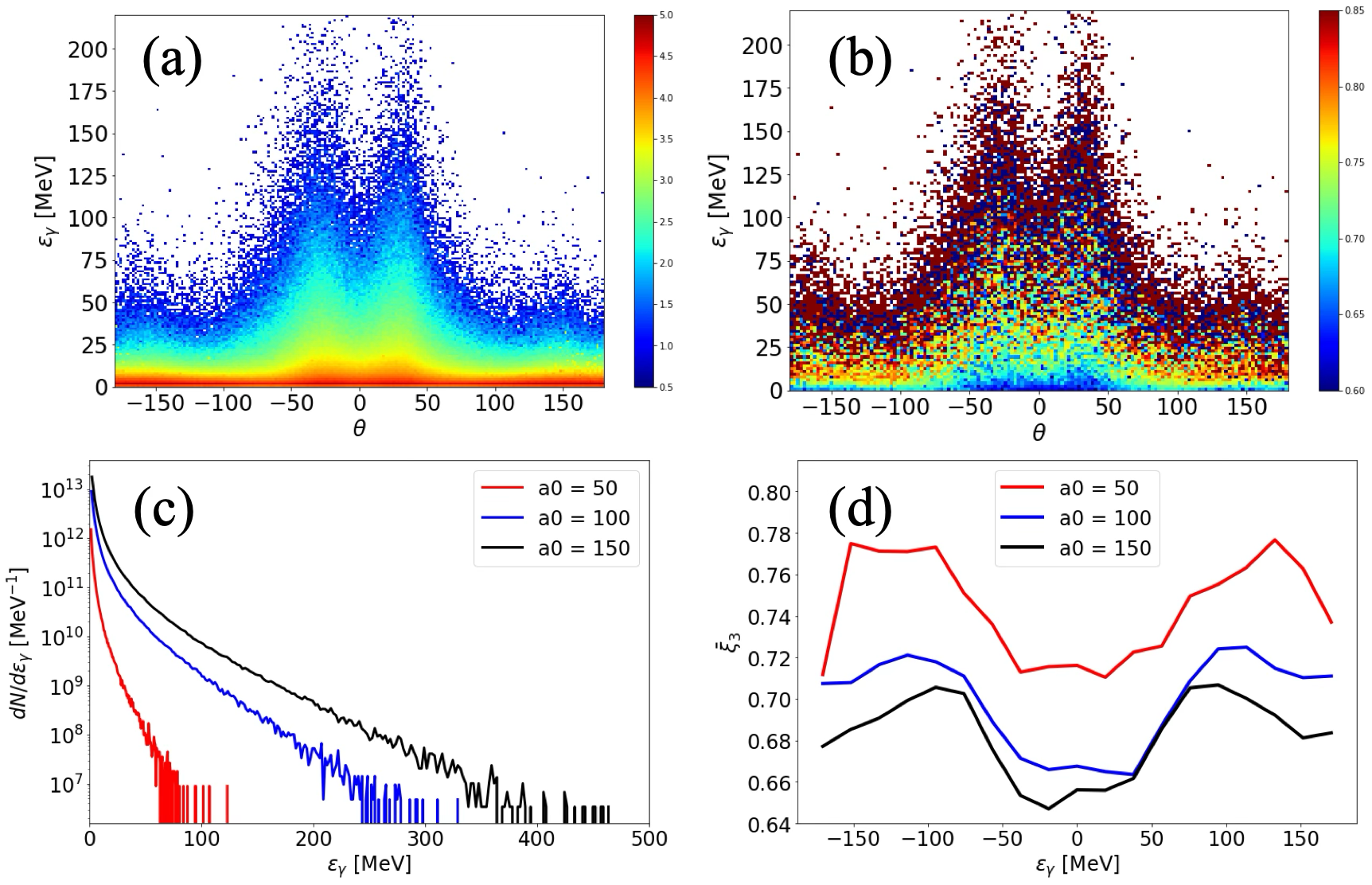}
\centering
\caption{Linear polarization gamma-ray in laser-solid target interaction, from the OSIRIS spin and polarization-resolved QED simulation. (a) Photon density distribution in the photon energy $\varepsilon_\gamma$ -deflection angle $\theta$ phase space. (b) Photon linear polarization distribution in the $\varepsilon_\gamma -\theta$ phase space. (c) Photon density distribution with respect to the photon energy $\varepsilon_\gamma$ for laser $a_0 = 50,\ 100,\ 150$. (d) Photon linear polarization distribution with respect to the deflection angle $\theta$ for lasers $a_0 = 50,\ 100,\ 150$.} 
\label{fig:linear_polarization_from_code}
\end{figure}

The main results from the reference paper \cite{Xue_MRE_2022} are illustrated in Fig.\ref{fig:linear-polarization_from_paper}: panels (a) and (b) show the spatial density and degree of linear polarization of the emitted gamma rays for $a_0 = 100$; panel (c) presents the gamma-ray energy spectra for $a_0 =$ 50, 100, and 150; and panel (d) displays the angular distribution of linear polarization. Our simulation results, shown in Fig.\ref{fig:linear_polarization_from_code}, closely reproduce these findings, even though fewer particles were used in our model, resulting in reduced computational cost. Nevertheless, we capture the essential conclusion: higher laser intensities generate more gamma-ray photons with higher energies, but at the expense of reduced linear polarization.

\subsubsection{Helicity Transfer in a Laser Field due to the Electron Anomalous Magnetic Moment}

Yan-fei Li et al. \cite{Li_PRL_2022} discovered a remarkable phenomenon: helicity can be transferred from a circularly polarized (CP) laser pulse to an ultra-relativistic electron beam via the non-linear Compton scattering process within the radiation reaction-dominated regime. Their study identifies the anomalous magnetic moment term in the classical spin precession as a critical factor in enabling this helicity transfer.

In their simulations \cite{Li_PRL_2022}, a right-handed CP, tightly focused Gaussian laser pulse is used with a peak intensity of $a_0 = 100\sqrt{2}$, wavelength $\lambda_0 = 1\ \mu\text{m}$, pulse duration $\tau_0 = 16.7\ \text{fs}$, and focal radius $w_0 = 5\ \mu\text{m}$. The counter-propagating electron bunch follows Gaussian distributions both transversely and longitudinally, with a width of $\sigma_y = 1\ \mu\text{m}$ and a length of $\sigma_x = 5\ \mu\text{m}$. The initial electron energy is 1 GeV, with a $10\%$ energy spread.

\begin{figure}[h!]
\includegraphics[width=\textwidth]{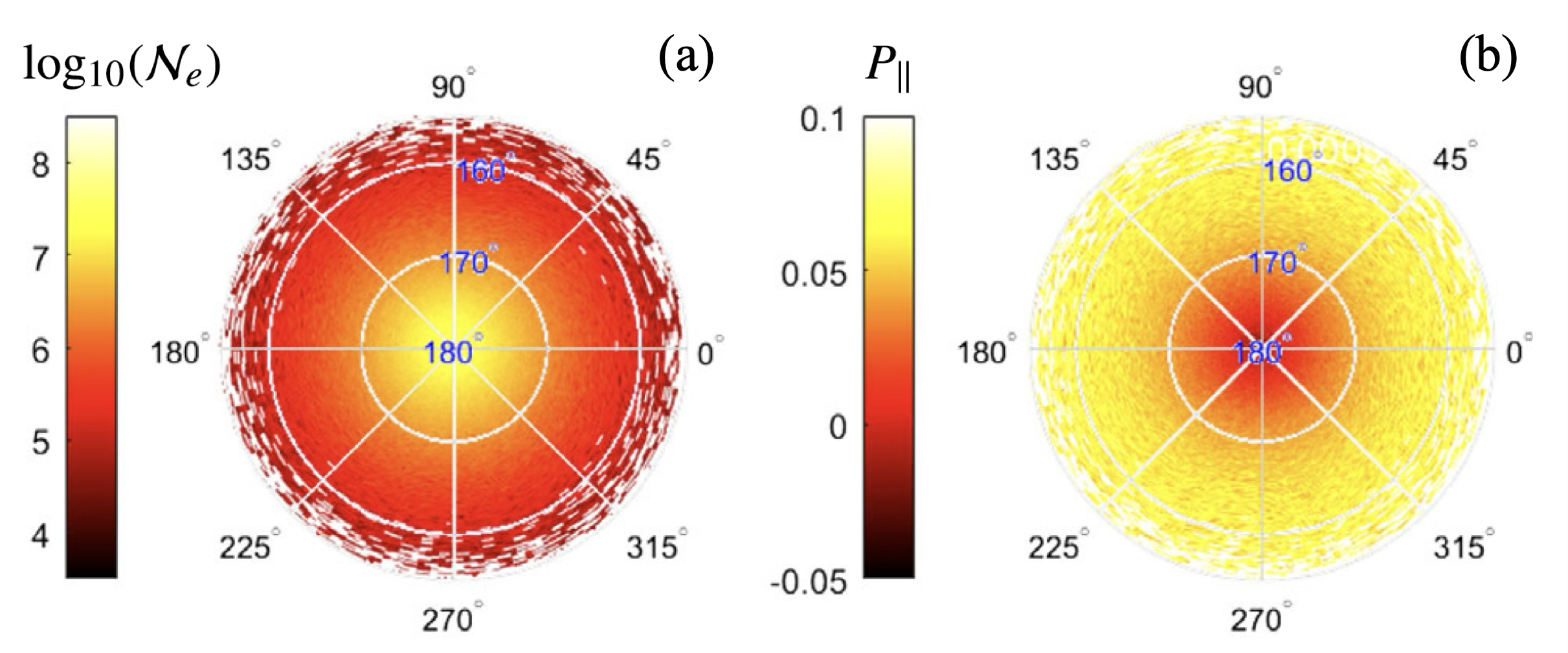}
\centering
\caption{Helicity transfer in a laser field due to the electron anomalous Magnetic Moment, from the paper \cite{Li_PRL_2022}. (a) Polar plot of electron density distribution (b) Polar plot of electron longitudinal spin distribution. Figure reproduced from Yan-Fei Li et al. Phys. Rev. Lett. \textbf{128}, 174801 (2022) {\textcopyright} American Physical Society.} 
\label{fig:helicity_transfer_from_paper}
\end{figure}

\begin{figure}[h!]
\includegraphics[width=\textwidth]{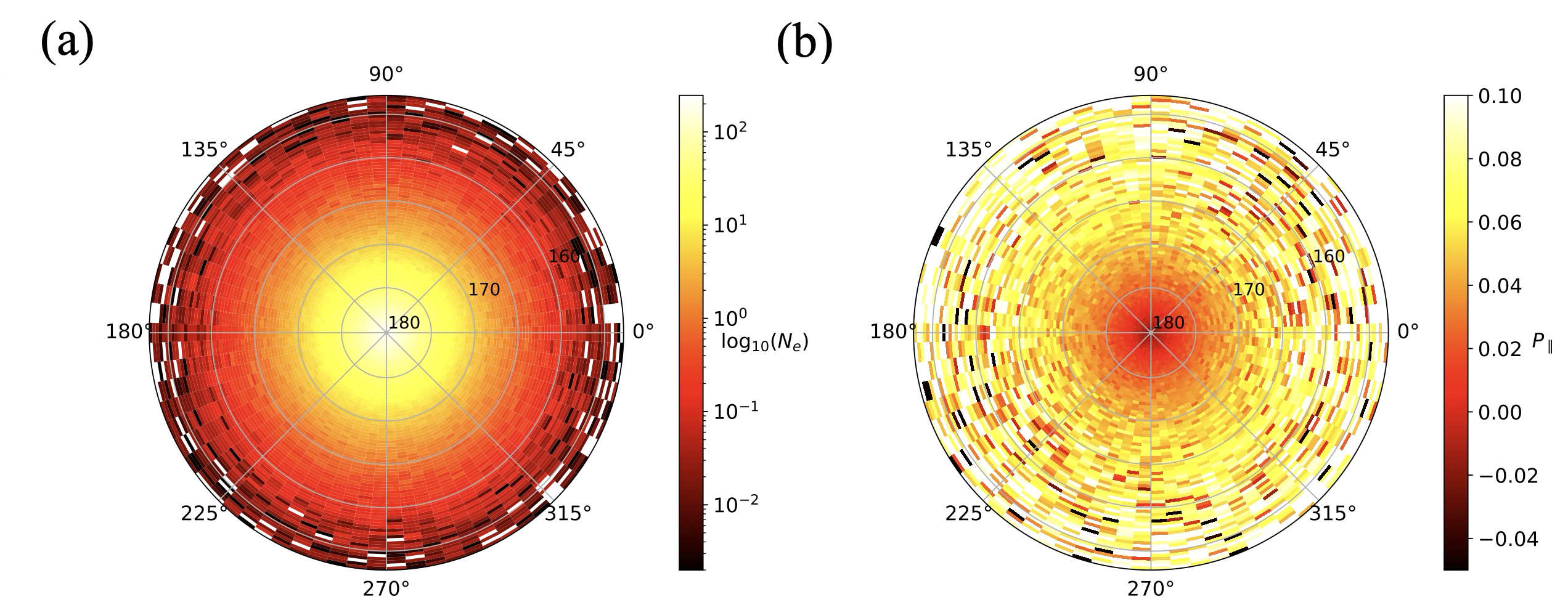}
\centering
\caption{Helicity transfer in laser field due to the electron Anomalous Magnetic Moment, from OSIRIS spin and polarization-resolved QED simulation. (a) Polar plot of electron density distribution (b) Polar plot of electron longitudinal spin distribution.} 
\label{fig:helicity_transfer_from_code}
\end{figure}

Figure~\ref{fig:helicity_transfer_from_paper} from their paper presents the post-collision electron density and longitudinal polarization distributions. Our simulation results, shown in Fig.\ref{fig:helicity_transfer_from_code}, closely match these findings. Specifically, our computed average longitudinal polarization is approximately $2.2\%$, aligning well with the $2.65\%$ reported in the paper. The data show that longitudinal polarization increases with the electron deflection angle, changing from about $-5\%$ at the beam center to around $10\%$ further away from the center.

\begin{figure}[h!]
\includegraphics[width=\textwidth]{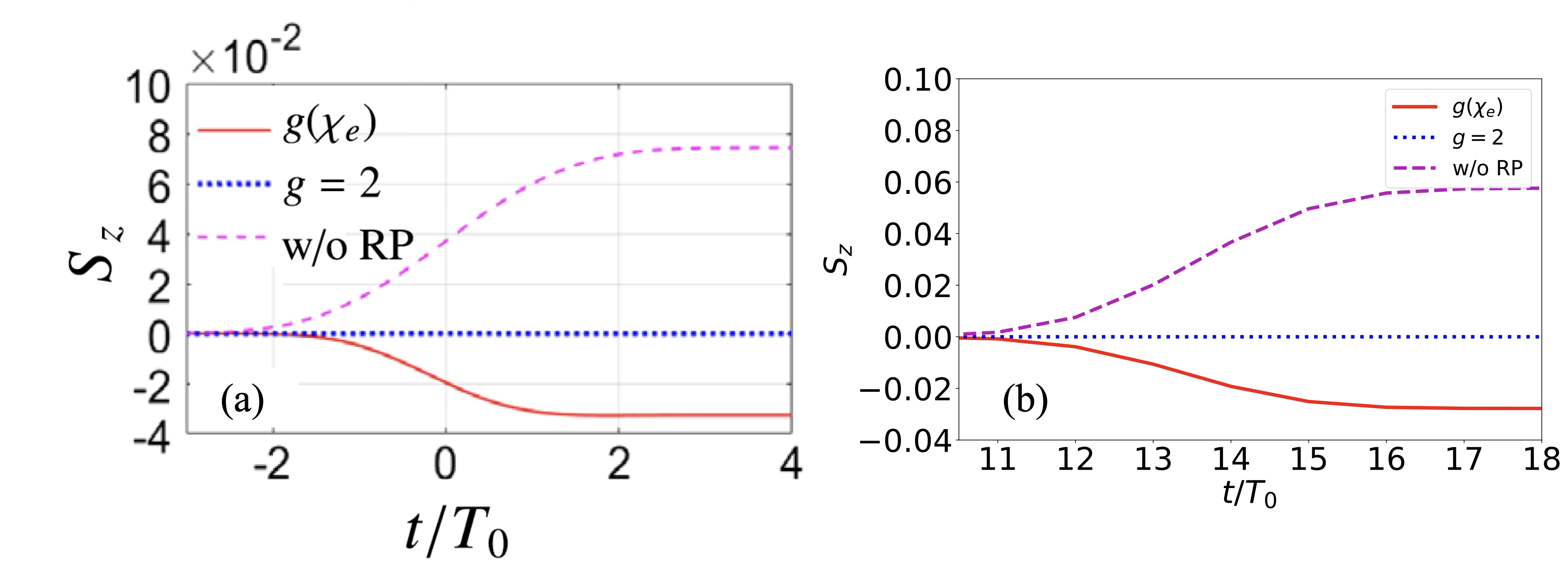}
\centering
\caption{Helicity transfer in laser field due to the electron anomalous magnetic moment, (a) from paper \cite{Li_PRL_2022}(b) from OSIRIS spin and polarization-resolved QED simulation. Subplot (a) is reproduced from Yan-Fei Li et al. Phys. Rev. Lett. \textbf{128}, 174801 (2022) {\textcopyright} American Physical Society.} 
\label{fig:helicity_transfer_g_factor}
\end{figure}

The role of the anomalous magnetic moment is highlighted in Fig.~\ref{fig:helicity_transfer_g_factor}, where panel (a) is taken from the reference and panel (b) is from our simulation. The red curve represents results with the true anomalous magnetic moment $g=g(\chi_e)$, while the blue dashed curve corresponds to a fixed $g = 2$. The purple curve shows $g=g(\chi_e)$ but only includes non-emission polarization, omitting radiative polarization. Our results are consistent with those in the paper: when $g = 2$, there is no helicity transfer; only when $g \neq 2$ does the effect appear. Furthermore, transverse spin polarization is also significant in this process—the presence or absence of radiative polarization results in opposite outcomes for longitudinal polarization.

\subsubsection{Circular-Polarized High-Energy Gamma-Ray Pulse via Laser-Electron Interaction}

Next, we examine helicity transfer arising purely from spin- and polarization-resolved non-linear Compton scattering (NLCS) and non-linear Breit–Wheeler (NBW) processes. Yan-fei Li et al. \cite{Li_PRL_2020_gamma_ray} investigate helicity transfer in the polarization-resolved NLCS process by colliding a linearly polarized laser with a fully longitudinally polarized electron beam, resulting in the generation of a high-energy, circularly polarized gamma-ray beam.

In their setup, a linearly polarized, tightly focused Gaussian laser pulse is used with a peak intensity $a_0 = 50$, wavelength $\lambda_0 = 1\ \mu\text{m}$, pulse duration $\tau_0 = 33\ \text{fs}$, and focal radius $w_0 = 5\ \mu\text{m}$. The counter-propagating electron bunch is fully longitudinally polarized with an initial mean energy of 10 GeV and a $6\%$ energy spread, and is modeled using a transverse width of $\sigma_y = 2\ \mu\text{m}$ and a longitudinal length of $\sigma_x = 3\ \mu\text{m}$.

\begin{figure}[h!]
\includegraphics[width=\textwidth]{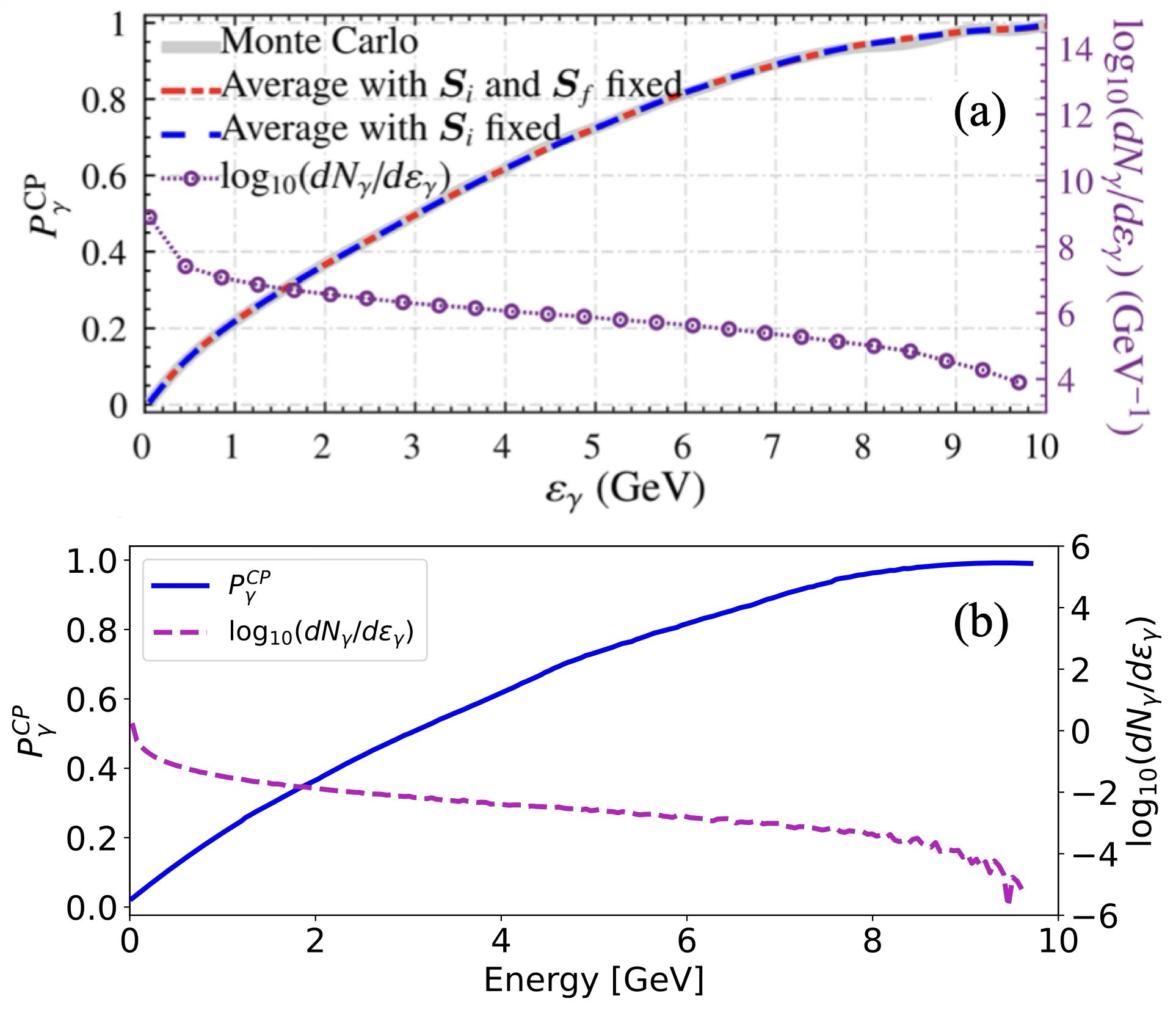}
\centering
\caption{Circular-polarized high-energy gamma-ray pulse via Laser-Electron Interaction (a) from paper \cite{Li_PRL_2020_gamma_ray}, (b) from OSIRIS spin and polarization-resolved QED simulation. Subplot (a) is reproduced from Yan-Fei Li et al. Phys. Rev. Lett. \textbf{125}, 044802  (2020) {\textcopyright} American Physical Society.} 
\label{fig:lon_to_cp}
\end{figure}

Figure~\ref{fig:lon_to_cp} (a) shows the generated gamma-ray energy spectrum and its degree of circular polarization, as reported in the paper. Our simulation results, presented in Fig.~\ref{fig:lon_to_cp}(b), closely reproduce these findings. Specifically, we observe that the degree of circular polarization of the emitted gamma-ray photons increases with photon energy, reaching nearly $100\%$ for the highest-energy photons.

\subsubsection{Production of Highly Polarized Positron Beams via Helicity Transfer from Polarized Electrons}

Helicity transfer is also observed in the polarization-resolved non-linear Breit–Wheeler (NBW) process, as demonstrated in Yan-Fei Li’s second PRL paper on this topic \cite{Li_PRL_2020_helicity_transfer}. In their study, a circularly polarized laser is collided with a fully longitudinally polarized electron beam. Here, the laser intensity is increased, enabling the highly circularly polarized gamma-ray photons generated in the NLCS process to be converted into longitudinally polarized positrons via the polarization-resolved NBW process.

The experimental setup uses a circularly polarized, tightly focused Gaussian laser pulse with a peak intensity $a_0 = 100$, wavelength $\lambda_0 = 1\ \mu\text{m}$, pulse duration $\tau_0 = 16.7\ \text{fs}$, and focal radius $w_0 = 5\ \mu\text{m}$. The counter-propagating electron bunch is fully longitudinally polarized with an initial mean energy of 10 GeV and a $6\%$ energy spread, described by a Gaussian profile with transverse width $\sigma_y = 0.6\ \mu\text{m}$ and longitudinal length $\sigma_x = 6\ \mu\text{m}$.

\begin{figure}[h!]
\includegraphics[width=\textwidth]{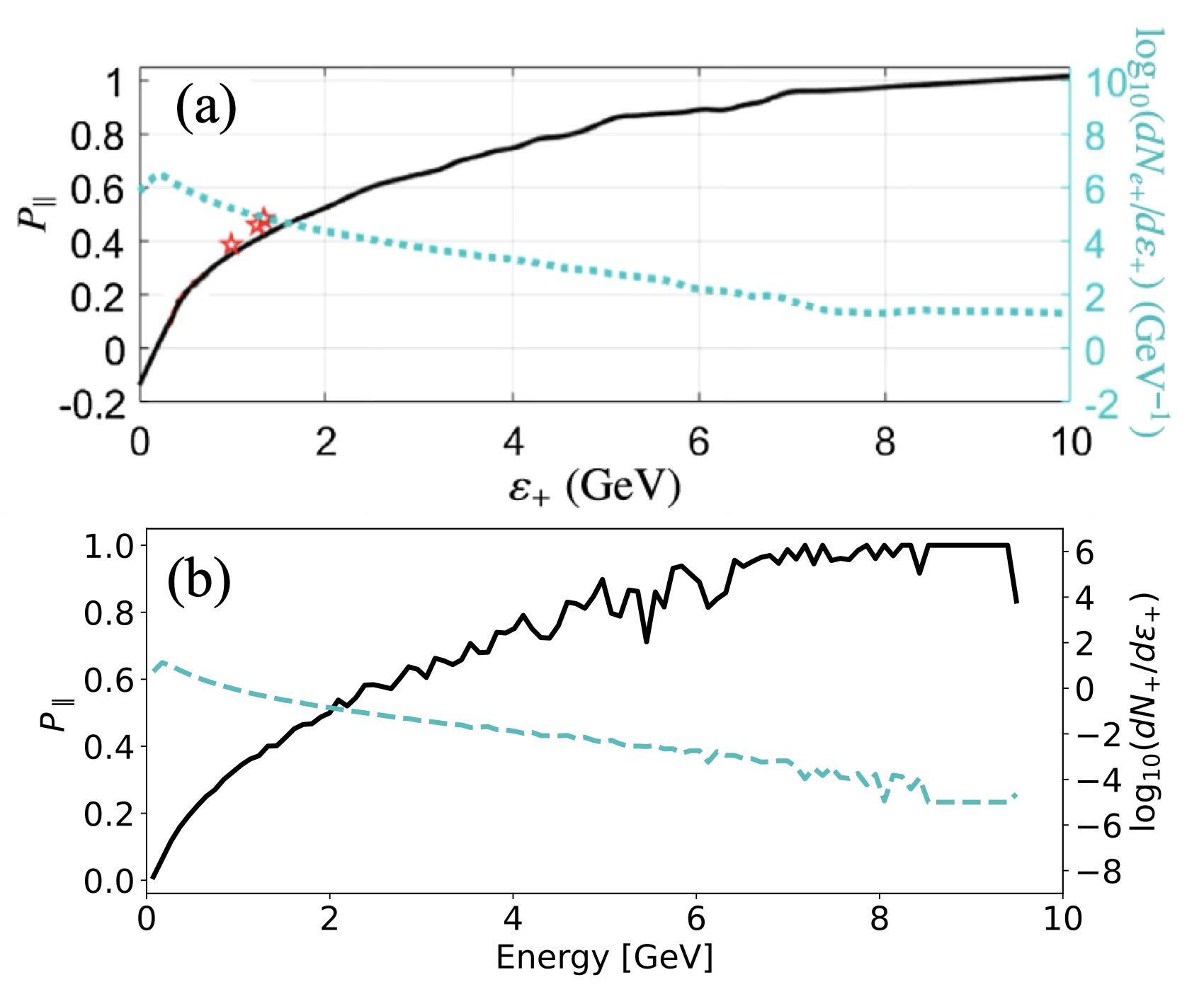}
\centering
\caption{Production of Highly Polarized Positron Beams via Helicity Transfer
from Polarized Electrons (a) from paper \cite{Li_PRL_2020_helicity_transfer}, (b) from OSIRIS spin and polarization-resolved QED simulation. Subplot (a) is reproduced from Yan-Fei Li et al.Phys. Rev. Lett. \textbf{124}, 014801 (2020) {\textcopyright} American Physical Society.} 
\label{fig:lon_to_cp_to_lon}
\end{figure}

Figure~\ref{fig:lon_to_cp_to_lon}(a) from the paper \cite{Li_PRL_2020_helicity_transfer} presents the generated positron energy spectrum and its degree of longitudinal polarization. Our simulation, shown in Fig.~\ref{fig:lon_to_cp_to_lon}(b), closely matches these results. We find that the degree of longitudinal polarization of the produced positrons increases with positron energy, reaching nearly $100\%$ for the highest-energy positrons, in agreement with the results reported in the paper.

\subsubsection{Polarized QED Cascade Implementation}
\label{cascade_test}
Finally, we use our polarization-resolved QED module to reproduce the results in Ref.~\cite{Seipt_NJP_2021}. This paper studied the seeded electron-positron pair cascade \cite{Bell_PRL_2008} in a rotating electric field calculated using a polarization-resolved Boltzmann-type solver. The polarized QED cascade is a sensitive process involving the interplay between the quantum emission rates and particle kinetics. Reproducing the main result in this paper can comprehensively test the performance of our spin and polarization-involved QED module. On the other hand, the intensity of the rotating electric field is uniform in space. This reduces the problem to 0-D, which greatly simplifies the simulations. Notice that the paper uses notation $\parallel$ and $\perp$ instead of $\pmb \sigma$ and $\pmb \pi$ for the photon polarization state, which has a similar meaning. 

\begin{figure}[h!]
\includegraphics[width=\textwidth]{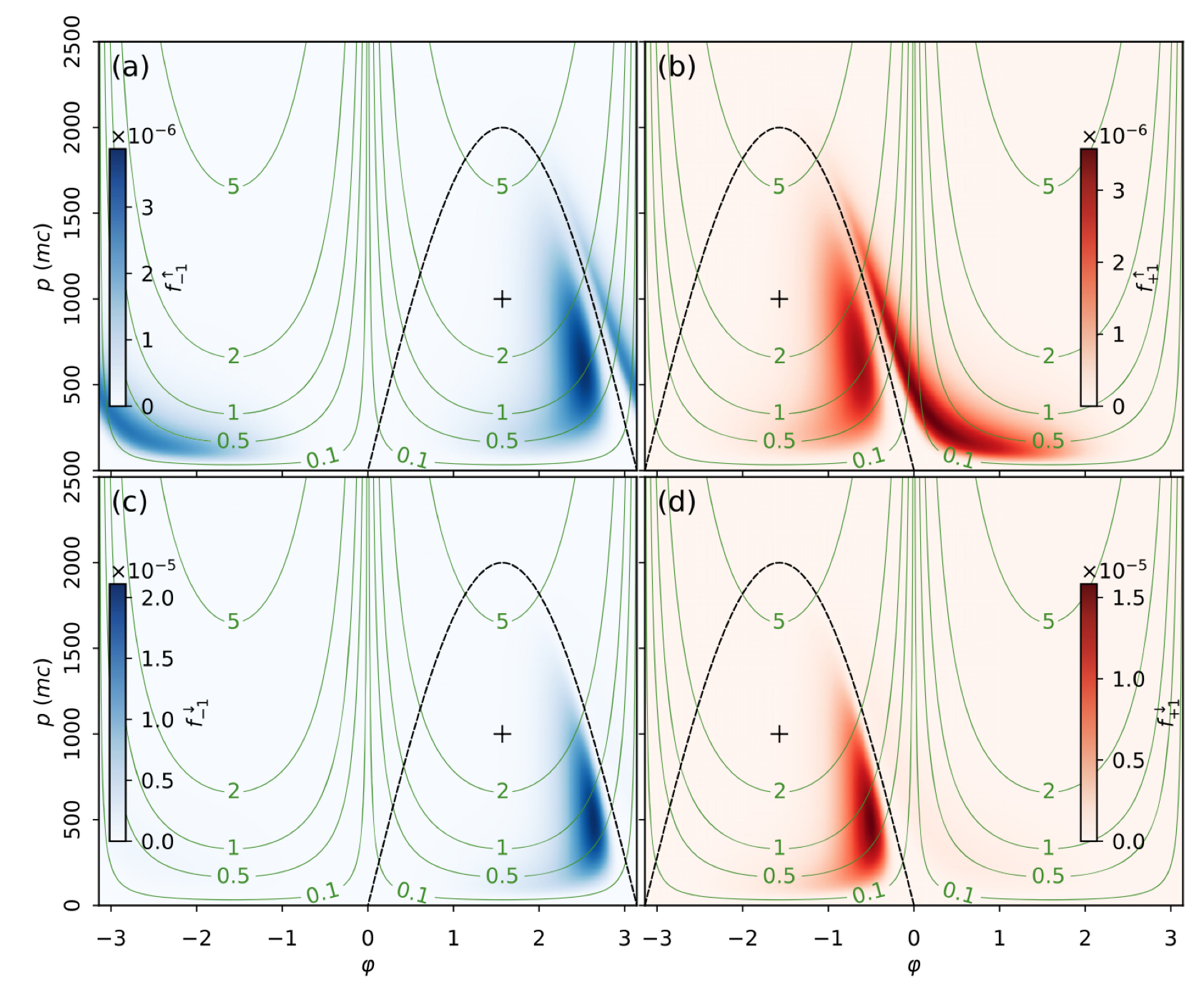}
\centering
\caption{Snapshot of the electron (a), (c) and positron (b), (d) distribution functions in an up (a), (b) or down (c), (d) spin state for $a_0 = 10^3$ and $\omega t = 10$ in a rotating radial frame. Green curves are $\chi$ isocontours. Black dashed curves represent the separatrix of the classical advection $p = -2a_0q \sin\varphi$, and black crosses are the corresponding fixed points at $\varphi = -q \pi/2.0$, $p = a_0$. This figure is from ``Polarized QED cascades'' New J. Phys. \textbf{23} 053025 (2022) by D. Seipt, C. P. Ridgers, D. Del Sorbo, and A. G. R. Thomas, which is licensed under CC BY 4.0.}
\label{fig:phase_space}
\end{figure}

\begin{figure}[h!]
\includegraphics[width=\textwidth]{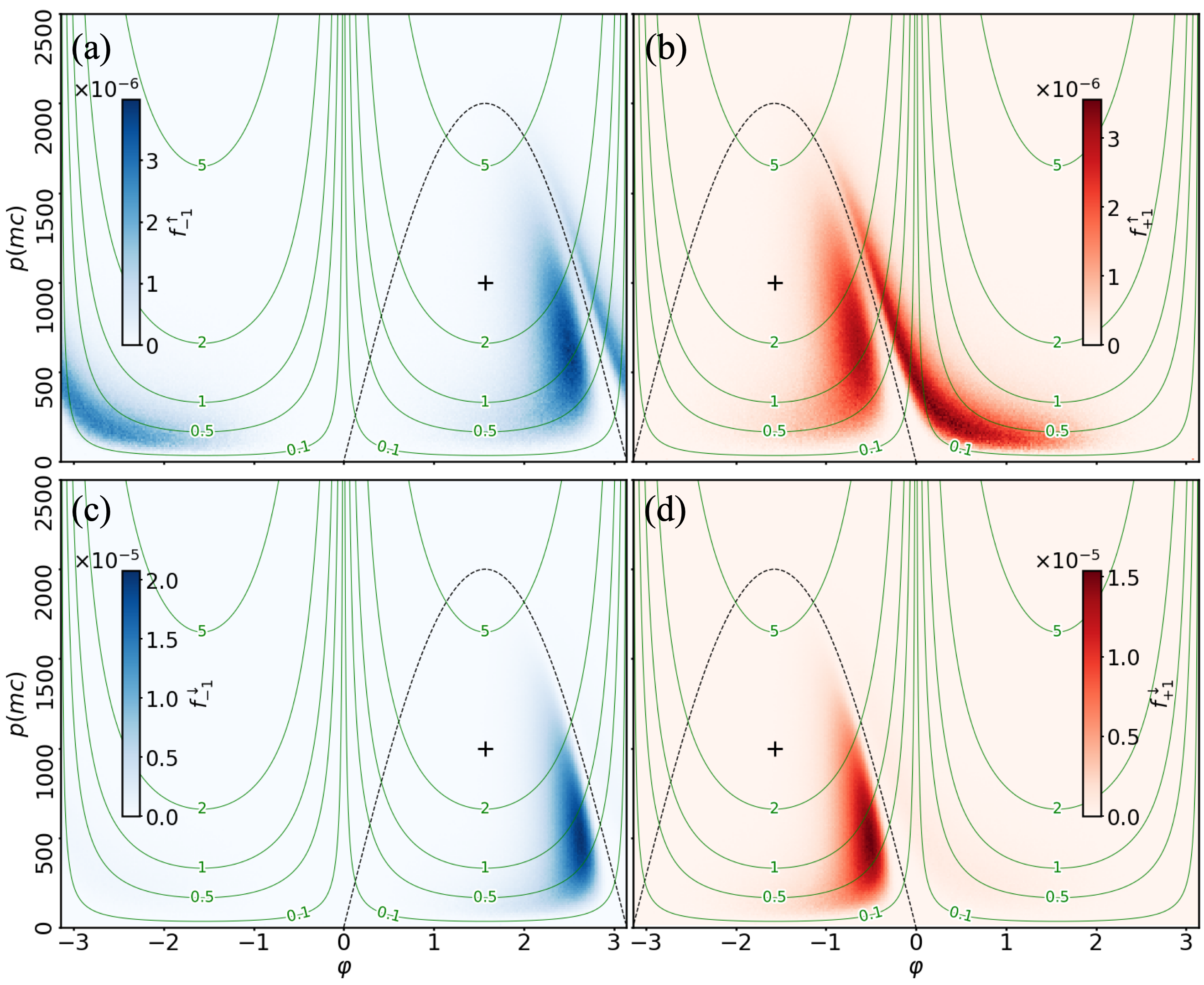}
\centering
\caption{Snapshot of the electron (a), (c) and positron (b), (d) distribution functions in an up (a), (b) or down (c), (d) spin state for the same conditions as the paper calculated using Osiris spin and polarization-dependent QED module.}
\label{fig:PIC_phase_space}
\end{figure}

The first part of the test is to reproduce the electron-seeded cascade result. Following the conditions given in the paper: laser intensity $a_0 = 1000$ and $\omega = 1.55\ eV$, I obtained the electron and positron distributions shown in Fig.~\ref{fig:PIC_phase_space} using our polarization resolved QED module. This is almost identical to Fig.~\ref{fig:phase_space} from the paper, which shows that my PIC code calculation result agrees with the calculation using the Boltzmann-type kinetic equations. For both Figs.~\ref{fig:phase_space} and~\ref{fig:PIC_phase_space}, the spin-down distributions for the electrons and positrons have the highest peak value inside the black dashed line separatrix. This separatrix is the classical advection for the leptons inside the rotation field without radiation energy loss. Due to the difference in the spin up-to-down and down-to-up transition rate of the quantum radiation process, spin-down leptons have a larger population and accumulate inside the separatrix. The particles outside the separatrix come from the pair production process initiated by photons generated from the oppositely charged particles due to their distribution being in different locations in phase space. The pair production process in this simulation generates a similar number of spin-up and spin-down leptons. The spin-up distribution inside the separatrix has a much lower peak than the spin-down distribution, so the distribution outside the separatrix for spin-up leptons is more significant. 


\begin{figure}[h!]
\includegraphics[width=\textwidth]{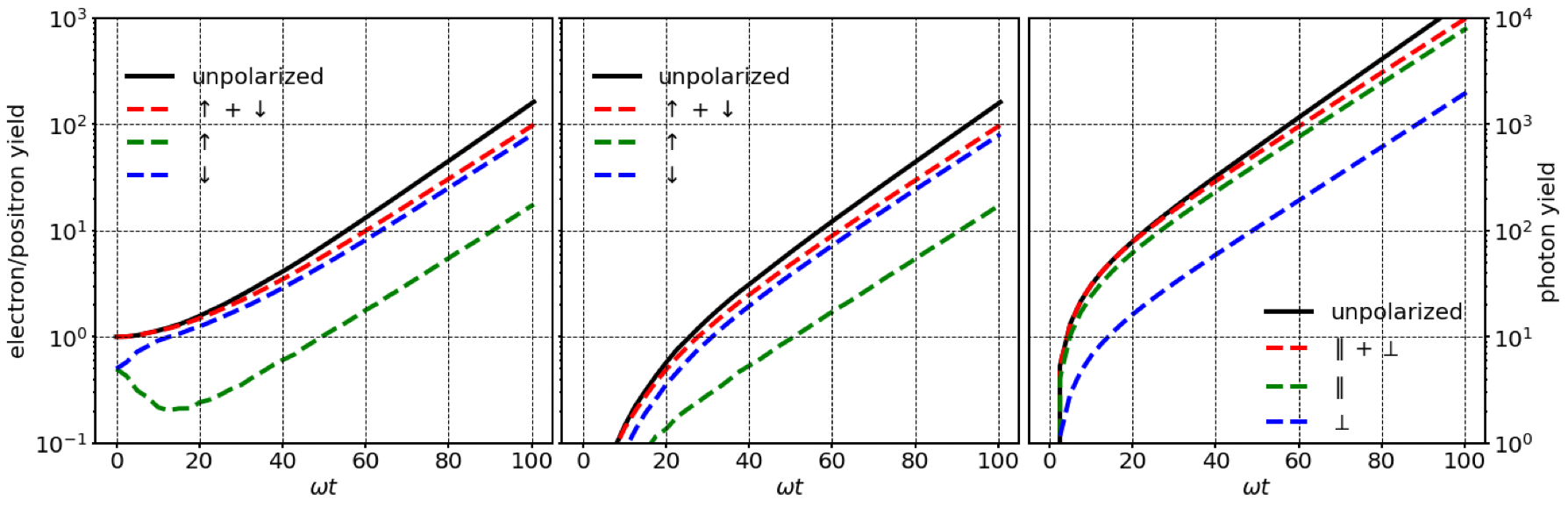}
\centering
\caption{Time evolution of electron (a), a positron (b), and photon (c) yields during a cascade seeded by unpolarized electrons calculated by the Osiris spin and polarization QED module.}
\label{yield_lepton_seeded}
\end{figure}

Fig.~\ref{yield_lepton_seeded} shows the time evolution of electrons, positrons, and photon yields during a cascade seeded with unpolarized electrons calculated using my QED module. Compared with the result in the paper calculated using the Boltzmann-type solver (see figure 3 in Ref.~\cite{Seipt_NJP_2021}), the PIC code gives a similar outcome. During the cascade process, the quantum radiation process controls the spin and polarization distribution of the leptons and photons, while the pair production process decides the growth rate of the leptons. Initially, the number of spin-up leptons decreases due to the asymmetry of the spin transition in the quantum radiation process. As the cascade process develops, the growth rate of the lepton becomes constant, and the spin-up-to-down and down-to-up transitions are balanced. There are five times more spin-down leptons than spin-up leptons. The ratio between the photons in different polarization states also becomes a constant in the exponential growth phase. A factor of four more $\parallel$-polarized photons is emitted compared to $\perp$-polarized photons. The particles produced in this QED cascade are highly polarized. 

\begin{figure}[h!]
\includegraphics[width=\textwidth]{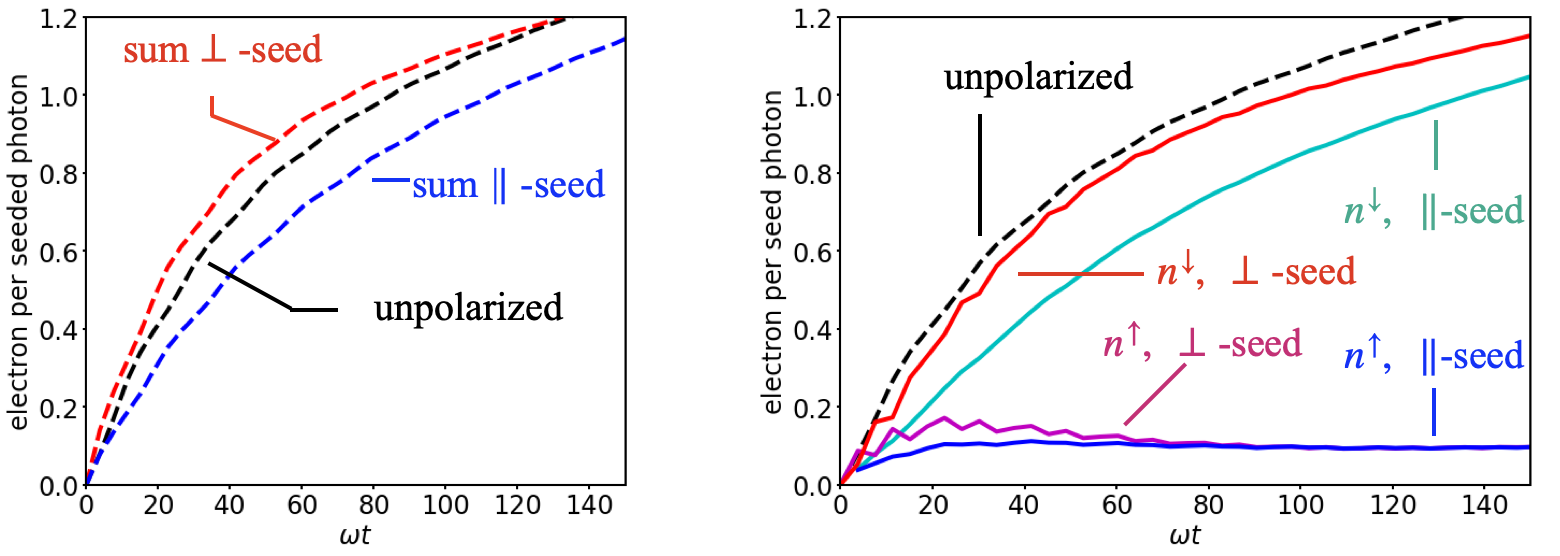}
\centering
\caption{ Time evolution of the electron yield of photon-seeded cascade calculated using Osiris spin and polarization QED module.}
\label{yield_photon_seeded}
\end{figure}

The second test is to reproduce the time evolution of the particle yields for QED-cascade seeded with the polarized photon. Following the same initial conditions in the paper, the simulation results from my polarization QED module are shown in Fig.~\ref{yield_photon_seeded}. This is similar to the result in the paper (figure 5 in Ref.~\cite{Seipt_NJP_2021}). The left plot shows that at the early stage, $\perp$-polarized photon-seeded cascade has almost two times higher yield than $\parallel$-polarized photon-seeded cascade. As the cascade process develops, the seeding photons are depleted, and the difference in yield becomes smaller. The plot on the right shows the population of spin-down and spin-up electrons. The $\perp$-polarized photon-seeded cascade initially generates more spin-down electrons than the $\parallel$-polarized photon-seeded cascade. The seeding photons are depleted as time passes, and the quantum radiation effect on the lepton spin distribution becomes dominant. The number of the spin-down and spin-up leptons for both $\perp$-polarized and $\parallel$-polarized photon-seeded cascades would finally become the same.

\section{Conclusions}
In conclusion, we have demonstrated our spin- and polarization-resolved QED module, built on a PIC code, which has been well tested for accuracy and reproduces many published results. This algorithm enables self-consistent simulations of processes that can occur in future high-intensity laser-plasma interaction experiments and under extreme astrophysical conditions. Looking ahead, we plan to explore the dynamics of relativistic plasmas in extreme fields, including relativistic magnetic reconnection, shocks, and current filamentation. These processes include gamma-ray bursts from neutron star mergers, black hole formation, and even extreme lab astrophysics experiments involving high-intensity lasers interacting with targets. The polarization signals in these extreme plasma processes can be unique: for instance, studies have shown that polarization features can uncover mechanisms responsible for turbulence transitions in relativistic collisionless shocks. We aim to delve into the unique information carried by photon polarization, and potentially by lepton spin as well, driven by our curiosity to understand some of the most energetic and luminous phenomena occurring billions of light-years away.

Currently, our Monte Carlo-based approach involves comparing a random number to the event rate, but its accuracy depends heavily on the time step. However, the optical depth method can maintain accuracy without being constrained by time step size. Implementing this method in the future could enhance the efficiency of our algorithm. The large number of particles generated by extremely strong-field QED events, such as QED cascades, consumes substantial memory. The memory requirements and run time increase exponentially with both laser duration and intensity, significantly limiting the parameter space we can explore for extreme QED processes. A potential solution is the particle merging algorithm. By combining particles that are close in both location and momentum phase space into a single particle, we can significantly reduce memory usage, particularly as large numbers of particles are generated by NLCS and NBW processes. This technique has already been applied to an unpolarized strong field QED model in OSIRIS. However, a detailed study is needed to determine whether we can apply particle merging similarly while accounting for spin and polarization states.

Future extensions of our algorithm could include additional physical models. For instance, incorporating vacuum birefringence effects for photons would be valuable for studying changes in photon polarization as they propagate through strong magnetic fields in extreme astrophysical environments, as well as for upcoming vacuum birefringence experiments at XFEL facilities. Including spin and polarization-resolved linear Compton scattering (LCS) and linear Breit-Wheeler (LBW) processes is also crucial. These processes occur more frequently in extreme astrophysical environments compared to NLC and NBW. For example, polarized synchrotron soft photons can undergo LCS with cold relativistic electrons in an outflow, producing hard gamma-ray photons—a process known as synchrotron self-Compton. These high-energy photons can then collide to generate electron-positron pairs through the LBW process. Including LCS and LBW would enable a more comprehensive simulation of the physics occurring in extreme astrophysical objects.

\appendix
\section{Spin and Polarization Resolved-QED Cross Section}
\label{appA}

According to equation 7 in \cite{Torgrimsson_NJP_2021}, the probability of the NLCS or NBW process can be expressed as:

\begin{equation}
\begin{split}
{\mathbb{P}} = & \langle {\mathbb{P}} \rangle + \textbf{n}_{\gamma}\cdot \textbf{P}_{\gamma} + \textbf{n}_{1}\cdot \textbf{P}_{1} + \textbf{n}_{0}\cdot \textbf{P}_{0} + \textbf{n}_{\gamma}\cdot \textbf{P}_{\gamma 1} \cdot \textbf{n}_{1} + \textbf{n}_{\gamma}\cdot \textbf{P}_{\gamma 0} \cdot \textbf{n}_{0}\\ & + \textbf{n}_{1}\cdot \textbf{P}_{1 0} \cdot \textbf{n}_{0}  + \textbf{P}_{\gamma 1 0, ijk} \textbf{n}_{\gamma i}  \textbf{n}_{1 j}  \textbf{n}_{0 k},
\end{split}
\end{equation}
which $\textbf{n}_0$ and $\textbf{n}_1$ are the Lepton's initial and final Stokes vector. $\textbf{n}_\gamma$ is the  stokes vector for the photon:
\begin{equation}
\textbf{n}_{\gamma} = \pmb{z} = \xi_1\pmb{\epsilon}_{EB} + \xi_2\pmb{\epsilon}_{2} + \xi_3\pmb{\epsilon}_{E}.
\end{equation}

\subsection{NLCS Process}
\label{appendix_NLCS}
Under LCFA approximation, the photon emission rate by an electron can be written as:

\begin{equation}
    \{ \langle \mathbb{P}^{\text{C}}\rangle, \textbf{P}_0^{\text{C}},\cdots\}=\frac{\alpha}{4}\int\frac{d\sigma}{b_0}\{ \langle \mathbb{\hat R}^{\text{C}}\rangle, \textbf{R}_0^{\text{C}},\cdots\},
\end{equation}

Here, $b_0 = kp$ is the dimensionless longitudinal momentum of the original particle that entered the laser. According to Torgrimsson \cite{Torgrimsson_NJP_2021} equation 43-50, we obtain the polarization-resolved emission spectrum:

\begin{equation}
\langle \hat {\mathbb{R}}^{\text{C}} \rangle  = -\text{Ai}_1(z) - \kappa \frac{\text{Ai}'(z)}{z},
\end{equation}

\begin{equation}
\hat{\textbf{R}}_0^{\text{C}} = \lambda\frac{\text{Ai}(z)}{\sqrt{z}}\hat {\pmb{\beta}},
\end{equation}

\begin{equation}
\hat{\textbf{R}}_1^{\text{C}}   = u\frac{\text{Ai}(z)}{\sqrt{z}}\hat {\pmb{\beta}},
\end{equation}

\begin{equation}
\hat{\textbf{R}}_\gamma^{\text{C}}  =- \frac{\text{Ai}'(z)}{z}\hat {\pmb{\varepsilon}}\cdot {\textbf{S}} \cdot \hat {\pmb{\varepsilon}},
\end{equation}

\begin{equation}
\hat{\textbf{R}}_{01}^{\text{C}} = -\text{Ai}_1(z)(\pmb{1}_{\perp}+[\kappa-1]\pmb{1}_{\parallel}) - \frac{\text{Ai}'(z)}{z}(2\pmb{1}_{\perp}+\kappa\pmb{1}_{\parallel}),
\end{equation}

\begin{equation}
\hat{\textbf{R}}_{\gamma 0}^{\text{C}} = -u \frac{\text{Ai}(z)}{\sqrt{z}}\textbf{S}\cdot\hat {\pmb{\beta}}-\lambda\left( \text{Ai}_1(z)+\left[1+\frac{1}{1-\lambda}\right]\frac{\text{Ai}'(z)}{z}\right)\pmb{\epsilon}_2\hat{\textbf{k}},
\end{equation}

\begin{equation}
\hat{\textbf{R}}_{\gamma 1}^{\text{C}} = -\lambda \frac{\text{Ai}(z)}{\sqrt{z}}\textbf{S}\cdot\hat {\pmb{\beta}}-u\left( \text{Ai}_1(z)+\left(2-\lambda\right)\frac{\text{Ai}'(z)}{z}\right)\pmb{\epsilon}_2\hat{\textbf{k}},
\end{equation}

\begin{equation}
\begin{split}
\hat{\textbf{R}}_{\gamma 01}^{\text{C}} = & -\frac{\lambda u}{2}\text{Ai}_1(z)\textbf{S} - \frac{\text{Ai}'(z)}{z}\hat{\pmb{\varepsilon}}\cdot\textbf{S}\cdot\hat{\pmb{\varepsilon}}\left(\frac{\kappa}{2}\pmb{1}_{\perp}+\pmb{1}_{\parallel}\right)\\
& -\frac{\tilde\kappa}{2}\frac{\text{Ai}'(z)}{z}\hat{\pmb{\beta}}\cdot\textbf{S}\cdot\hat{\pmb{\varepsilon}}\  i\pmb{\sigma}^{(3)}_2  + \frac{\text{Ai}(z)}{\sqrt{z}}\pmb{\epsilon}_2\left(u{\hat{\textbf{k}}\hat{\pmb{\beta}}}-\lambda{\hat{\pmb{\beta}}\hat{\textbf{k}}}\right).
\end{split}
\end{equation}

Here, $\lambda$ is the momentum ratio between the radiated photon and the initial energy of the electron, $u = \lambda/(1-\lambda)$ is the momentum ratio between the radiated photon and the final energy of the electron. $\kappa = 1/(1-\lambda)+ (1-\lambda)$, $\tilde \kappa =  1/(1-\lambda) - (1-\lambda)$. $\textbf{S} = \pmb{\epsilon}_1\pmb{\sigma}_1^{(3)}+\pmb{\epsilon}_3\pmb{\sigma}_3^{(3)} = \pmb{\epsilon}_E(\hat{\pmb{\varepsilon}}\hat{\pmb{\varepsilon}}-\hat{\pmb{\beta}}\hat{\pmb{\beta}})+\pmb{\epsilon}_{EB}(\hat{\pmb{\varepsilon}}\hat{\pmb{\beta}}+\hat{\pmb{\beta}}\hat{\pmb{\varepsilon}})$, $z=(u/\chi)^{2/3}$, $\chi$ is the quantum parameter. $\hat{\pmb{\varepsilon}}$ and $\hat{\pmb{\beta}}$ are unit vectors (anti)parallel to the local electric and magnetic fields. 

We can rearrange the polarization-resolved NLC spectrum so that the expected spin and polarization state after one NLC event can be easily observed:

\begin{equation}
F = F_0 + \textbf{n}_\gamma \cdot \textbf{F} = F_0 + \xi_1F_1 + \xi_2F_2 + \xi_3F_3,
\label{eq:spin_pol_spectrum}
\end{equation}

which $\textbf{F} = \pmb{\epsilon}_{EB} F_1+\pmb{\epsilon}_{2} F_2+\pmb{\epsilon}_{E} F_3$. We start by solving the lepton spin term $F_0$. Note here, $ \textbf{s}_i =\textbf{n}_0$, $\textbf{s}_f = \textbf{n}_1$. 

\begin{equation}
\begin{split}
F_0  &  = \langle \mathbb{\hat R}^{\text{C}} \rangle + \textbf{s}_i \cdot  \hat{\textbf{R}}_0^{\text{C}} +  \textbf{s}_f \cdot  \hat {\textbf{R}}_1^{\text{C}}  +\textbf{s}_f \cdot  \hat {\textbf{R}}_{01}^{\text{C}} \cdot \textbf{s}_i  \\
& = -\text{Ai}_1(z) - \kappa \frac{\text{Ai}'(z)}{z} + \lambda\frac{\text{Ai}(z)}{\sqrt{z}} ({\textbf{s}}_i \cdot \hat{\pmb{\beta}}) + u\frac{\text{Ai}(z)}{\sqrt{z}} ({\textbf{s}}_f \cdot \hat{\pmb{\beta}}) \\ &
+ {\textbf{s}}_f \cdot \left[ -\text{Ai}_1(z)(\textbf{1}+(\kappa-2)\textbf{1}_{\parallel}) - \frac{\text{Ai}'(z)}{z}(2\textbf{1}+(\kappa-2)\textbf{1}_{\parallel})\right]\cdot {\textbf{s}}_i.
\end{split}
\end{equation}

We use $\textbf{1}_{\perp} = \textbf{1} - \textbf{1}_{\parallel}$ to replace $\textbf{1}_{\perp}$.  We can also express this spectrum in terms of the modified Bessel function: $\text{Ai}_1{(z)} = \frac{1}{\pi\sqrt{3}}\text{IntK}_{1/3}(u')$, $\text{Ai}'{(z)}/z = -\frac{1}{\pi\sqrt{3}}\text{K}_{2/3}(u')$, $\text{Ai}{(z)}/\sqrt{z} = -\frac{1}{\pi\sqrt{3}}\text{K}_{1/3}(u')$, which we have $u' = 2u/3\chi= \frac{2}{3}z^{3/2}$. We then get, ignoring the common constant factor:

 \begin{equation}
 \begin{split}
 F_0 = & -\text{IntK}_{1/3}(u')+\kappa\text{K}_{2/3}(u')+\lambda(\textbf{s}_i\cdot \hat{\pmb{\beta}})\text{K}_{1/3}(u')\\ & +u(\textbf{s}_f\cdot \hat{\pmb{\beta}})\text{K}_{1/3}(u')  + \left(-\text{IntK}_{1/3}(u') + 2\text{K}_{2/3}(u')\right)(\textbf{s}_f\cdot \textbf{s}_i) \\ &  +\lambda u\left(-\text{IntK}_{1/3}(u') + \text{K}_{2/3}(u')\right)(\textbf{s}_f\cdot\hat{\textbf{k}}) (\textbf{s}_i\cdot\hat{\textbf{k}}).
\end{split}
\end{equation}

Rearrange the terms related to the final spin state, and we get:

 \begin{equation}
 \begin{split}
 F_0 = & -\text{IntK}_{1/3}(u')+\kappa\text{K}_{2/3}(u')+\lambda(\textbf{s}_i\cdot \hat{\pmb{\beta}})\text{K}_{1/3}(u')\\ & +\textbf{s}_f\cdot \left(u \text{K}_{1/3}(u') \hat{\pmb{\beta}} + \left(-\text{IntK}_{1/3}(u') + 2\text{K}_{2/3}(u')\right)\textbf{s}_i \right. \\ &  \left. + \lambda u\left(-\text{IntK}_{1/3}(u') + \text{K}_{2/3}(u')\right)(\textbf{s}_i\cdot\hat{\textbf{k}}) \hat{\textbf{k}}\right). 
\end{split}
\label{eq:NLC_stokes_F0}
\end{equation}

Notice that the spin-resolved radiation rate can be written as: 

\begin{equation}
F_0 =w + \textbf{s}_f \cdot \textbf{g}.
\end{equation}

\begin{equation}
    w =-\text{IntK}_{1/3}(u')+\kappa\text{K}_{2/3}(u')+\lambda(\textbf{s}_i\cdot \hat{\pmb{\beta}})\text{K}_{1/3}(u'),
\end{equation}

\begin{equation}
\begin{split}
    \textbf{g} = &u \text{K}_{1/3}(u') \hat{\pmb{\beta}} + \left(-\text{IntK}_{1/3}(u') + 2\text{K}_{2/3}(u')\right)\textbf{s}_i  \\ &  + \lambda u\left(-\text{IntK}_{1/3}(u') + \text{K}_{2/3}(u')\right)(\textbf{s}_i\cdot\hat{\textbf{k}}) \hat{\textbf{k}}
\end{split}
\end{equation}

Which $w$ is the unpolarised, spin-averaged NLCS rate. We can also obtain the expected spin polarisation vector:
\begin{equation}
\langle\pmb{s}\rangle = \frac{\textbf{g}}{w}.
\label{eq:SQA_spin}
\end{equation}

Then, we calculate terms related to photon polarization. We do this by separate terms along $\pmb{\epsilon}_{EB}$, $\pmb{\epsilon}_{2}$, and $\pmb{\epsilon}_{B}$, as $F_1$, $F_2$ and $F_3$. Notice that here, $\textbf{S}  = \pmb{\epsilon}_E\cdot(\hat{\pmb{\varepsilon}}\hat{\pmb{\varepsilon}}-\hat{\pmb{\beta}}\hat{\pmb{\beta}}) +\pmb{\epsilon}_{EB}\cdot(\hat{\pmb{\varepsilon}}\hat{\pmb{\beta}}+\hat{\pmb{\beta}}\hat{\pmb{\varepsilon}})$. We start with the spectrum related to the diagonal polarization component $F_1$:
 \begin{equation}
 \begin{split}
 F_1  &  =  \pmb{\epsilon}_{EB}\cdot\left(\hat{\textbf{R}}_\gamma^{\text{C}} + \hat{\textbf{R}}_{\gamma 0}^{\text{C}}\cdot \textbf{s}_i + \hat{\textbf{R}}_{\gamma 1}^{\text{C}}\cdot \textbf{s}_f + \textbf{s}_{f} \cdot \hat{\textbf{R}}_{\gamma 01}^{\text{C}} \cdot \textbf{s}_{i} \right) \\
 & = -\frac{\text{Ai}(z)}{\sqrt{z}}\left(u(\textbf{s}_i\cdot \hat{\pmb{\varepsilon}}) +\lambda(\textbf{s}_f\cdot \hat{\pmb{\varepsilon}})\right)
 \\ & - \frac{\lambda u}{2}\text{Ai}_1(z)\left((\textbf{s}_f \cdot \hat{\pmb{\varepsilon}})(\textbf{s}_i \cdot \hat{\pmb{\beta}})+(\textbf{s}_f \cdot \hat{\pmb{\beta}})(\textbf{s}_i \cdot \hat{\pmb{\varepsilon}})\right) \\& - \frac{\tilde\kappa}{2}\frac{\text{Ai}'(z)}{z}\varepsilon_{abc}(\textbf{s}_{f})_a(\textbf{s}_{i})_b\hat{\textbf{k}}_c
\\ & = -\frac{\text{Ai}(z)}{\sqrt{z}}\left(u (\textbf{s}_i\cdot \hat{\pmb{\varepsilon}}) +\lambda (\textbf{s}_f\cdot \hat{\pmb{\varepsilon}})\right)
 \\ & - \frac{\lambda u}{2}\text{Ai}_1(z)\left((\textbf{s}_f \cdot \hat{\pmb{\varepsilon}})(\textbf{s}_i \cdot \hat{\pmb{\beta}})+(\textbf{s}_f \cdot \hat{\pmb{\beta}})(\textbf{s}_i \cdot \hat{\pmb{\varepsilon}})\right) \\& - \frac{\tilde\kappa}{2}\frac{\text{Ai}'(z)}{z}(\textbf{s}_{f}\times \textbf{s}_{i})\cdot \hat{\textbf{k}}.
\end{split}
\end{equation}

By replacing the Airy function with the modified Bessel function, ignoring the common constant factor. We get the following:

\begin{equation}
 \begin{split}
F_1  =  &  -\text{K}_{1/3}(u')
 \left(u(\textbf{s}_i\cdot \hat{\pmb{\varepsilon}}) + \lambda(\textbf{s}_f\cdot \hat{\pmb{\varepsilon}})\right) \\ & - \frac{\lambda u}{2}\text{IntK}_{1/3}(u')\left((\textbf{s}_f \cdot \hat{\pmb{\varepsilon}})(\textbf{s}_i \cdot \hat{\pmb{\beta}})+(\textbf{s}_f \cdot \hat{\pmb{\beta}})(\textbf{s}_i \cdot \hat{\pmb{\varepsilon}})\right) \\ & 
+ \frac{\tilde \kappa}{2}\text{K}_{2/3}(u')(\textbf{s}_{f}\times \textbf{s}_{i})\cdot \hat{\textbf{k}}.
\end{split}
\label{eq:NLC_stokes_F1}
\end{equation}

We then look at the term related to circular polarized $F_2$
\begin{equation}
 \begin{split}
 F_2  = &  \ \pmb{\epsilon}_{2}\cdot\left(\hat{\textbf{R}}_\gamma^{\text{C}} + \hat{\textbf{R}}_{\gamma 0}^{\text{C}}\cdot \textbf{s}_i + \hat{\textbf{R}}_{\gamma 1}^{\text{C}}\cdot \textbf{s}_f + \textbf{s}_{f} \cdot \hat{\textbf{R}}_{\gamma 01}^{\text{C}} \cdot \textbf{s}_{i} \right)
 \\ = & -\lambda\left( \text{Ai}_1(z)+\left[1+\frac{1}{1-\lambda}\right]\frac{\text{Ai}'(z)}{z}\right)(\textbf{s}_i\cdot \hat{\textbf{k}}) \\ & -u\left( \text{Ai}_1(z)+\left(2-\lambda\right)\frac{\text{Ai}'(z)}{z}\right)(\textbf{s}_f\cdot \hat{\textbf{k}}) \\ &
 +\frac{\text{Ai}(z)}{\sqrt{z}}\left(u(\textbf{s}_f\cdot \hat{\textbf{k}})(\textbf{s}_i\cdot \hat{\pmb{\beta}})-\lambda(\textbf{s}_f\cdot \hat{\pmb{\beta}})(\textbf{s}_i\cdot \hat{\textbf{k}})\right)
 \\ = & -\lambda\left( \text{Ai}_1(z)+\left[1+\frac{1}{1-\lambda}\right]\frac{\text{Ai}'(z)}{z}\right)(\textbf{s}_i\cdot \hat{\textbf{k}}) \\ & -u\left( \text{Ai}_1(z)+\left(2-\lambda\right)\frac{\text{Ai}'(z)}{z}\right)(\textbf{s}_f\cdot \hat{\textbf{k}}) \\ &
 +\frac{\text{Ai}(z)}{\sqrt{z}}\left(\frac{u}{2} + \frac{\lambda}{2}\right)\left((\textbf{s}_f\cdot \hat{\textbf{k}})(\textbf{s}_i\cdot \hat{\pmb{\beta}}) - (\textbf{s}_f\cdot \hat{\pmb{\beta}})(\textbf{s}_i\cdot \hat{\textbf{k}})\right)
 \\ &
 +\frac{\text{Ai}(z)}{\sqrt{z}}\left(\frac{u}{2} - \frac{\lambda}{2}\right)\left((\textbf{s}_f\cdot \hat{\textbf{k}})(\textbf{s}_i\cdot \hat{\pmb{\beta}}) + (\textbf{s}_f\cdot \hat{\pmb{\beta}})(\textbf{s}_i\cdot \hat{\textbf{k}})\right).
\end{split}
\end{equation}

Using the vector identity $(\textbf{A} \times \textbf{B})\cdot (\textbf{C} \times \textbf{D}) = (\textbf{A} \cdot \textbf{C}) (\textbf{B}\cdot \textbf{D}) - (\textbf{A} \cdot \textbf{D}) (\textbf{B}\cdot \textbf{C})$, also, $\hat{\textbf{k}}\times\hat{\pmb{\beta}} = \hat{\pmb{\varepsilon}}$, replacing the Airy function with the modified Bessel function, we get:
\begin{equation}
 \begin{split}
 F_2 = &\ \frac{\tilde \kappa }{2}\text{K}_{1/3}(u')\left((\textbf{s}_f\times\textbf{s}_i)\cdot\hat{\pmb{\varepsilon}}\right)\\&
 -\left( \lambda\text{IntK}_{1/3}(u')-\tilde \kappa \text{K}_{2/3}(u')\right)(\textbf{s}_i\cdot \hat{\textbf{k}}) 
 \\&
 -\left( u\text{IntK}_{1/3}(u')-\tilde \kappa \text{K}_{2/3}(u')\right)(\textbf{s}_f\cdot \hat{\textbf{k}})  \\&
  +\frac{\lambda u}{2}\text{K}_{1/3}(u')\left((\textbf{s}_f\cdot \hat{\textbf{k}})(\textbf{s}_i\cdot \hat{\pmb{\beta}}) + (\textbf{s}_f\cdot \hat{\pmb{\beta}})(\textbf{s}_i\cdot \hat{\textbf{k}})\right).
 \end{split}
 \label{eq:NLC_stokes_F2}
\end{equation}
Finally, we look at the term related to linear polarized $F_3$

\begin{equation}
 \begin{split}
 F_3  &  = \pmb{\epsilon}_{E}\cdot\left(\hat{\textbf{R}}_\gamma^{\text{C}} + \hat{\textbf{R}}_{\gamma 0}^{\text{C}}\cdot \textbf{s}_i + \hat{\textbf{R}}_{\gamma 1}^{\text{C}}\cdot \textbf{s}_f + \textbf{s}_{f} \cdot \hat{\textbf{R}}_{\gamma 01}^{\text{C}} \cdot \textbf{s}_{i} \right)\\
 & = -\frac{\text{Ai}'(z)}{z}+\frac{\text{Ai}(z)}{\sqrt{z}}\left(\frac{q_1}{s_1} (\textbf{s}_i\cdot \hat{\pmb{\beta}}) +\frac{q_1}{s_0}(\textbf{s}_f\cdot \hat{\pmb{\beta}})\right)
 \\ & - \frac{\lambda u}{2}\text{Ai}_1(z)\left((\textbf{s}_f \cdot \hat{\pmb{\varepsilon}})(\textbf{s}_i \cdot \hat{\pmb{\varepsilon}})-(\textbf{s}_f \cdot \hat{\pmb{\beta}})(\textbf{s}_i \cdot \hat{\pmb{\beta}})\right)\\
 & -  \frac{\kappa}{2}\frac{\text{Ai}'(z)}{z}(\textbf{s}_f\cdot \textbf{s}_i) -  \frac{\lambda u}{2}\frac{\text{Ai}'(z)}{z}(\textbf{s}_f\cdot \hat{\textbf{k}})(\textbf{s}_i\cdot \hat{\textbf{k}}).
\end{split}
\end{equation}

Replace the Airy function with the modified Bessel function, we get:

\begin{equation}
 \begin{split}
 F_3  =  &\ \text{K}_{2/3}(u')+\left(u (\textbf{s}_i\cdot \hat{\pmb{\beta}}) +\lambda(\textbf{s}_f\cdot \hat{\pmb{\beta}})\right)\text{K}_{1/3}(u')
 \\ & - \frac{\lambda u}{2}\text{IntK}_{1/3}(u')\left((\textbf{s}_f \cdot \hat{\pmb{\varepsilon}})(\textbf{s}_i \cdot \hat{\pmb{\varepsilon}})-(\textbf{s}_f \cdot \hat{\pmb{\beta}})(\textbf{s}_i \cdot \hat{\pmb{\beta}})\right) \\
 & + \frac{\kappa}{2}\text{K}_{2/3}(u')(\textbf{s}_f\cdot \textbf{s}_i) - \frac{\lambda u}{2}\text{K}_{2/3}(u')(\textbf{s}_f\cdot \hat{\textbf{k}})(\textbf{s}_i\cdot \hat{\textbf{k}}).
\end{split}
 \label{eq:NLC_stokes_F3}
\end{equation}

We can find that if we replace  $\hat{\pmb{\varepsilon}}$ as $-\textbf{s}$, $\hat{\pmb{\beta}}$ as $-\pmb{\beta}=-\hat{\textbf{v}}\times \textbf{s}$, and $\hat{\textbf{k}}$ as $\hat{\textbf{v}}$, equation \ref{eq:NLC_stokes_F0}, \ref{eq:NLC_stokes_F1}, \ref{eq:NLC_stokes_F2}, \ref{eq:NLC_stokes_F3} will be the same as the equation (13) and (14) in  Chen PRD 2022. Notice that in Chen's equations, $F_0 = dW_{11}+dW_{22}$, $F_1 = dW_{12}+dW_{21}$, $F_2 = i(dW_{12}-dW_{21})$, $F_3 = dW_{11}-dW_{22}$.

One important special case is the spectrum radiated by an unpolarized electron/positron beam. We obtain this by taking $\textbf{s}_i =0 $ and $\textbf{s}_f = 0$ in Eq.~(\ref{eq:spin_pol_spectrum}):

\begin{equation}
 \begin{split}
 F^{unpol}  =  -\text{IntK}_{1/3}(u')+\kappa\text{K}_{2/3}(u') + \xi_3\text{K}_{2/3}(u')
\end{split}
 \label{eq:nlc_spectrum_unpol}
\end{equation}

The radiated photon polarization state under this scenario can only have $\xi_3$ component, which is the linear polarization, with the expected degree of linear polarization:

\begin{equation}
 \begin{split}
\langle\xi_3^{unpol}\rangle  =  \frac{\text{K}_{2/3}(u')}{-\text{IntK}_{1/3}(u')+\kappa\text{K}_{2/3}(u')} 
\end{split}
 \label{eq:NLC_unpol}
\end{equation}

In addition to the photon emission contribution at $O(\alpha)$, we also need the $O(\alpha)$ contribution from the one-loop electron self-energy. More precisely, these contributions are coming from the interference of the one-loop self-energy at $O(\alpha)$ with the “free” propagation of the electron. The expression is given by:

\begin{equation}
R_{0}^{L} = \int d\lambda \alpha\left[\text{Ai}_1(z)+2g\frac{\text{Ai}’(z)}{z}\right]
\end{equation}

\begin{equation}
\pmb{R}_{i}^{L} = \pmb{R}_{f}^{L} = \alpha\int d\lambda \alpha\frac{\text{Ai}(z)}{\sqrt{z}}\hat{B}
\end{equation}

\begin{equation}
{R}_{if}^{L} = {R}_{0}^{L} + \alpha\int d\lambda \alpha\frac{\text{Gi}(z)}{\sqrt{z}}(\hat K\hat E-\hat \varepsilon \hat K)
\end{equation}

where $\text{Gi}$ is the Scorer function.Airy and Scorer functions are the real and imaginary parts of the integral:

\begin{equation}
\text{Ai}(z) + i\text{Gi}(z) = \frac{1}{\pi}\int_0^{\infty}dt e^{i\left(zt+\frac{t^3}{3}\right)}
\end{equation}

The Muller matrix $M^L$ for the loop is constructed analogously to:
\begin{equation}
M^{L} =\begin{pmatrix}
R_{0}^{L} & \pmb{R}_{f}^{L}\\
\pmb{R}_{i}^{L}& {R}_{if}^{L}
\end{pmatrix}
\end{equation}

\subsection{NBW Pair Production Process}
\label{appendix_NBW}
The nonlinear Breit-Wheeler pair production rate can be written as:

\begin{equation}
    \{ \langle \mathbb{P}^{\text{BW}}\rangle, \textbf{P}_0^{\text{BW}},\cdots\}=\frac{\alpha}{4}\int\frac{d\sigma}{b_0}\{ \langle \mathbb{\hat R}^{\text{BW}}\rangle, \textbf{R}_0^{\text{BW}},\cdots\},
\end{equation}

According to Torgrimsson \cite{Torgrimsson_NJP_2021} equation 43-50, this gives the polarization-resolved emission spectrum:

\begin{equation}
\langle \hat {\mathbb{R}}^{\text{BW}} \rangle  = \text{Ai}_1(z) - \kappa \frac{\text{Ai}'(z)}{z},
\end{equation}

\begin{equation}
\hat{\textbf{R}}_2^{\text{BW}} = \frac{1}{1-\tilde \lambda} \frac{\text{Ai}(\tilde z)}{\sqrt{\tilde z}}\hat {\pmb{\beta}},
\end{equation}

\begin{equation}
\hat{\textbf{R}}_3^{\text{BW}}   = \frac{1}{\tilde \lambda} \frac{\text{Ai}(\tilde z)}{\sqrt{\tilde z}}\hat {\pmb{\beta}},
\end{equation}

\begin{equation}
\hat{\textbf{R}}_\gamma^{\text{BW}}  = \frac{\text{Ai}'(\tilde z)}{\tilde z}\hat {\pmb{\varepsilon}}\cdot {\textbf{S}} \cdot \hat {\pmb{\varepsilon}},
\end{equation}

\begin{equation}
\hat{\textbf{R}}_{23}^{\text{BW}} = -\text{Ai}_1(\tilde z)(\pmb{1}_{\perp}+[\kappa+1]\pmb{1}_{\parallel}) - \frac{\text{Ai}'(\tilde z)}{\tilde z}(2\pmb{1}_{\perp}+\kappa\pmb{1}_{\parallel}),
\end{equation}

\begin{equation}
\hat{\textbf{R}}_{\gamma 2}^{\text{BW}} = \frac{1}{\tilde \lambda} \frac{\text{Ai}(\tilde z)}{\sqrt{\tilde z}}\textbf{S}\cdot\hat {\pmb{\beta}}+\frac{1}{1-\tilde \lambda}\left( \text{Ai}_1(\tilde z)+\left[1-\frac{1-\tilde \lambda}{\tilde \lambda}\right]\frac{\text{Ai}'(\tilde z)}{\tilde z}\right)\pmb{\epsilon}_2\hat{\textbf{k}},
\end{equation}

\begin{equation}
\hat{\textbf{R}}_{\gamma 3}^{\text{BW}} = \frac{1}{1-\tilde \lambda} \frac{\text{Ai}(\tilde z)}{\sqrt{\tilde z}}\textbf{S}\cdot\hat {\pmb{\beta}}-\frac{1}{\tilde \lambda}\left( \text{Ai}_1(\tilde z)+\left[1-\frac{\tilde \lambda}{1-\tilde \lambda}\right]\frac{\text{Ai}'(\tilde z)}{\tilde z}\right)\pmb{\epsilon}_2\hat{\textbf{k}},
\end{equation}

\begin{equation}
\begin{split}
\hat{\textbf{R}}_{\gamma 23}^{\text{BW}} = &\ \frac{\tilde u}{2}\text{Ai}_1(\tilde z)\textbf{S} +\frac{\text{Ai}'(\tilde z)}{z}\hat{\pmb{\varepsilon}}\cdot\textbf{S}\cdot\hat{\pmb{\varepsilon}}\left(\frac{\rho}{2}\pmb{1}_{\perp}+\pmb{1}_{\parallel}\right)\\
& +\frac{\tilde\rho}{2}\frac{\text{Ai}'(\tilde z)}{\tilde z}\hat{\pmb{\beta}}\cdot\textbf{S}\cdot\hat{\pmb{\varepsilon}}\  i\pmb{\sigma}^{(3)}_2  + \frac{\text{Ai}(\tilde z)}{\sqrt{\tilde z}}\pmb{\epsilon}_2\left(\frac{\hat{\textbf{k}}\hat{\pmb{\beta}}}{\tilde \lambda}-\frac{\hat{\pmb{\beta}}\hat{\textbf{k}}}{1-\tilde \lambda}\right).
\end{split}
\end{equation}

Here, $\tilde\lambda$ are the momentum ratios between the positron and the initial photon. $\rho =  (1-\tilde\lambda)/\tilde\lambda+\tilde\lambda/(1-\tilde\lambda) $, $\tilde \rho = (1-\tilde\lambda)/\tilde\lambda -\tilde\lambda/(1-\tilde\lambda)$, and $\tilde z=(\tilde u/\chi)^{2/3}$, with $\tilde u = 1/\tilde\lambda+1/(1-\tilde\lambda)$. 

Following the way the as how we rearrange the NLC spectrum, we can also rewrite the polarization-resolved NBW spectrum as:

\begin{equation}
G = G_0 + \textbf{n}_\gamma \cdot \textbf{G} = G_0 + \xi_1G_1 + \xi_2G_2 + \xi_3G_3,
\label{eq:NBW_spectrum_total}
\end{equation}

which $\textbf{G} = \pmb{\epsilon}_{EB} G_1+\pmb{\epsilon}_{2} G_2+\pmb{\epsilon}_{E} G_3$. We start by solving the Lepton spin term $G_0$. Note here, $ \textbf{s}_e =\textbf{n}_2$, $\textbf{s}_p = -\textbf{n}_3$. 

\begin{equation}
\begin{split}
G_0  &  = \langle \mathbb{\hat R}^{\text{BW}} \rangle + \textbf{s}_e \cdot  \hat{\textbf{R}}_2^{\text{BW}} -  \textbf{s}_p \cdot  \hat {\textbf{R}}_3^{\text{BW}}  -\textbf{s}_p \cdot  \hat {\textbf{R}}_{23}^{\text{BW}} \cdot \textbf{s}_e  \\
& = \text{Ai}_1(\tilde z) - \rho \frac{\text{Ai}'(\tilde z)}{\tilde z} + \frac{1}{1-\tilde \lambda}\frac{\text{Ai}(\tilde z)}{\sqrt{\tilde z}} ({\textbf{s}}_e \cdot \hat{\pmb{\beta}}) - \frac{1}{\tilde \lambda}\frac{\text{Ai}(\tilde z)}{\sqrt{\tilde z}} ({\textbf{s}}_p \cdot \hat{\pmb{\beta}}) \\ &
+ {\textbf{s}}_p \cdot \left[ \text{Ai}_1(\tilde z)(\textbf{1}+\kappa\textbf{1}_{\parallel}) +\frac{\text{Ai}'(\tilde z)}{\tilde z}(2\textbf{1}+(\rho-2)\textbf{1}_{\parallel})\right]\cdot {\textbf{s}}_e.
\end{split}
\end{equation}

Similar as the NLCS spectrum, we can also replace the Airy function with the modified Bessel function: $\text{Ai}_1{(\tilde z)} = \frac{1}{\pi\sqrt{3}}\text{IntK}_{1/3}(\tilde u')$, $\text{Ai}'{(\tilde z)}/\tilde z = -\frac{1}{\pi\sqrt{3}}\text{K}_{2/3}(\tilde u')$, $\text{Ai}{(\tilde z)}/\sqrt{\tilde z} = -\frac{1}{\pi\sqrt{3}}\text{K}_{1/3}(\tilde u')$, which $\tilde u' = \frac{2}{3\chi_{\gamma}}\tilde u= \frac{2}{3}{\tilde z}^{3/2}$. We then get, ignoring the common constant factor:
 
 \begin{equation}
 \begin{split}
 G_0 = &\ \text{IntK}_{1/3}(\tilde u')+\rho\text{K}_{2/3}(\tilde u')+\frac{1}{1-\tilde\lambda}(\textbf{s}_e\cdot \hat{\pmb{\beta}})\text{K}_{1/3}(\tilde u')\\ & -\frac{1}{\tilde \lambda}(\textbf{s}_p\cdot \hat{\pmb{\beta}})\text{K}_{1/3}(\tilde u')  + \left(\text{IntK}_{1/3}(\tilde u') - 2\text{K}_{2/3}(\tilde u')\right)(\textbf{s}_p\cdot \textbf{s}_e) \\ &  +\left(\rho\text{IntK}_{1/3}(\tilde u') - (\rho-2)\text{K}_{2/3}(\tilde u')\right)(\textbf{s}_p\cdot\hat{\textbf{k}}) (\textbf{s}_e\cdot\hat{\textbf{k}}).
\end{split}
\label{eq:NBW_G0}
\end{equation}

Then, we calculate terms related to photon polarization. We do this by separate terms along $\pmb{\epsilon}_{EB}$, $\pmb{\epsilon}_{2}$, and $\pmb{\epsilon}_{B}$, as $G_1$, $G_2$ and $G_3$. We start with the spectrum related to diagonal polarization component $G_1$:

 \begin{equation}
 \begin{split}
 G_1  &  =  \pmb{\epsilon}_{EB}\cdot\left(\hat{\textbf{R}}_\gamma^{\text{BW}} + \hat{\textbf{R}}_{\gamma 2}^{\text{BW}}\cdot \textbf{s}_e - \hat{\textbf{R}}_{\gamma 3}^{\text{BW}}\cdot \textbf{s}_p - \textbf{s}_{p} \cdot \hat{\textbf{R}}_{\gamma 23}^{\text{BW}} \cdot \textbf{s}_{e} \right) \\
 & = \frac{\text{Ai}(\tilde z)}{\sqrt{\tilde z}}\left(\frac{1}{\tilde \lambda} (\textbf{s}_e\cdot \hat{\pmb{\varepsilon}}) -\frac{1}{1-\tilde \lambda}(\textbf{s}_p\cdot \hat{\pmb{\varepsilon}})\right)
 \\ & - \tilde u\text{Ai}_1(\tilde z)\left((\textbf{s}_p \cdot \hat{\pmb{\varepsilon}})(\textbf{s}_e \cdot \hat{\pmb{\beta}})+(\textbf{s}_p \cdot \hat{\pmb{\beta}})(\textbf{s}_e \cdot \hat{\pmb{\varepsilon}})\right) \\& - \frac{\tilde\rho}{2}\frac{\text{Ai}'(\tilde z)}{\tilde z}\varepsilon_{abc}(\textbf{s}_{p})_a(\textbf{s}_{e})_b\hat{\textbf{k}}_c
\\ & = \frac{\text{Ai}(\tilde z)}{\sqrt{\tilde z}}\left(\frac{1}{\tilde \lambda} (\textbf{s}_e\cdot \hat{\pmb{\varepsilon}}) -\frac{1}{1-\tilde \lambda}(\textbf{s}_p\cdot \hat{\pmb{\varepsilon}})\right)
 \\ & - \tilde u\text{Ai}_1(\tilde z)\left((\textbf{s}_p \cdot \hat{\pmb{\varepsilon}})(\textbf{s}_e \cdot \hat{\pmb{\beta}})+(\textbf{s}_p \cdot \hat{\pmb{\beta}})(\textbf{s}_e \cdot \hat{\pmb{\varepsilon}})\right) \\& - \frac{\tilde\rho}{2}\frac{\text{Ai}'(\tilde z)}{\tilde z}(\textbf{s}_{p}\times \textbf{s}_{e})\cdot \hat{\textbf{k}}.
\end{split}
\end{equation}

Apply the notation used in the other group and replace the Airy function with the modified Bessel function, ignoring the common constant factor. We get:

\begin{equation}
 \begin{split}
G_1  =  &\  \text{K}_{1/3}(\tilde u')
 \left(\frac{1}{\tilde \lambda} (\textbf{s}_e\cdot \hat{\pmb{\varepsilon}}) - \frac{1}{1-\tilde \lambda}(\textbf{s}_p\cdot \hat{\pmb{\varepsilon}})\right) \\ & - \frac{\tilde u}{2}\text{IntK}_{1/3}(\tilde u')\left((\textbf{s}_p \cdot \hat{\pmb{\varepsilon}})(\textbf{s}_e \cdot \hat{\pmb{\beta}})+(\textbf{s}_p \cdot \hat{\pmb{\beta}})(\textbf{s}_e \cdot \hat{\pmb{\varepsilon}})\right) \\ & 
+ \frac{\tilde \rho}{2}\text{K}_{2/3}(\tilde u')(\textbf{s}_{p}\times \textbf{s}_{e})\cdot \hat{\textbf{k}}.
\end{split}
\label{eq:NBW_G1}
\end{equation}
We then look at the term related to circular polarized $G_2$

\begin{equation}
 \begin{split}
 G_2  = &  \ \pmb{\epsilon}_{2}\cdot\left(\hat{\textbf{R}}_\gamma^{\text{BW}} + \hat{\textbf{R}}_{\gamma 2}^{\text{BW}}\cdot \textbf{s}_e - \hat{\textbf{R}}_{\gamma 3}^{\text{BW}}\cdot \textbf{s}_p - \textbf{s}_{p} \cdot \hat{\textbf{R}}_{\gamma 23}^{\text{BW}} \cdot \textbf{s}_{e} \right)
 \\ = &\ \frac{1}{1-\tilde \lambda}\left( \text{Ai}_1(\tilde z)+\left[1-\frac{1-\tilde \lambda}{\tilde \lambda}\right]\frac{\text{Ai}'(\tilde z)}{\tilde z}\right)(\textbf{s}_e\cdot \hat{\textbf{k}}) \\ & +\frac{1}{\tilde \lambda}\left( \text{Ai}_1(\tilde z)+\left[1-\frac{\tilde \lambda}{1-\tilde \lambda}\right]\frac{\text{Ai}'(\tilde z)}{\tilde z}\right)(\textbf{s}_p\cdot \hat{\textbf{k}}) \\ &
 -\frac{\text{Ai}(\tilde z)}{\sqrt{\tilde z}}\left(\frac{1}{\tilde \lambda}(\textbf{s}_p\cdot \hat{\textbf{k}})(\textbf{s}_e\cdot \hat{\pmb{\beta}})-\frac{1}{1-\tilde \lambda}(\textbf{s}_p\cdot \hat{\pmb{\beta}})(\textbf{s}_e\cdot \hat{\textbf{k}})\right)
 \\ = &\ \frac{1}{1-\tilde \lambda}\left( \text{Ai}_1(\tilde z)+\left[1-\frac{1-\tilde \lambda}{\tilde \lambda}\right]\frac{\text{Ai}'(\tilde z)}{\tilde z}\right)(\textbf{s}_e\cdot \hat{\textbf{k}}) \\ & +\frac{1}{\tilde \lambda}\left( \text{Ai}_1(\tilde z)+\left[1-\frac{\tilde \lambda}{1-\tilde \lambda}\right]\frac{\text{Ai}'(\tilde z)}{\tilde z}\right)(\textbf{s}_p\cdot \hat{\textbf{k}}) \\ &
 -\frac{\text{Ai}(\tilde z)}{\sqrt{\tilde z}}\left(\frac{1}{2\tilde \lambda} + \frac{1}{2(1-\tilde \lambda)}\right)\left((\textbf{s}_p\cdot \hat{\textbf{k}})(\textbf{s}_e\cdot \hat{\pmb{\beta}}) - (\textbf{s}_p\cdot \hat{\pmb{\beta}})(\textbf{s}_e\cdot \hat{\textbf{k}})\right)
 \\ &
 -\frac{\text{Ai}(\tilde z)}{\sqrt{\tilde z}}\left(\frac{1}{2\tilde \lambda} - \frac{1}{2(1-\tilde \lambda)}\right)\left((\textbf{s}_p\cdot \hat{\textbf{k}})(\textbf{s}_e\cdot \hat{\pmb{\beta}}) + (\textbf{s}_p\cdot \hat{\pmb{\beta}})(\textbf{s}_e\cdot \hat{\textbf{k}})\right).
\end{split}
\end{equation}
Using the vector identity $(\textbf{A} \times \pmb{\beta})\cdot (\textbf{C} \times \textbf{D}) = (\textbf{A} \cdot \textbf{C}) (\pmb{\beta} \cdot \textbf{D}) - (\textbf{A} \cdot \textbf{D}) (\pmb{\beta} \cdot \textbf{C})$, also, $\hat{\textbf{k}}\times\hat{\pmb{\beta}} = \hat{\pmb{\varepsilon}}$ (laser propagates in $-\hat{\textbf{k}}$ direction), replacing the Airy function with the modified Bessel function, we get:
\begin{equation}
 \begin{split}
 G_2 = &-\frac{\tilde u}{2}\text{K}_{1/3}(\tilde u')\left((\textbf{s}_p\times\textbf{s}_e)\cdot\hat{\pmb{\varepsilon}}\right)\\&
 +\left( \frac{1}{1-\tilde\lambda}\text{IntK}_{1/3}(\tilde u')+\tilde\rho\text{K}_{2/3}(\tilde u')\right)(\textbf{s}_e\cdot \hat{\textbf{k}}) 
 \\&
 +\left( \frac{1}{\tilde\lambda}\text{IntK}_{1/3}(\tilde u')-\tilde\rho\text{K}_{2/3}(\tilde u')\right)(\textbf{s}_p\cdot \hat{\textbf{k}})  \\&
  -\frac{\tilde\rho}{2}\text{K}_{1/3}(\tilde u')\left((\textbf{s}_p\cdot \hat{\textbf{k}})(\textbf{s}_e\cdot \hat{\pmb{\beta}}) + (\textbf{s}_p\cdot \hat{\pmb{\beta}})(\textbf{s}_e\cdot \hat{\textbf{k}})\right).
 \end{split}
 \label{eq:NBW_G2}
\end{equation}

Finally, we look at the term related to linear polarized $G_3$

\begin{equation}
 \begin{split}
 G_3  &  = \pmb{\epsilon}_{E}\cdot\left(\hat{\textbf{R}}_\gamma^{\text{C}} + \hat{\textbf{R}}_{\gamma 2}^{\text{C}}\cdot \textbf{s}_e - \hat{\textbf{R}}_{\gamma 3}^{\text{C}}\cdot \textbf{s}_p - \textbf{s}_{p} \cdot \hat{\textbf{R}}_{\gamma 23}^{\text{C}} \cdot \textbf{s}_{e} \right)\\
 & = \frac{\text{Ai}'(\tilde z)}{\tilde z}+\frac{\text{Ai}(\tilde z)}{\sqrt{\tilde z}}\left(-\frac{1}{\tilde \lambda} (\textbf{s}_e\cdot \hat{\pmb{\beta}}) +\frac{1}{1-\tilde \lambda}(\textbf{s}_p\cdot \hat{\pmb{\beta}})\right)
 \\ & - \frac{\tilde u}{2}\text{Ai}_1(\tilde z)\left((\textbf{s}_p \cdot \hat{\pmb{\varepsilon}})(\textbf{s}_e \cdot \hat{\pmb{\varepsilon}})-(\textbf{s}_p \cdot \hat{\pmb{\beta}})(\textbf{s}_e \cdot \hat{\pmb{\beta}})\right)\\
 & -  \frac{\rho}{2}\frac{\text{Ai}'(\tilde z)}{\tilde z}(\textbf{s}_p\cdot \textbf{s}_e) -  \left(1-\frac{\rho}{2}
 \right)\frac{\text{Ai}'(\tilde z)}{\tilde z}(\textbf{s}_p\cdot \hat{\textbf{k}})(\textbf{s}_e\cdot \hat{\textbf{k}}).
\end{split}
\end{equation}
Replace the Airy function with the modified Bessel function, and we get:

\begin{equation}
 \begin{split}
 G_3  =  &-\text{K}_{2/3}(\tilde u')+\left(-\frac{1}{\tilde \lambda} (\textbf{s}_e\cdot \hat{\pmb{\beta}}) +\frac{1}{1-\tilde \lambda}(\textbf{s}_p\cdot \hat{\pmb{\beta}})\right)\text{K}_{1/3}(\tilde u')
 \\ & - \frac{\tilde u}{2}\text{IntK}_{1/3}(\tilde u')\left((\textbf{s}_p \cdot \hat{\pmb{\varepsilon}})(\textbf{s}_e \cdot \hat{\pmb{\varepsilon}})-(\textbf{s}_p \cdot \hat{\pmb{\beta}})(\textbf{s}_e \cdot \hat{\pmb{\beta}})\right) \\
 & + \frac{\rho}{2}\text{K}_{2/3}(\tilde u')(\textbf{s}_p\cdot \textbf{s}_e) - \left(1-\frac{\rho}{2}\right)\text{K}_{2/3}(\tilde u')(\textbf{s}_p\cdot \hat{\textbf{k}})(\textbf{s}_e\cdot \hat{\textbf{k}}).
\end{split}
\label{eq:NBW_G3}
\end{equation}
We can find that if we replace  $\hat{\pmb{\varepsilon}}$ as $-\textbf{s}$, $\hat{\pmb{\beta}}$ as $-\pmb{\beta}=-\hat{\textbf{v}}\times \textbf{s}$, and $\hat{\textbf{k}}$ as $\hat{\textbf{v}}$, equation \ref{eq:NBW_G0}, \ref{eq:NBW_G1}, \ref{eq:NBW_G2}, \ref{eq:NBW_G3} will be the same as the equation (23) in  Chen \cite{Chen_PRD_2022}.

The expected spin direction for the generated electron can be obtained by averaging over the spin state for the positron in Eq.~(\ref{eq:NBW_spectrum_total}):

\begin{equation}
 \begin{split}
 G = &\ \text{IntK}_{1/3}(\tilde u')+\rho\text{K}_{2/3}(\tilde u') - \xi_3\text{K}_{2/3}(\tilde u') \\ & 
 + \textbf{s}_e\cdot\left[\left(\frac{1}{1-\tilde \lambda}-\xi_3\frac{1}{\tilde\lambda}\right)\text{K}_{1/3}(\tilde u')\hat{\pmb{\beta}}+\xi_1\frac{1}{\tilde \lambda}\text{K}_{1/3}(\tilde u')\hat{\pmb{\varepsilon}} \right. \\ & \left.
 +\xi_2\left( \frac{1}{1-\tilde\lambda}\text{IntK}_{1/3}(\tilde u')+\tilde\rho\text{K}_{2/3}(\tilde u')\right)\hat{\pmb{k}}\right]
\end{split}
\label{eq:electron_epect_NBW}
\end{equation}

We can rewrite the spectrum as:
\begin{equation}
G = w_0 + \pmb{j}_e
\end{equation}

which:
\begin{equation}
w_0 =  \text{IntK}_{1/3}(\tilde u')+\rho\text{K}_{2/3}(\tilde u') - \xi_3\text{K}_{2/3}(\tilde u')
\end{equation}

\begin{equation}
\begin{split}
\pmb{j}_e = & \left(\frac{1}{1-\tilde \lambda}-\xi_3\frac{1}{\tilde\lambda}\right)\text{K}_{1/3}(\tilde u')\hat{\pmb{\beta}}+\xi_1\frac{1}{\tilde \lambda}\text{K}_{1/3}(\tilde u')\hat{\pmb{\varepsilon}} \\ &
 +\xi_2\left( \frac{1}{1-\tilde\lambda}\text{IntK}_{1/3}(\tilde u')+\tilde\rho\text{K}_{2/3}(\tilde u')\right)\hat{\pmb{k}}
 \end{split}
\end{equation}

The expected spin vector for the generated electron is then:

\begin{equation}
\langle\textbf{s}_e\rangle = \frac{\pmb{j}_e}{w_0}
\end{equation}

Similarly, we can obtain the expected spin vector for the generated positron. 

\begin{equation}
G = w_0 + \pmb{j}_p
\end{equation}

which:
\begin{equation}
w_0 =  \text{IntK}_{1/3}(\tilde u')+\rho\text{K}_{2/3}(\tilde u') - \xi_3\text{K}_{2/3}(\tilde u')
\end{equation}

\begin{equation}
\begin{split}
\pmb{j}_p = & \left(-\frac{1}{\tilde \lambda}+\xi_3\frac{1}{1-\tilde\lambda}\right)\text{K}_{1/3}(\tilde u')\hat{\pmb{\beta}}-\xi_1\frac{1}{1-\tilde \lambda}\text{K}_{1/3}(\tilde u')\hat{\pmb{\varepsilon}} \\ &
 +\xi_2\left( \frac{1}{\tilde\lambda}\text{IntK}_{1/3}(\tilde u')-\tilde\rho\text{K}_{2/3}(\tilde u')\right)\hat{\pmb{k}}
 \end{split}
\end{equation}

The expected spin vector for the generated electron is then:

\begin{equation}
\langle\textbf{s}_p\rangle = \frac{\pmb{j}_p}{w_0}
\end{equation}

One important special case here is the dependence of the NBW spectrum on photon polarization only, ignoring the dependence on the generated electron and positron spins. We obtain this by setting $\textbf{s}_e = 0$ and $\textbf{s}_p = 0$:

\begin{equation}
G^{unpol} = \text{IntK}_{1/3}(\tilde u')+\rho\text{K}_{2/3}(\tilde u') - \xi_3\text{K}_{2/3}(\tilde u')
\label{eq:nbw_spectrum_unpol}
\end{equation}

And the corresponding NBW rate:

\begin{equation}
R^{unpol} = C_0 \int_0^1 d\tilde\lambda\ \frac{\alpha m_e^2c^4}{\sqrt{3}\pi\hbar^2\omega^2}\left[ \text{IntK}_{1/3}(\tilde u')+\rho\text{K}_{2/3}(\tilde u') - \xi_3\text{K}_{2/3}(\tilde u')\right]
\label{eq:nbw_rate_unpol}
\end{equation}

We can see that the NBW spectrum depends only on the linear degree of polarization $\xi_3$. The pair production yield from the NBW process only depends on $\xi_3$.











\bibliographystyle{unsrt} 
\bibliography{cas-refs}
 
\end{document}